\documentclass[11pt]{article}

\usepackage{epsfig}
\usepackage{amssymb}
\usepackage{amsmath}
\usepackage{amsfonts}
\usepackage{feynmp}
\usepackage{url}
\usepackage{hyperref}
\def\wup{{W^+}}

\def\half{\hbox{${1\over 2}$}}
\newcommand{\be}{\begin{equation}}
\newcommand{\ee}{\end{equation}}
\newcommand{\beq}{\begin{equation}}
\newcommand{\eeq}{\end{equation}}
\newcommand{\bea}{\begin{eqnarray}}
\newcommand{\eea}{\end{eqnarray}}

\newcommand{\lc}{\left[}
\newcommand{\rc}{\right]}

\newcommand{\xuno}{\frac{{\omega_{11}^{(2)}}}
{1+e^{{\theta_1^{(2)}}-x{\omega_{11}^{(1)}}}}}
\newcommand{\xdue}{\frac{{\omega_{12}^{(2)}}}
{1+e^{{\theta_2^{(2)}}-x{\omega_{21}^{(1)}}}}}
\def\epm#1#2{\hbox{${\lower1pt\hbox{$\scriptstyle +#1$}}
\atop {\raise1pt\hbox{$\scriptstyle -#2$}}$}}
\newcommand{\as}{\alpha_s}

\def\tozero#1{\mathrel{\mathop{\sim}\limits_{\scriptscriptstyle
{#1\rightarrow0 }}}}
\def\toone#1{\mathrel{\mathop{\sim}\limits_{\scriptscriptstyle
{#1\rightarrow1 }}}}
\def\frac#1#2{{{#1}\over {#2}}}
\def\gsim{\mathrel{\rlap{\lower4pt\hbox{\hskip1pt$\sim$}}
    \raise1pt\hbox{$>$}}}         
\def\lsim{\mathrel{\rlap{\lower4pt\hbox{\hskip1pt$\sim$}}
    \raise1pt\hbox{$<$}}}         

\newcommand{\draft}[1]{}

\def\Aslash{\not{\hbox{\kern-4pt $A$}}}
\def\Eslash{\not{\hbox{\kern-4pt $E$}}}

\newcommand{\bi}{\begin{itemize}}
\newcommand{\ei}{\end{itemize}}
\newcommand{\ben}{\begin{enumerate}}
\newcommand{\een}{\end{enumerate}}
\def\href#1{(\ref{#1})}
\def \as {\alpha_s}
\def\gsim{\mathrel{\rlap{\lower4pt\hbox{\hskip1pt$\sim$}}
    \raise1pt\hbox{$>$}}}         
\def\lsim{\mathrel{\rlap{\lower4pt\hbox{\hskip1pt$\sim$}}
    \raise1pt\hbox{$<$}}}         

\begin{document}

\pagestyle{empty}

\begin{flushright}

IFUM-967-FT
\end{flushright}

\begin{center}
\vspace*{0.5cm}
{\Large \bf Parton distributions at the dawn of the LHC}

\vspace*{1.cm}
Stefano~Forte
\\
\vspace{1.cm}  {\it
Dipartimento di Fisica, Universit\`a di Milano and
INFN, Sezione di Milano,\\
Via Celoria 16, I-20133 Milano, Italy}\\
\vspace*{1.5cm}

{\bf Abstract}
\end{center}
\bigskip
We review basic ideas and recent developments on the determination of
the parton substructure of the nucleon, in view of applications to
precision hadron collider physics. We review the way information on
PDFs is extracted from the data exploiting QCD factorization, and
discuss the current main two
approaches to parton determination (Hessian and Monte Carlo) and their
use in conjunction with different kinds of parton parametrization. We
summarize the way different physical processes can be used to
constrain different aspects of PDFs.
We discuss the meaning, determination and use of
parton uncertainties. We briefly summarize the current state of the art 
on PDFs for LHC physics.

\noindent

\vfill
\begin{center}
{\it To the memory of Wu-Ki Tung}
\end{center}
\vfill
\noindent

\begin{flushleft} November 2010 \end{flushleft}
\eject

\setcounter{page}{1} \pagestyle{plain}

\eject

\section{QCD in the LHC era}
\label{sec:intro}
The theory and phenomenology of the strong
interactions~\cite{Altarelli:2008zz} 
have witnessed
an impressive development in the last two decades, driven first by the
availability of HERA~\cite{Klein:2008di} --- a QCD machine --- and
then by the needs of present (Tevatron) and especially upcoming (LHC)
hadron colliders~\cite{Mangano:2008ag}. The LHC will be looking for
new physics in hadronic collisions. 

\begin{figure}[htb]
\begin{center}
\includegraphics[width=.8\linewidth]{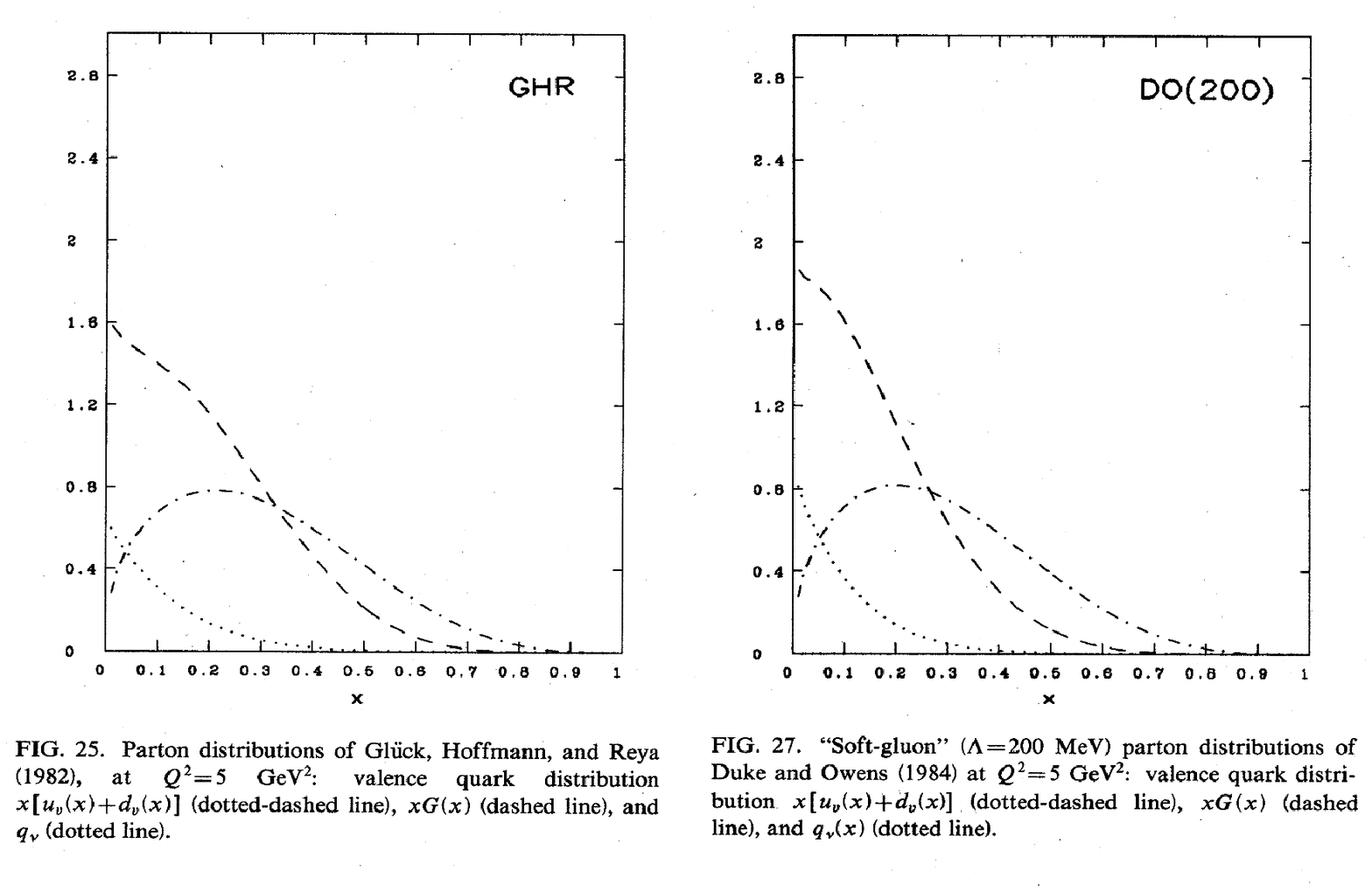}
\caption{Comparison of two parton distribution
  sets~\cite{Gluck:1980cp,Duke:1983gd}  from the early
  eighties (From Ref.~\cite{Eichten:1984eu}).}
\label{oldpdfs}
\end{center}
\end{figure}

The last time this happened was
back in the early eighties, when the $W$ and $Z$ were discovered at
the SPS collider~\cite{WZ} --- and, of course, one may argue to which extent the
$W$ and $Z$ then were genuinely ``new'' physics. At the time, QCD was
at best a semi-quantitative theory: for example, in
Ref.~\cite{Arnison:1985rs} a measured $W$ cross section of
$0.63\pm0.10$~nb (at $\sqrt s=630$~GeV) 
was described as ``in agreement with the theoretical
expectation''~\cite{AEM} of $0.47\epm{0.14}{0.08}$~nb. One reason why
at that time a NLO 
calculation couldn't be expected to agree with the data to better than
20\% is that the knowledge of nucleon structure was at the time
extremely sketchy: a parton set consisted of three parton
distributions (valence, quark sea, and gluon), differences at the 30\%
level between sets would be standard, and, of course, there would be
no idea on the associate uncertainty (see Fig.~\ref{oldpdfs}).

\begin{figure}[htb]
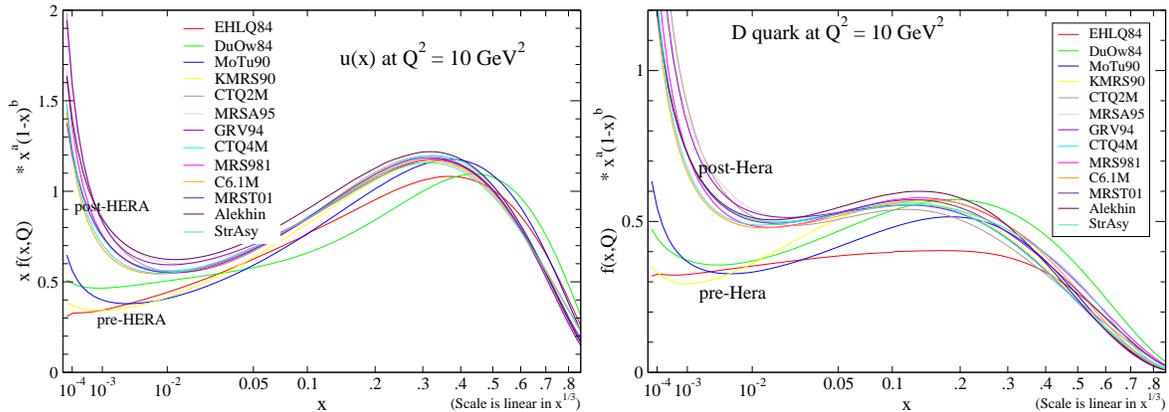

\begin{center}
\includegraphics[width=.45\linewidth,clip]{Uqk1.eps}
\includegraphics[width=.45\linewidth,clip]{Dqk1.eps}
\caption{Historical evolution of the up (left) and down (right) quark
  distributions (from Ref.~\cite{Tung:2004rw}).}
\label{tdeppdfs}
\end{center}
\end{figure}
The evolution in time of parton distributions (see
Fig.~(\ref{tdeppdfs})) since then shows that it is only during the HERA age
that predictions from different groups converged: this is both a
consequence and a cause of the fact that perturbative QCD has now turned into a
quantitative theory, which leads to predictions for hard processes
with typical accuracies below 10\%, and often of a few percent.
Perturbative QCD today is an integral part of the Standard Model, and
it is tested to an accuracy which is comparable to that of the
electroweak sector: in fact, HERA has played for QCD a similar role as
LEP for electroweak theory. In the last decade,  theoretical and
phenomenological progress has been impressive: at the LHC we can
envisage
 quantitative control of
QCD contribution to collider signal and background processes at the
percent level, as will be necessary for discovery at the
LHC~\cite{Mangano:2008ag}.  

Progress in QCD has taken place in (at
least) five distinct directions, namely (listing from the bottom beam
nucleons up to the final state): First, the understanding of the structure of
the nucleon in terms of parton distributions  has now become a 
quantitative science. Second, perturbative computations
are being pushed to  hard processes with 
increasingly high numbers of  particles and at increasingly high
orders, thanks to the development of a variety of techniques which
include twistor methods, analyticity techniques, and the use of exact
results from supersymmetric QCD and the AdS/CFT duality. Third,
all-order resummation of perturbation theory is being extended in various
kinematic regimes (small $x$ and large $x$) to new classes of
observables (typically less inclusive), to higher logarithmic orders,
and it is being accomplished using perturbative renormalization-group
methods, path integral techniques, and effective field theory
methods. Fourth, definitions of jet observables 
which are both consistent with perturbative factorization to all
orders and numerically efficient have been constructed theoretically 
and implemented in computer interfaces. Fifth, new collinear subtraction 
algorithms have been developed which make the development and
implementation of
next-to-leading order Monte Carlo codes possible.

The lectures at the Zakopane school on which this paper is based, ambitiously
entitled ``QCD at the dawn of the LHC'', covered the first three of these
topics: parton distributions (PDFs), perturbative computations, and
resummation. Here we will concentrate on PDFs; recent
good reviews of progress in perturbative computations are in
Refs.~\cite{Berger:2009zb,Weinzierl:2010ps}, while a 
comprehensive overview of resummation is
unfortunately not available yet. 
At Zakopane, jets and Monte Carlos where discussed by other
speakers; excellent recent reviews of these topics are
in Refs.~\cite{Salam:2009jx,Nason:2010ap} respectively.

The purpose of this overview of PDFs is both to provide an
elementary introduction to the subject, and also a summary of recent
developments, several of which are little known outside a small group
of practitioners. Progress in this field has been largely driven by two
series of HERA-LHC workshops 2004-1005 and 2006-2007, 
which have organized and stimulated the
transfer of know-how from deep-inelastic scattering to hadron collider
physics, and whose results are collected in the respective 
reports Refs.~\cite{Alekhin:2005dx,Jung:2009eq}. 
Since 2007, the PDF4LHC working group has been
formed~\cite{deroeck} with a mandate from the CERN directorate to steer and
coordinate research on PDFs for the LHC community: many of the more recent ideas
discussed here were developed in the context
of this working group.

This review is organized as follows. We will start with the more basic
concepts, then work our way to somewhat more advanced developments.
First,
we will very briefly review
some basic (mostly kinematic) facts on QCD factorization. We will then present
the two main existing  approaches (Hessian and Monte Carlo)
to the determination of PDFs and the way they are used in conjunction
with various forms of parton parametrization. Next, we will 
review standard ideas on how
information on PDFs can be extracted from the data. We will then 
discuss in some detail the problem of PDF uncertainties --- what they
mean and how they are determined. In the final section we will briefly
summarize  the state of the art: the role of theoretical
uncertainties, and the current understanding of 
standard candle processes at the LHC.

These lectures are dedicated to the memory of Wu-Ki Tung, who
pioneered this field, pursued it for more than 30 years, and shaped much
of our current understanding of it.

\section{Factorization}
\label{fact}
Factorization of cross sections into hard (partonic) cross sections
and universal parton distributions is the basic property of QCD which
makes it predictive in the perturbative regime, and which enables a
determination of parton distributions. Here we only review some
basic facts which will be useful for our subsequent discussion, while
referring to standard textbooks~\cite{Ellis:1991qj} and recent
reviews~\cite{Altarelli:2008zz} 
for a detailed treatment.

\subsection{Electro- and hadro-production kinematics}
\label{factkin}
The basic factorization for hadroproduction processes has the structure
\bea
\sigma_X(s,M_X^2) &=& \sum_{a,b} \int_{x_{\rm min}}^1 dx_1 \, dx_2 \, f_{a/h_1}(x_1,M_X^2)
f_{b/h_2}(x_2,M_X^2){\hat \sigma_{ab\to X}}\left(
x_1x_2 s,M_X^2\right)\nonumber \\
&=& \sigma^0_{ab}\sum_{a,b} \int_{\tau}^1
\frac{dx_1}{x_1} \int_{\tau/x_1}^1 \frac{dx_2}{x_2} 
  f_{a/h_1}(x_1,M_X^2)
f_{b/h_2}(x_2,M_X^2)C\left(\frac{\tau}{x_1x_2},\alpha_s(M^2_X)\right)\\
&=& \int_{\tau}^1\frac{dx}{x}
{\cal L}\left(x\right)C\left(\frac{\tau}{x},\alpha_s(M^2_X)\right)
,\nonumber
\label{hadrfact}
\eea
where $f_{a/h_i}(x_i)$  is the
distribution of partons of type $a$ in the $i$-th incoming hadron;
${\hat \sigma_{q_aq_b\to X}}$ is the parton-level cross section for
the production of the desired (typically inclusive) final state $X$;
the minimum value of $x_i$ is  $x_{\rm min}= \tau$, with
\be\tau\equiv\frac{M^2_X}{s}\label{taudef}
\ee
the scaling variable of the hadronic process, and in the last step we
defined the parton luminosity
\be
   {\cal L}(x)\equiv \int_{x}^1 \frac{dz}{z} f_{a/h_1}(z,M_X^2)
f_{b/h_2}\left(\frac{x}{z},M_X^2\right)=\int_{x}^1 \frac{dz}{z} f_{a/h_2}(z,M_X^2)
f_{b/h_1}\left(\frac{x}{z},M_X^2\right).
\label{lumdef}
\ee

The hard coefficient function $C\left(\tau,\alpha_s(M^2_X)\right)$ is 
defined by viewing the parton-level cross section as a function of the
hard scale $M_X^2$ and the dimensionless ratio  of this scale to the
center-of-mass energy $\hat s$ of the partonic subprocess:
\beq
\frac{M_X^2}{\hat s}=\frac{\tau}{x_1x_2},
\label{convvar}
\eeq
in terms of the scaling variable.
At the lowest order in the strong interaction, the partonic cross
section is then either zero (for partons that do not couple to the
given final state at leading order), or else just a function fixed by
dimensional analysis  times a
Dirac delta, and the hard coefficient function is thus defined as
\beq 
\hat \sigma_{ab\to X}(s,M^2_X)=\sigma_0
  C_{ab}\left(\tau,\alpha_s(M^2_X)\right);\qquad
C_{ab}\left(x,\alpha_s(M^2_X)\right)=c_{ab}\delta(1-x)+ O(\alpha_s),
\label{cfdef}
\eeq
where $c_{ab}$ is a matrix with non-vanishing entries only between
quark and antiquark states, which will be discussed explicitly in
Sect.~\ref{pdfconstr} below (see in particular Eqs.~(\ref{lody}-\ref{loplum}).
For example, for virtual photon (Drell-Yan) production
$c_{ab}$
is nonzero when $ab$ is a pair of a quark and an antiquark of the same
flavour, and in such case $\sigma_0=\frac{4}{9}\pi\alpha \frac{1}{s}$.

The factorized result Eq.~(\ref{hadrfact}) holds both for inclusive
cross sections and for rapidity distributions.:
\beq
\frac{d\sigma}{dM_X^2 dY}(\tau,Y,M_X^2) =
\sum_{i,j} \int_{x_1^0}^1 dx_1 \int_{x_2^0}^1 dx_2\,
f_i^1(x_1,M_X^2)\, f_j^2(x_2,M_X^2)\,
\frac{d\hat\sigma_{ij}}{dM_X^2 dy}
\left(\frac{\tau}{x_1x_2},y,\as(M_X^2)\right),
\label{rapfact}
\eeq
where the
hadronic cross section is differential with respect to the rapidity
$Y$ of
the final state $X$, while the partonic cross section is differential
in  partonic rapidity 
\beq
y = Y-\frac{1}{2}\ln\frac{x_1}{x_2}:
\label{parrap}
\eeq
the effect of parton emission from the
incoming hadrons is to perform a Lorentz boost from the hadronic
center-of-mass frame to a frame in which the energy of each of the two
incoming hadrons are rescaled by $x_1$ and $x_2$ respectively.
 The lower limit of
integration $x_{\rm min}$ is then fixed by requiring that the rapidity
of the incoming partons be at least sufficient to yield the observed
final state rapidity:
\beq
x_1^0=\sqrt{\tau}e^Y,\qquad
x_2^0=\sqrt{\tau}e^{-Y}.
\label{lowerlim}
\eeq
At leading order, the two partons couple directly to the final state
so $y=0$ and 
\be
Y_{\rm LO}=-\frac{1}{2}\ln\frac{x_1}{x_2}.
\label{lorapkin}
\ee

\begin{figure}[t]\begin{center}
\includegraphics[width=.4\linewidth,clip]{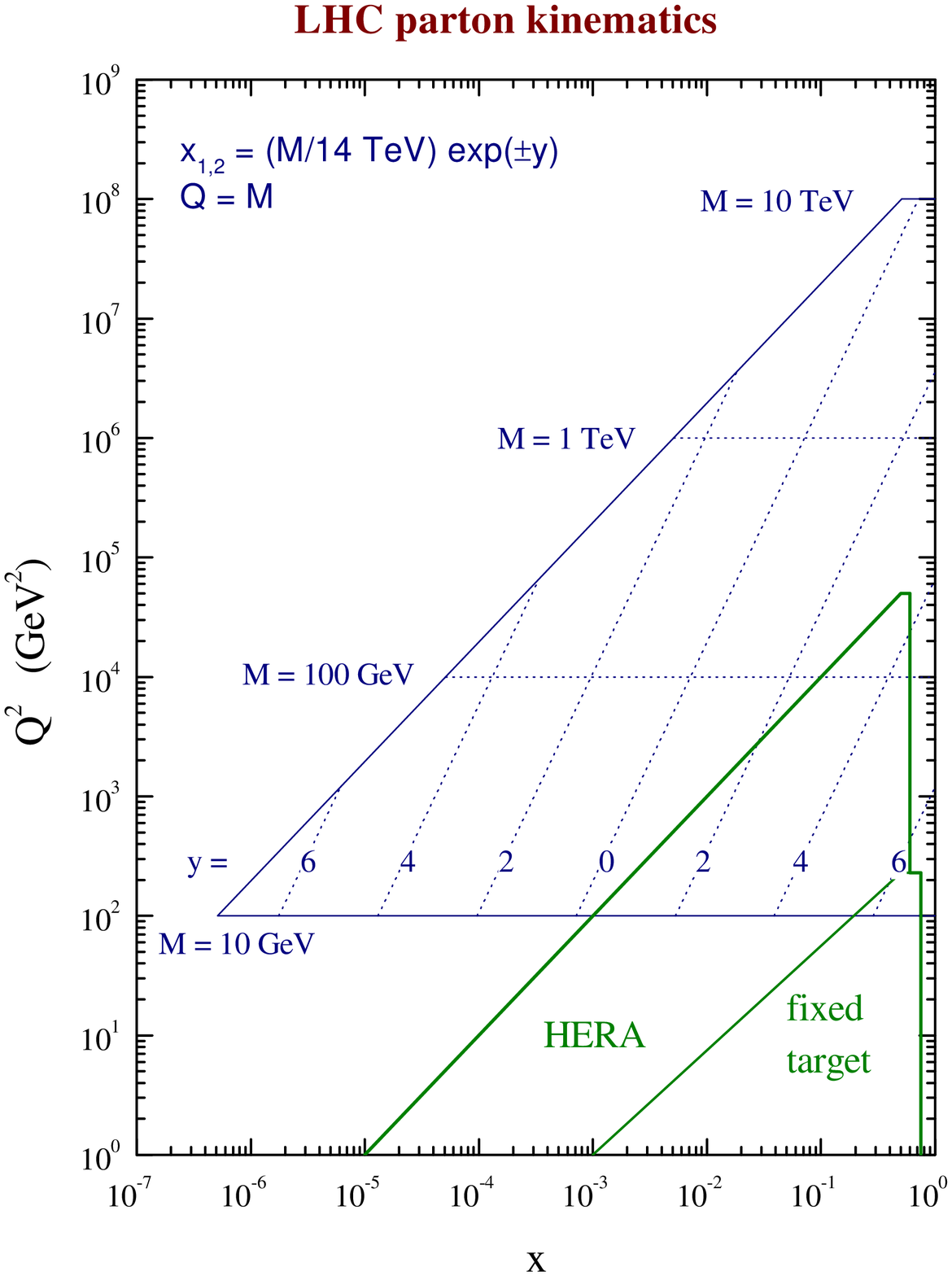}
\includegraphics[width=.4\linewidth,clip]{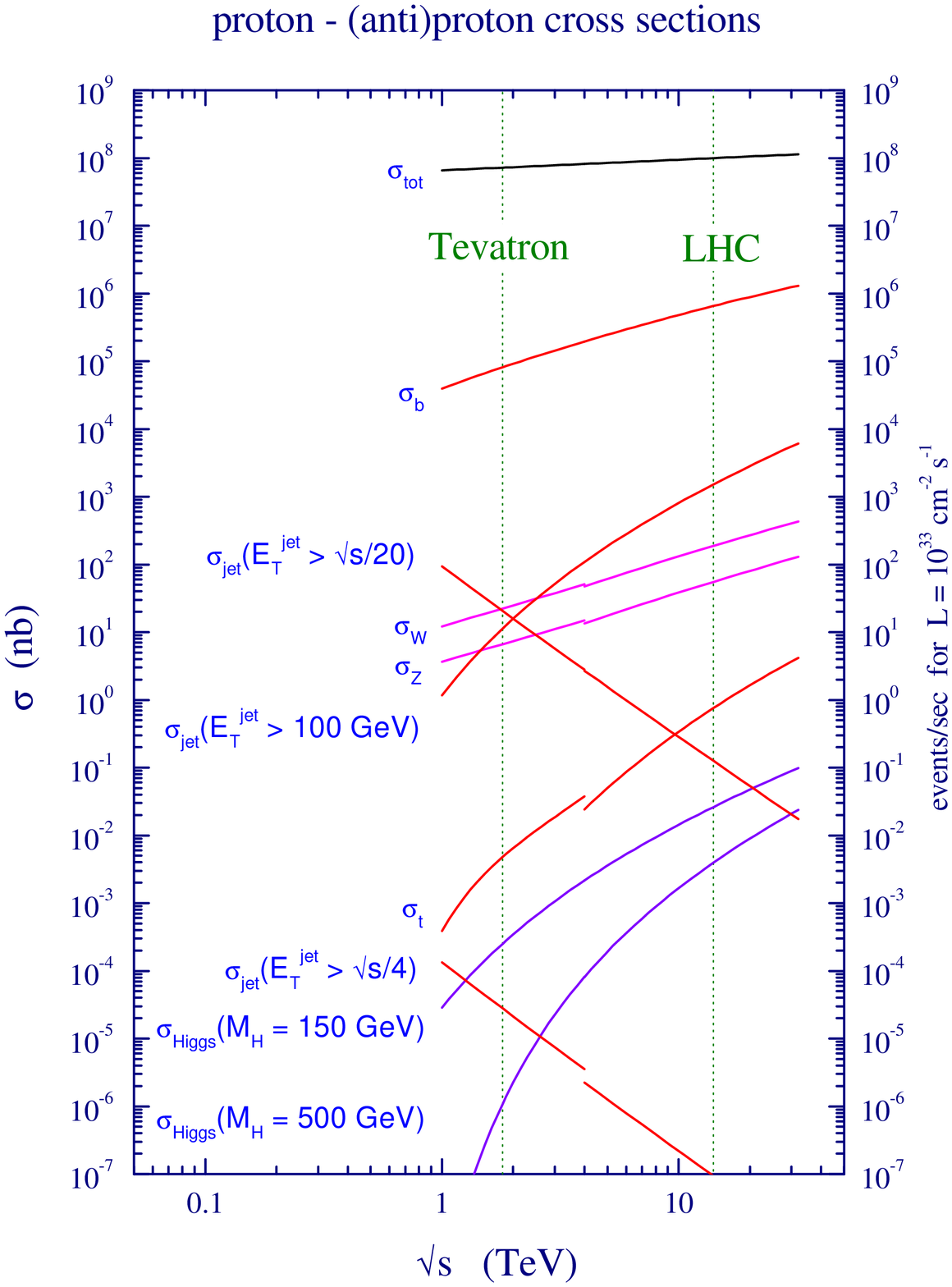}
\caption{LHC kinematics (left) and processes (right) (from
  Ref.~\cite{Alekhin:2005dx}).} 
\label{lhckinxs}
\end{center}
\end{figure}
Equation~(\ref{hadrfact}) is to be contrasted with the standard factorization for
the deep-inelastic structure functions $F_i(x,Q^2)$:
\beq
F_i(x,Q^2)=x\sum_i\int_x^1\frac{dz}{z} C_i\left(\frac{x}{z},
\alpha_s(Q^2)\right)f_i(z,Q^2).
\label{disfac}
\eeq
Here in the argument of the structure function $x=\frac{Q^2}{2p\cdot
  q}$ is the standard Bjorken variable
and the hard coefficient function is the structure function computed
with an incoming parton and $f_i(z, Q^2)$ is the
distribution of the $i$-th parton in
the only incoming hadron. Also in this case at lowest
$O(\alpha_s^0)$ it is either zero (for incoming gluons) or a constant
(an electroweak charge) times a Dirac delta. 

Note that the structure functions are related to the cross section
which is actually measured in lepton-hadron scattering by 
\be
\frac{d^2\sigma^{\rm NC,\ell^{\pm}}}{dxdQ^2} (x,y,Q^2)=\frac{2\pi \alpha^2}{ x Q^4}
 \lc
Y_+ F_2^{NC}(x,Q^2) \mp Y_- x F_3^{NC}(x,Q^2)-y^2 F_L^{NC}(x,Q^2)\rc \ ,
\label{eq:ncxsect}
\ee
for neutral-current charged lepton $\ell^{\pm}$ DIS,  where 
the longitudinal structure function is defined as 
\begin{equation}
F_L(x,Q^2)\equiv F_2(x,Q^2)- 2 xF_1(x,Q^2) ,
\label{eq:fdef}
\end{equation}
and 
\begin{equation}
 Y_{\pm}\equiv 1\pm (1-y)^2,
\label{eq:ypmdef}
\end{equation}
in terms of the electron momentum fraction
\beq
y\equiv\frac{p\cdot q}{p\cdot k}=\frac{Q^2}{x s},
\label{ydef}
\eeq
(not to be confused with the partonic rapidity Eq.~(\ref{parrap}))
where $p$ and $k$ are respectively the incoming proton and lepton
momenta,  $q$ is the virtual photon momentum ($q^2=-Q^2$) and in the
last step, which holds neglecting the proton mass, $s$ is the
center-of-mass energy of the lepton-proton collision.
Similar expressions hold for charged-current charged and neutral
lepton scattering.

The set of values of 
$y$ over which the PDF is probed is of course the same in the hadro-
and leptoproduction cases, and it ranges
from the scaling variable of the hadronic process to one: $x\le y\le 1$
in Eq.~(\ref{disfac}), and  $\tau\le x_1,x_2 \le 1$ in
  Eq.~(\ref{hadrfact}). The kinematic region which is typical of 
of collider
    (HERA) or fixed-target DIS experiments is compared in
Fig.~\ref{lhckinxs} 
to that of LHC
    processes, whose typical cross sections are also shown.

There is  an important kinematic difference when comparing the hadronic and
deep-inelastic factorization formulae, Eqs.~(\ref{hadrfact})
and~(\ref{disfac}) 
respectively. This is related to the fact that the leading order
coefficient function is proportional to a Dirac delta. For DIS, this
implies that at leading order, the value of the structure function at
given $x$ determines  the quark distributions at the same value of
$x$, and it is only at next-to-leading order, where the coefficient
function has a nontrivial dependence on $x$, that the PDF is probed
for all values $x\le y\le 1$. But for hadronic processes, because there
are two partons in the initial state, even at leading order, for
inclusive cross sections the delta
kills one but not both of the convolution integrals in
Eq.~(\ref{hadrfact}), so all values $\tau\le x_i\le 1$ are
probed. However, for rapidity distributions because of the further
kinematic constraint Eq.~(\ref{lorapkin}) the leading order
kinematics is also fixed, and for given $Y$ and $M_X^2$ the momentum
fractions of both partons are fixed.

\subsection{Constraints on PDFs}
\label{pdfconstr}

The kinematics of the factorized expressions Eqs.~(\ref{hadrfact})
and~(\ref{disfac}) immediately implies that, as discussed in
  Sect.~\ref{factkin},  at leading order deep-inelastic structure
  functions and rapidity distributions provide a direct handle on
  individual quark and antiquark PDFs (DIS), or pairs of PDFs
  (Drell-Yan). It is possible to understand what is dominantly 
 measured by each individual process by looking at the leading order
 expressions, bearing in mind that, of course, beyond leading order
 all other contributions turn on (and that NLO corrections can be
 quite large, in fact of the same order of magnitude as the LO for
 Drell-Yan).

The leading order contributions to the DIS  structure functions $F_1$ and
$F_3$ are the following (at leading order $F_2=2 x F_1$):
\be
\begin{tabular}[c]{lc}
NC\qquad&${ F_1}^{\gamma} =\sum_{i}  e^2_i\left(q_i+\bar
q_i\right)$\quad\qquad\\
NC\qquad&${ F_1}^{Z,\,{\rm int.}} =\sum_{i}  B_i\left(q_i+\bar
q_i\right)$\quad\qquad\\
NC\qquad&${ F_3}^{Z,\,{\rm int.}} =\sum_{i}  D_i\left(q_i+\bar
q_i\right)$\quad\qquad\\
CC\qquad&{ $F_1^\wup =\bar u + d + s + \bar c$}\quad\qquad\\
CC&${ -F_3^\wup/2 = \bar u - d - s +\bar c }$,\quad\qquad \\
\end{tabular}
\label{strfun}
\ee
where $NC$ and $CC$ denotes neutral or charged current scattering and
we have lumped together the contributions coming from $Z$ exchange and
from $\gamma Z$ interference, with couplings given by
\bea
B_q(M^2_X)&=& -2e_qV_\ell V_qP_Z+(V_\ell^2+A_\ell^2)(V_q^2+A_q^2)P_Z^2;\\
  D_q(M^2_X)&=&-2e_qA_\ell A_qP_Z+4V_\ell A_\ell V_qA_q P_Z^2
\label{bdcoup}
\eea
in terms of the electroweak couplings of quarks and leptons listed in
Table~\ref{ewcoup} and the propagator correction $P_Z= M^2_X/(M^2_X+M_Z^2)$. 
\begin{table}[ht]
\vskip0.5cm
  \begin{center}
    \begin{tabular}{|c|c|c|c|}
      \hline
      fermions  & $e_f$ & $V_f$ & $A_f$  \\
      \hline
      u,c,t  & +2/3 & $(+1/2-4/3\sin^2\theta_W)$ & +1/2  \\
      d,s,b  & -1/3 & $(-1/2+2/3\sin^2\theta_W)$ & -1/2  \\
      $\nu_e,\nu_{\mu},\nu_{\tau}$  & 0 & +1/2 & +1/2  \\
      e,$\mu,\tau$  & -1 & $(-1/2+2\sin^2\theta_W)$ & -1/2  \\
      \hline
    \end{tabular}
    \caption{Electroweak couplings of fermions.
                                    \label{ewcoup}} 
\end{center}
\end{table}

The leading order contribution to Drell-Yan 
is given by
\be
\begin{tabular}[c]{lc}
$\gamma$\qquad & $\frac{d  \sigma}{d M_X^2 d y}(M_X^2,y) = \frac{4\pi \alpha^2}{9 M_X^2 s}\sum_i e_i^2 L^{ii}(x^0_1,x^0_2)$\\
$W$\qquad& $\frac{d \sigma}{d y} =  \frac{\pi G_F M_V^2\sqrt{2}}{3 s}
    \sum_{i,j} |V^{\rm CKM}_{ij}| L^{ij}(x^0_1,x^0_2)$\\
$Z$\qquad&$\frac{d \sigma}{d y} =  \frac{\pi G_F M_V^2\sqrt{2}}{3 s}
    \sum_i \left(V^2_{i}+A^2_i\right) L^{ii}(x^0_1,x^0_2)$
\end{tabular}
\label{lody}
\ee
in terms of the differential leading order parton luminosity
\be
L^{ij}\left(x_1,x_2\right)\equiv q_i(x_1,M_X^2) \bar q_j(x_2,M_X^2)+q_i(x_2,M_X^2) \bar q_j(x_1,M_X^2)
\label{loplum}
\ee
and the CKM matrix elements $V_{ij}$, with $x_i^0$ given by
Eq.~(\ref{lowerlim}). This shows explicitly that, as already
mentioned, for a rapidity distribution the leading order parton
kinematics (i.e. the values of $x_i$) is completely fixed by the
hadronic kinematics (i.e. the values of $y$ and $M_X^2$).

 Note that while at a $pp$
collider (or when a $p$ beam collides with a $p$ fixed target)
such as the LHC it makes no difference whether the incoming
quark and antiquark come from either of the initial-state hadrons, at a
$p\bar p$ collider such as the Tevatron (or when a $p$ beam collides
with a deuterium fixed target) there are  two different 
contributions, according to whether each of the incoming partons is
extracted from either of the initial-state hadrons.

\begin{figure}
\begin{center}
\includegraphics[width=.3\linewidth]{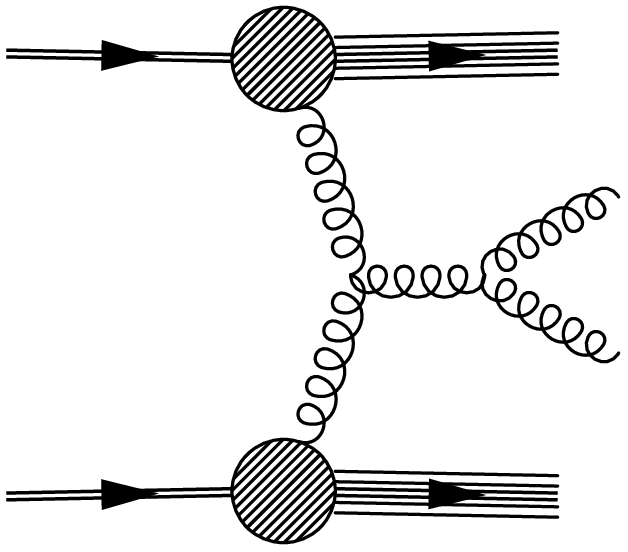}
\includegraphics[width=.3\linewidth]{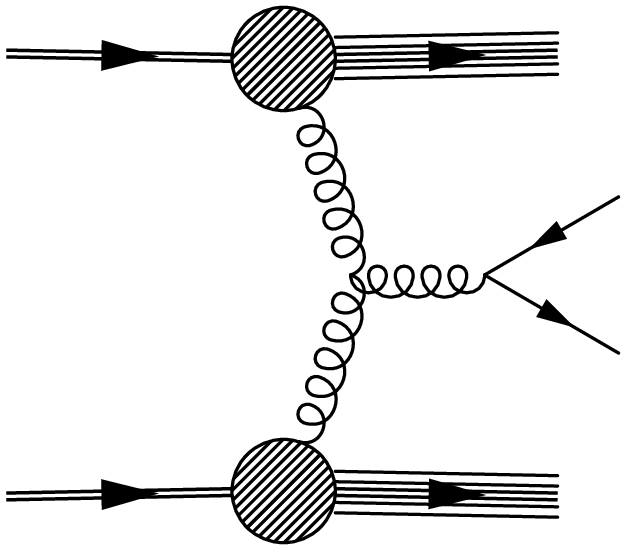}
\end{center}
\caption{Leading order diagrams for inclusive jet production from
  gluons}
\label{jetfig}
\end{figure}
Leading-order information on the gluon can be extracted
from jet production (see Fig.~\ref{jetfig}), or from scaling
violations, as measured for instance by the $Q^2$ dependence of
deep-inelastic structure functions. The latter are coupled to the
gluon even at leading order through the singlet QCD evolution
equations, which in terms of Mellin moments
\be
f_i(N,Q^2)\equiv \int_0^1 \!dx\, x^{N-1} f_i(x,Q^2)
\label{mellindef}
\ee
of parton distributions take the form
\bea
\frac{d}{dt}\ \left (\begin{matrix} \Sigma(N,Q^2) \\  g(N,Q^2)\end{matrix}\right)
&=&\frac{\alpha_s(t)}{2\pi}
\left (\begin{matrix}\gamma_{qq}^S(N,\as(t)) & 2n_f\gamma_{qg}^S (N,\as(t))\\ \gamma_{gq}^S(N,\as(t))
& \gamma_{gg}^S(N,\as(t)) \end{matrix}\right) \otimes \left (\begin{matrix}\Sigma(N,Q^2) \cr g(N,Q^2)\end{matrix}\right),\\
\frac{d}{dt} q^{NS}_{ij}(N,Q^2)
&=& \frac{\alpha_s(t)}{2\pi} \gamma^{NS}_{ij}(N,\as(t))  q^{NS}_{ij}(N,Q^2)
\label{dglap}
\eea
where the singlet combination of
quark distributions is defined as 
\be
\Sigma(x,Q^2)\equiv\sum_{i=1}^{n_f}
  \left(q_i(x,Q^2)+\bar q_i(x,Q^2)\right),
\label{singdef}
\ee
 and the remaining
  nonsinglet combinations 
can be taken as any linearly independent set of $2n_f-1$ 
differences of quark and antiquark distributions,
$q^{NS}_{ij}(N,Q^2)=q^{NS}_i(N,Q^2)-q^{NS}_j(N,Q^2)$ which all
evolve
  according to individual, decoupled equations. 

\begin{figure}
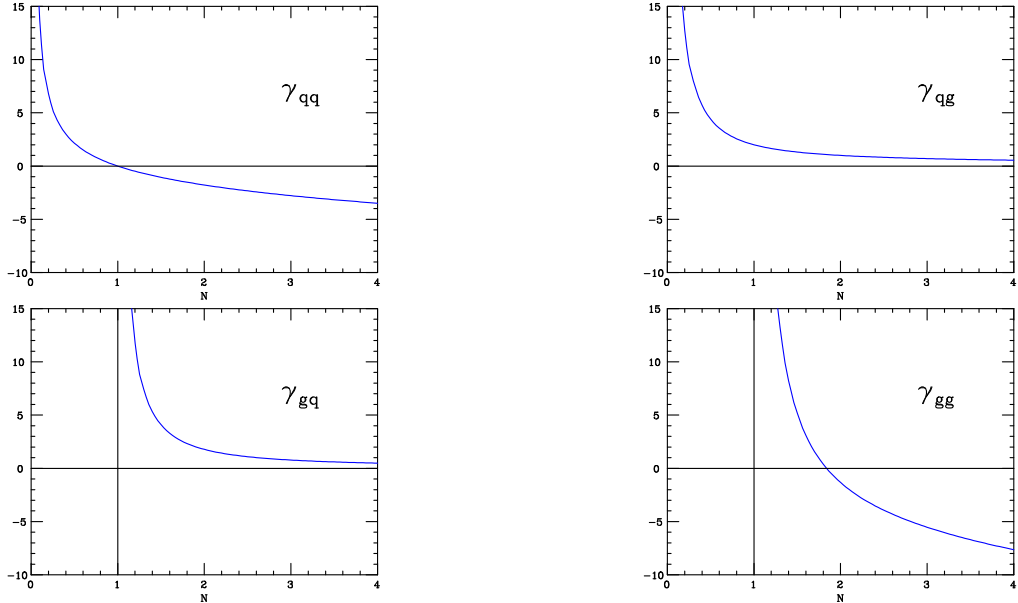

\begin{center}
\begin{minipage}{.49\linewidth}
\begin{center}
\includegraphics[width=.6\linewidth]{g0qq.ps}
\end{center}
\end{minipage}
\begin{minipage}{.49\linewidth}
\begin{center}
\includegraphics[width=.6\linewidth]{g0qg.ps}
\end{center}
\end{minipage}\\
\begin{minipage}{.49\linewidth}
\begin{center}
\includegraphics[width=.6\linewidth]{g0gq.ps}
\end{center}
\end{minipage}
\begin{minipage}{.49\linewidth}
\begin{center}
\includegraphics[width=.6\linewidth]{g0gg.ps}
\end{center}
\end{minipage}\\
\caption{The matrix of leading order anomalous dimensions shown as a
  function  of the Mellin variable $N$ Eq.~(\ref{mellindef}).}
\label{lodglap}
\end{center}
\end{figure}

The leading order
  anomalous dimensions are shown in Fig.~\ref{lodglap}, while at
  leading order all nonsinglet $\gamma^{NS}_{ij}$ are equal to each
  other and are also equal to $\gamma_{qq}$. The qualitative behaviour
  of perturbative evolution is then deduced
 recalling that
  Mellin transformation maps the large (small ) $x\to 1$ ($x\to0$) region into the
  large (small) $N\to\infty$ ($N\to0)$ region. A first relevant
  feature is that as the scale increases all PDFs decrease at
  large $x$ and increase at small $x$. A second important feature is
  that because the gluon has the rightmost singularity at small $N$ it
  drives small $x$ scaling violations, and thus in particular at
  sufficiently small $x$ and large $Q^2$ all PDFs have the same shape,
  driven by the gluon. Finally,
the  evolution of the gluon (driven by $\gamma_{gg}$) 
is strongest  at either large or
  small $x$ but its coupling to the quark (driven by $\gamma_{qg}$) is
  only large at small $x$, so  it is only at not too large $x$
  that
scaling violations provide leading  constraints on the gluon. 

Finally, it is important to note that constraints on PDFs come from
their cross-talk imposed by sum rules: specifically the conservation
of baryon number
\be
\int_0^1 \! dx\left(u^p(x,Q^2)-\bar
    u^p(x,Q^2)\right)= 2=
2 \int_0^1 \! dx \left(d^p(x,Q^2)-\bar
    d^p(x,Q^2)\right)
\label{baryonsr}
\ee
and the conservation of total energy-momentum
\be
\int_0^1 \! dx x \left[\sum_{i=1}^{N_f} \left(q^i(x,Q^2)+\bar
    q_i(x,Q^2)\right)+ g (x,Q^2)\right]= 1
\label{momsr}
\ee
Clearly, these sum rules provide  constraints on the behaviour of
parton distributions even in the region where there are no data.

\section{Statistics}
\label{statsec}
A determination of parton distributions is a determination of at least
seven independent functions: three light quark and antiquark
distributions and the gluon at some initial scale, from which PDFs at
all other scales can be obtained solving evolution equations. 
More functions must be determined if one wishes to keep open the
possibility~\cite{Brodsky:1980pb} 
that heavy quarks PDFs are at least in
part of ``intrinsic'' nonperturbative origin, rather than being
determined radiatively from gluons by QCD evolution. A determination of PDFs with
uncertainties thus involves determining a probability distribution in a
space of several independent functions. Because experimental data used
for this determination will always be finite in number, this is in
principle an ill-posed (or unsolvable) problem. 

The
time-honored~\cite{Gluck:1976iz,Johnson:1976xc} 
method to make this problem tractable is to assume a
specific functional form for parton distributions, which  projects
the infinite dimensional problem onto a finite-dimensional parameter
space. This method is justified because PDFs are
expected to be smooth functions of the scaling variable $x$. Because
$0\le x\le 1$, a representation of these functions with finite
accuracy must be possible on a finite basis of functions: hence,
a representation of PDFs must be
possible in terms of a finite number of
parameters. The problem is then reduced to the choice of an optimal
parametrization, namely, one that for given accuracy minimizes the
number of parameters without introducing a bias. We will discuss below two such parametrizations.

Whatever the parametrization, determining a set of PDFs involves computing a number of
physical processes with a given set of input PDFs, and extremizing a
suitable figure of merit, such as a $\chi^2$ or likelihood function in
order to determine a best-fit set of PDFs. Existing
sets of parton distributions which are made available for
the computation of LHC processes through standard interfaces 
are determined and delivered following two main strategies: a ``Hessian'' approach,
in which the best-fit result is given in the form of an optimal set of
parameters and an error matrix centered on this optimal fit to compute
uncertainties, and a Monte Carlo approach, in which 
the best fit is determined from the Monte Carlo sample by
averaging and uncertainties are obtained as variances of the sample. 

It turns out that available
Hessian PDF sets are mostly based on a ``standard''
parametrization, inspired by various QCD arguments. On the other hand,
the only available
full Monte Carlo PDF set is based on a rather different form of
parametrization, which adopts neural networks as interpolating
functions in an attempt to reduce the bias related to the choice of
functional form~\cite{Forte:2002fg}. However, Monte
Carlo studies based on other standard~\cite{FGR} 
and non-standard~\cite{Glazov:2010bw} parametrizations have also been
presented. 

Here we will summarize the  main features of both the Hessian and
Monte Carlo approach, and in each case also discuss the parton
parametrization which is most commonly used with each approach, and
the way the best fit is determined in each case --- which in turn
requires a peculiar
algorithm within the neural network approach.

\subsection{Hessian uncertainties and the ``standard'' approach}
\label{sechessian}
The standard approach to PDF determination is based on assuming for
PDFs at some reference scale $Q_0$ a
functional form inspired by counting rules~\cite{Brodsky:1973kr},
which suggest that PDFs should behave as $f_i(x)\toone{x}
(1-x)^\beta_i$, and Regge theory, which suggest~\cite{Abarbanel:1969eh} 
that they should behave
as $f_i(x)\tozero{x}x^{\alpha_i}$. Note that these limiting behaviours are
necessarily approximate, because  even if they hold at some scale, 
at any other scale perturbative evolution
will  correct them by logarithmic terms which behave as $\ln (1-x)$
as $x\to1$ and as $\ln x$ as $x\to 1$. Therefore, even if counting
rules and Regge theory actually provide predictions for the values of
the exponents $\beta_i$ and $\alpha_i$ respectively (for given parton
and parent hadron), they are taken as free fit parameters.

Based on this, typically PDFs are assumed to have the form
\be
f_i(x,Q_0^2)=x^{\alpha_i}(1-x)^{\beta_i} g_i(x),
\label{pdfparmgen}
\ee
 where $g_i(x)$ tends to a
constant for both $x\to 0$ and $x\to 1$. For instance, the CTEQ/TEA
collaboration assumes generally~\cite{Nadolsky:2008zw,Lai:2010vv}
\beq
x  f(x,Q^2_{0}) = a_{0}  x^{a_{1}}  (1-x)^{a_{2}}
\exp\left(a_{3} x+a_4 x^2 +a_5 \sqrt{x}+ a_6x^{-a_7}\right),
\label{pdfparmct}
\eeq
with different parameters $a_i$ for each PDF, but some parameters
fixed or set to zero for some PDFs --- for example, parameters $a_6$
and $a_7$ are nonzero only for the gluon distribution. Other groups
assume that  $g_i(x)$ is
a  polynomial in $x$ or in $\sqrt{x}$: for instance HERAPDF~\cite{:2009wt}
assumes $g_i(x)=1+ \epsilon_i \sqrt{x}+ D_i x+ E_i x^2$.

Different choices are  possible for
the set of linearly independent combinations of PDFs
for which the parametrization Eq.~(\ref{pdfparmgen}) is adopted, and
for the total number of free parameters to be used. For instance
CTEQ/CT parametrizes the ``valence'' light combinations $u_v=u-\bar u$,
$d_v=d-\bar d$, the antiquark distributions $\bar u$ and $\bar d$, 
the two strangeness combinations $s^\pm=s\pm\bar s$ (but in the
CTEQ6.6~\cite{Nadolsky:2008zw} and CT10~\cite{Lai:2010vv} fits it is
assumed that $s-\bar s=0$) and the gluon, with 22 (CTEQ6.6) or 26
(CT10) free parameters; MSTW08~\cite{Martin:2009iq} parametrizes also $u_v$, $d_v$,
$s^\pm$ and the gluon, and then the two combinations $\bar u\pm\bar d$
with a total of 28 parameters, and so forth.

Given a  parametrization of PDFs, the problem is reduced to that of
determining best fit values and uncertainty ranges for the
parameters. In a Hessian approach, this is done by minimizing a
figure of merit such as
\be
 \chi^2(\vec a)=\frac{1}{N_{\rm dat}}\sum_{i,j}
  (d_i-\bar d_i(\vec a)){\rm cov}_{ij}(d_j-\bar d_j(\vec a))
\label{chisq}
\ee
where the sum runs over all data points, $d_i$ are experimental
data with experimental covariance matrix ${\rm cov}_{ij}$ (including
all correlated and uncorrelated statistical and systematic
uncertainties),  $\bar d_i(\vec a)$ are theoretical predictions which are
obtained by evolving the starting PDFs at any scale $Q^2$ using the
evolution equations Eq.~(\ref{dglap}), and then folding the result
with known partonic cross sections according to the factorization
theorems
Eqs.~(\ref{hadrfact},\ref{rapfact},\ref{disfac}), and $\vec a$
denotes the full  set of parameters on which the PDFs at scale $Q_0$
depend, which we may view as a vector in parameter space
(which is 26-dimensional for CT10, and so on).  
The $\chi^2$ thus is a function of the $\vec a$  through the
predictions  $\bar d_i(\vec a)$.

 Note that the
$\chi^2$ Eq.~(\ref{chisq}) is normalized to the number of data points: this is
conventionally done in order to allow for approximate comparisons of
fit quality between fits with different numbers of data points; in
practice, this is likely to be close to the $\chi^2$ per degree of
freedom because typical datasets include thousands of data, while the
total number of parameters needed to describe accurately all PDFs with
functional forms like Eq.~(\ref{pdfparmgen}), though of
course unknown, is likely to be rather lower than a hundred. For the
sake of future discussions it is convenient to also introduce an
unnormalized 
\be
\bar\chi^2=N_{\rm dat}\chi^2.
\label{unnormchisqdef}
\ee

 It is
important to note that there  are subtleties in the 
definition of the $\chi^2$, which may make the comparison of $\chi^2$
values from different groups only qualitatively significant, because
slightly different definitions are used. The main subtlety is related to
the inclusion of normalization uncertainties, which cannot be simply
introduced in the covariance matrix, as this would bias the 
fit~\cite{D'Agostini:1993uj}: a full unbiased
solution~\cite{Ball:2009qv} 
requires an iterative construction of the covariance matrix, but other
approximate solutions are also adopted. 

Once the $\chi^2$ is defined, for given data
$\chi^2$ is a function of the PDF parameters through the predictions
$\bar d_i$ which in turn depend on the PDFs. Hence, the best fit set of
parameters can be identified with the absolute minimum of the
$\chi^2$ in parameter space. Furthermore,
the variance of any
observable  $X$ which depends on parameters $\vec a$ (such as a
physical cross section, or indeed the PDFs
themselves), if we  assume linear error propagation 
$X(\vec a)\approx X_0+ a_i\partial_i X(\vec
  z)$, is given by 
\be
\sigma^2_X=\sigma_{ij}\partial_iX\partial_jX .
\label{varx}
\ee
Here $\sigma_{ij}$ is the covariance matrix of the parameters which,
in turn, 
assuming that the $\chi^2$ is a quadratic function
of the parameters in the vicinity of the minimum, is given by
(see
e.g.~\cite{Cowan:1998ji,PDGstat})
\be
\sigma_{ij}=\partial_i\partial_j\bar \chi^2|_{\rm min}
\label{hessian}
\ee
i.e.  it is the (Hessian) matrix of second derivatives of  the
unnormalized $\bar\chi^2$ Eq.~(\ref{unnormchisqdef}),
evaluated at its minimum.

The Hessian method for the determination of uncertainties thus
in particular implies that the one-$\sigma$ (i.e. 68\% confidence
level) for the parameters themselves is the ellipsoid in parameter space
which is fixed by the condition $\bar\chi^2=\bar\chi^2_{\rm min}+1$ .
As we will discuss in Sect.~\ref{tol}
in practice this argument may have to be modified in realistic cases, in order to
account for various effects (such as incorrect estimation of the
covariance matrix of the data). 

However, for the time being let us
stick to the textbook argument, and make a couple of observations on it.
The first observation is that we are always free to adjust the
parametrization in such a way that all eigenvalues of the Hessian
matrix $\sigma_{ij}$ are equal to one, by simply diagonalizing the
matrix and rescaling the eigenvectors by the eigenvalues, i.e. by
looking for new parameters $a^\prime_j(a_i)$ such that
\be
\sigma_{ij}(a_i-a_i^{\rm min})(a_j-a_j^{\rm min})=\sum_{i=1}^{N_{\rm
    par}} \left(a^\prime_i\right)^2,
\label{diaghessian}
\ee
which  immediately implies that
\be
\sigma^2_X=|{\vec \nabla}^\prime X|^2 ,
\label{varxgrad}
\ee
where the gradient is computed with
respect to $\vec a^\prime$. Equation~(\ref{varxgrad}) has the immediate interesting
consequence that the total contribution to the uncertainty due to two
different sources, being the length of a vector, 
is simply found by adding the components i.e. the different
uncertainties 
 in quadrature (even
when the two uncertainties are correlated). This has been emphasized recently in
Ref.~\cite{Lai:2010nw}, where it is shown explicitly that, contrary to what one
may naively think, the total uncertainty due to PDF parameters and
some other parameter (such as the value of the strong coupling
constant) is simply found adding the two uncertainties in quadrature.

The second comment has to do with the fact that the one-$\sigma$
interval in parameter space corresponds to the contour
$\Delta N_{\rm dat}\chi^2=1$ about the minimum. This is identical to
the statement that the Hessian Eq.~(\ref{hessian}) is the covariance
matrix in parameter space. This simple fact is sometimes source of
confusion because it seems to contradict the observation that 
the standard deviation of the (unnormalized) $ \chi^2$ distribution with
$N_{\rm dof}$ degrees of freedom is $\sqrt{2 N_{\rm dof}}$: in fact, sometimes
(see e.g.~\cite{Hirai:2003qk}) it is incorrectly stated
that one-$\sigma$ contours correspond to $\Delta \bar\chi^2\sim N_{\rm dof}$. However, the 
contradiction is only apparent: $\Delta \bar\chi^2\sim N_{\rm dof}$ sets a hypothesis-testing
criterion~\cite{Collins:2001es}, namely, it gives the size of
fluctuations of $\Delta \bar\chi^2$ upon repetition of the experiment, and thus the range of
$\bar\chi^2$ values away from the mean $\langle \bar \chi^2\rangle=N_{\rm
  dat}$
which are acceptable for a given theory (experiment). One the
other hand $\Delta \bar\chi^2=1$ provides a parameter-fitting
criterion~\cite{Collins:2001es}: it gives the range of 
parameter values which are compatible at one sigma
for a given experimental result (and theory).

\begin{figure}\begin{center}
\includegraphics[width=.6\linewidth,clip]{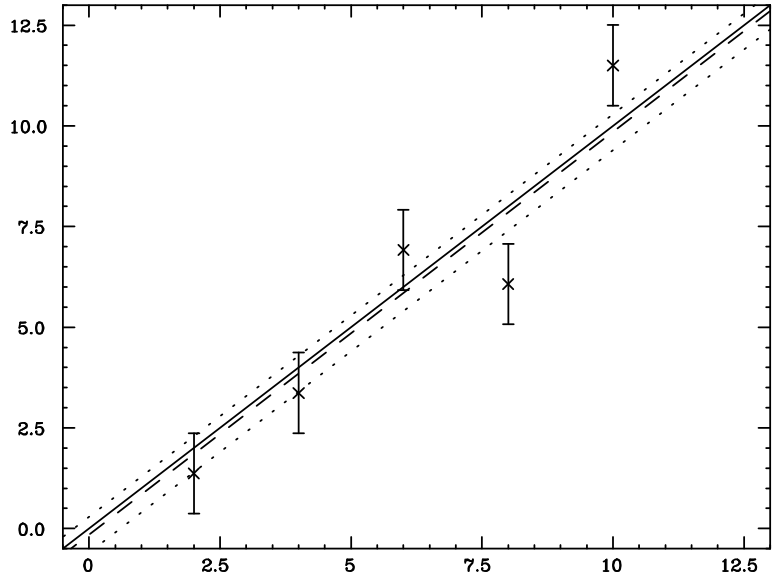}
\caption{Fit to data Gaussianly distributed about a linear law. The
  solid line is the true underlying law, the dashed line is the
  best-fit, and the dotted lines give the one-$\sigma$
  ($\Delta\chi^2=1$)
error band, which is by $\frac{1}{N_{\rm dat}}$ smaller than the error
on each point.} 
\label{linfit}
\end{center}
\end{figure}
A simple example may help in understanding the distinction. Consider
the case of a simple linear fit, in which one has a set of data which
are expected to satisfy a linear law $y=x+k$, with unknown intercept
$k$ that one wishes to determine by fitting to data (see
Fig.~\ref{linfit}). Define the deviation $\Delta_i\equiv d_i-\bar
d_i(x_i)$ between the $i$-th data point $d_i$ and the linear
prediction $\bar d_i=x_i+k$. If $\Delta_i$ are gaussianly distributed
with  standard deviation $\sigma$ about their true values, 
then clearly the average square
deviation $\sigma^2_\Delta=\langle \Delta^2\rangle=N_{\rm dat}
\sigma^2$. 
This is the
``hypothesis testing'' fluctuation range of the $\bar \chi^2$. However, the
best-fit intercept $k$ is just the average deviation $k=\langle
\Delta\rangle$, and the square uncertainty on it is
$\sigma^2_k=\frac{\sigma^2_\Delta}{N_{\rm dat}}$: so the ``parameter
fitting'' range for $k$ is indeed by a factor ${N_{\rm dat}}$ smaller
than the expected total square fluctuation, because the best-fit value
is
determined as a mean, whose square fluctuation is by a factor ${N_{\rm
    dat}}$ smaller than the fluctuation of the individual data.

 \subsection{Monte Carlo uncertainties and the NNPDF approach}

A Monte Carlo approach differs from the Hessian approach in the way
the uncertainty on the observable is determined in terms of the
uncertainties in parameter space: the distinction  Hessian vs. Monte
Carlo thus has to do only with the way uncertainties are propagated from
parameters to observables. However, the Monte Carlo way of propagating
uncertainties is especially convenient when used together with a
parametrization whose functional form is less manageable, for instance
because the number of parameters is particularly large, or because the
functional form is less simple than that of Eq.~(\ref{pdfparmgen}),
or, more, in general, whenever linear error propagation and the
quadratic approximation to the $\chi^2$ in parameter space are not
advisable, for reasons of principle or of practice. Therefore, we
will first discuss the distinction between Hessian and Monte Carlo {\it
  per se}, then turn a brief review of the way a Monte Carlo approach has been
used by the NNPDF group together with a choice of basic underlying
functional form for PDFs which differs from that of
Eq.~(\ref{pdfparmgen}), and finally address  some issues related to the
determination of the best fit PDFs when such functional forms
are adopted.

\subsubsection{Monte Carlo uncertainties}

Whereas in a Hessian approach parameters are assumed to be gaussianly
distributed with covariance matrix $\sigma_{ij}$ given by the Hessian
Eq.~(\ref{hessian}), in a Monte Carlo approach the probability
distribution in parameter space
is given by assigning a Monte Carlo sample of replicas  of
the total parameter set. For example, if one uses the parametrization
Eq.~(\ref{pdfparmgen}) one would then simply give a list of $N_{\rm
  rep}$ replica copies of the vector of parameters $\vec s$. Any
observable 
$X$ is then computed by repeating its
determination $N_{\rm
  rep}$ times, each time using a different parameter replica: the
central value for $X$ is  the average of these $N_{\rm
  rep}$ results, the standard deviation is the variance, and in fact any
moment of the probability distribution can be determined from the
sample of $N_{\rm
  rep}$ values of $X$ thus obtained. 

\begin{figure}\begin{center}
\includegraphics[width=.6\linewidth,angle=-90]{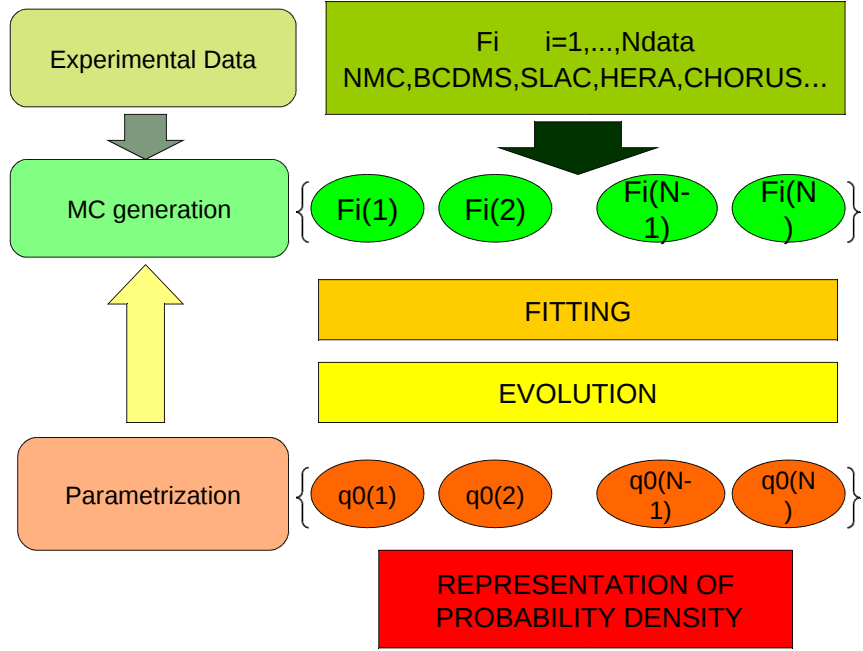}
\caption{Schematic representation of the construction of a Monte Carlo
  representation of parton distributions.} 
\label{nnpdfscheme}
\end{center}
\end{figure}
Of course, this begs the question
of how the distribution of parameter values, i.e. the distribution of
parameter replicas is determined in the first
place. In fact, this may look like a hopeless task: let's say that for
each parameter the probability distribution in parameter space
is given for each parameter 
as a histogram with three bins, one
corresponding to the one-$\sigma$ region about the central value of the
given parameter, and two for
the two outer regions. Then, for $N_{\rm par}$ parameters
the total number of bins is equal to $3^{N_{\rm par}}\gsim 3\cdot
10^{9}$ with $N_{\rm par}=20$ parameters. Hence it looks like the total
number  of replicas must be hopelessly large in order to have sufficient
statistics. This, however, is not necessarily the case, because it may
well turn out that most of the bins
are actually empty. To understand this, recall the Hessian computation
of the uncertainty on $X$ Eq.~(\ref{varxgrad}): it is clear that in
order to determine the uncertainty on $X$, it is sufficient to know
the distribution in parameter space along the direction of $\vec\nabla^\prime
X$. Hence, for this specific observable only one parameter is
relevant. Even if one wants to determine the uncertainty on
observables which probe any direction in parameter space, for any
reasonably smooth function the number of bins which is needed in order
to get an accurate representation of the probability distribution is
likely to be much smaller than $10^{9}$. This then again raises the
question of how one should sample the replica distribution in parameter space.

The answer is found by noting that the maximum likelihood method gives
a way of mapping the probability distribution in data space onto the
probability distribution in parameter space.
Namely, assume one has data
$d_i$ with covariance 
matrix
${\rm cov}_{ij}$. Then, generate $N_{\rm rep}$ {\it data} replicas
$d^\alpha_i$,
with $\alpha=1,2,\dots,N_{\rm rep}$. For each value of $\alpha$,
i.e. for each replica, the whole set of data $i=1,2,\dots N_{\rm dat}$
is replicated,  in such a way
that if one takes the average over  the $N_{\rm rep}$
replicas $d^\alpha_n$ of the $n$-th data point,
then in the limit $N_{\rm rep}\to\infty$ this average tends to the
original data value $d_n$; if one computes the variance of these
$N_{\rm rep}$ values in the same limit it tends to to the the standard
deviation of the data; and if one computes the covariance of the
$n$-th and $m$-th data replicas it tends to the covariance matrix
element  ${\rm cov}_{nm}$. Now, for each data replica,  determine a
best-fit parameter vector $\vec a^\alpha$ by minimizing the $\chi^2$
Eq.~(\ref{chisq}), but of the fit to the replica data  $d^\alpha_i$,  rather than the
original data. We end up with a Monte Carlo set of best-fit parameter
vectors $\vec a^\alpha$: again, the average over these
$\alpha+1,2,\dots,N_{\rm rep}$ vectors $\vec a^\alpha$
gives us the best-fit parameters  $\vec a^\alpha$, and the
covariance of the $n$-th and $m$-th components of the parameter vector
gives us the covariance matrix $\sigma_{m n}$. In fact, it is easy to
check (see e.g.~\cite{Ball:2009qv}) that for
gaussianly distributed data the results coincides with the Hessian
covariance matrix Eq.~(\ref{hessian}). We will see an explicit
numerical check of this equivalence in Fig.~\ref{gaussvsnormal} below.

The procedure is summarized in Fig.~\ref{nnpdfscheme}: one starts with
experimental data (denoted as $F_I$ in the Figure),
generates data replicas (denoted as $F_I(1)\dots F_I(N)$) and fits a
set of PDFs to each data replica (denoted as $q_0(i))$). The PDFs can
be parametrized in any desired way at some reference scale, 
and they are fitted to the data
replicas in the way discussed in Sect.~\ref{sechessian}, namely by
evolving them to the scale of the data, using them to compute
observables, and minimizing the $\chi^2$ of the comparison to the data
with respect to the parameters.

\begin{figure}\begin{center}
\includegraphics[width=.45\linewidth]{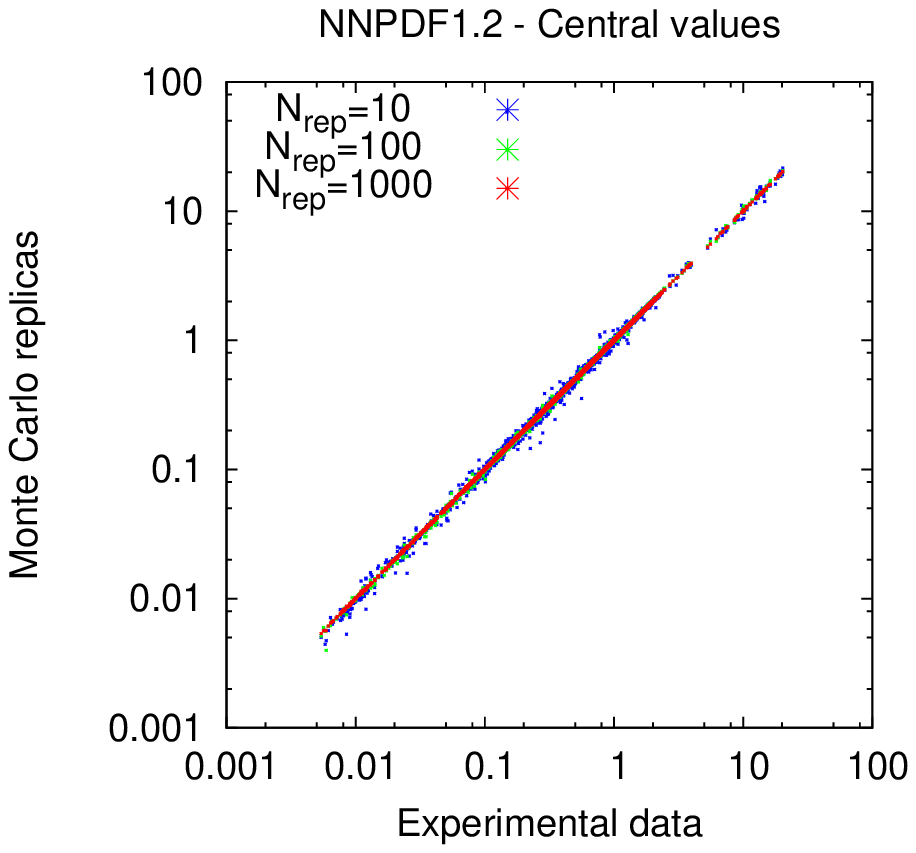}
\includegraphics[width=.45\linewidth]{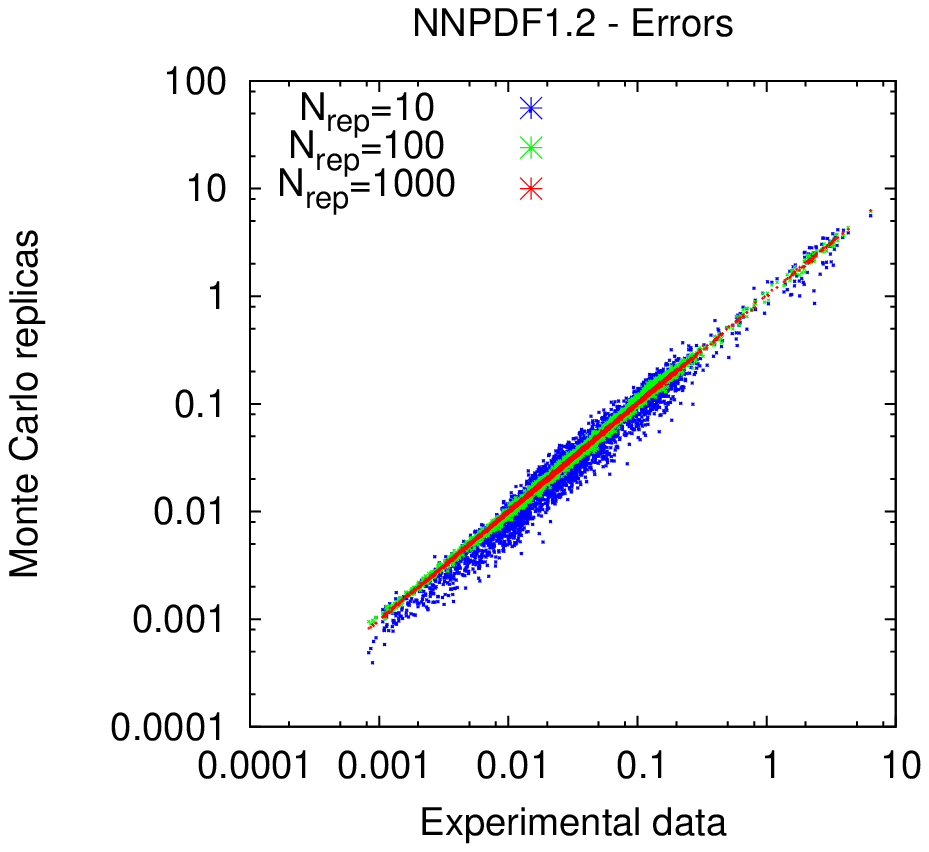}
\caption{Scatter plot of central values and uncertainties of a Monte
  Carlo sample compared to the data for the dataset of the NNPDF1.2
  parton fit~\cite{Ball:2009mk}.} 
\label{scatnn}
\end{center}
\end{figure}
But then, the problem of  constructing an adequate sampling of
parameter space has been reduced to that of constructing an adequate
Monte Carlo representation of the original data: i.e. the space of parameters
is sampled in a way which is determined by the distribution of the
data (``importance sampling''). Whether a given set of replicas
provides an accurate enough representation of the data is then
something that may be checked explicitly for a given sample, by
comparing means, variances and covariances from the sample with the
desired features of the data.
For a typical set of data used in a parton fit the numbers of replicas
required turn out to be surprisingly small: for instance, in
Fig.~\ref{scatnn} we show a scatter plot of the averages {\it vs.} central
values and variances {\it vs.} standard deviations for the set of $N_{\rm
  dat}=3372$ data points included in the NNPDF1.2~\cite{Ball:2009mk}
parton fit, computed using $N_{\rm rep}=10,\>100,\>1000$ Monte Carlo
Replicas. It is clear that the scatter
plot deviates by just a few percent from a straight line already for
$N_{\rm rep}=10$ for central values, and for
$N_{\rm rep}=100$; $N_{\rm rep}=1000$ replicas turn out to be
only necessary in order to get percent accuracy on correlation
coefficients.

It should finally be mentioned that within a Monte Carlo approach it
is possible to sidestep the problem of choosing an adequate
parametrization by using Bayesian
inference~\cite{Giele:2001mr}. Namely, one starts from some prior
Monte Carlo representation of the probability distribution based on
some initial subset of data, or even on assumptions. Then, 
the initial Monte Carlo set is updated by including the information
contained in new data  through Bayes'
theorem. Without entering into details, it is clear that this can be
done by changing the distribution of replicas: more or less copies are taken
for those replicas which agree or respectively do not agree with the
new data, in a way which is specified by Bayes' theorem. To the
extent that results do not depend too much on the choice of prior,
which is often the case if the information used through Bayes' theorem
is sufficiently abundant, final results are then free of bias. Whereas
the construction of a parton set fully  based on this method has so
far not been completed, preliminary results have been
presented~\cite{Ubialitrento} on the inclusion of new data in an
existing Monte Carlo fit using this methodology.

\subsubsection{Neural Network parametrization and cross-validation}
\label{mcnnpdf}
The Monte Carlo approach has been recently used for the determination
of a PDF set in conjunction with the use of neural networks as a
parton parametrization. Neural networks are just another functional
form. In analogy to polynomial forms, they have the feature that any
function (with suitable assumptions of continuity) may be fitted in
the limit of infinite number of parameters; unlike polynomial forms
they are nonlinear, and they are  ``unbiased'' in that a finite-dimensional
truncation of the neural network parametrization is adequate to fit a
very wide class of functions (for instance, both periodic and non
periodic) without the need to adjust the form of the parametrization
to the desired problem. 

\begin{figure}\begin{center}
\includegraphics[width=.29\linewidth]{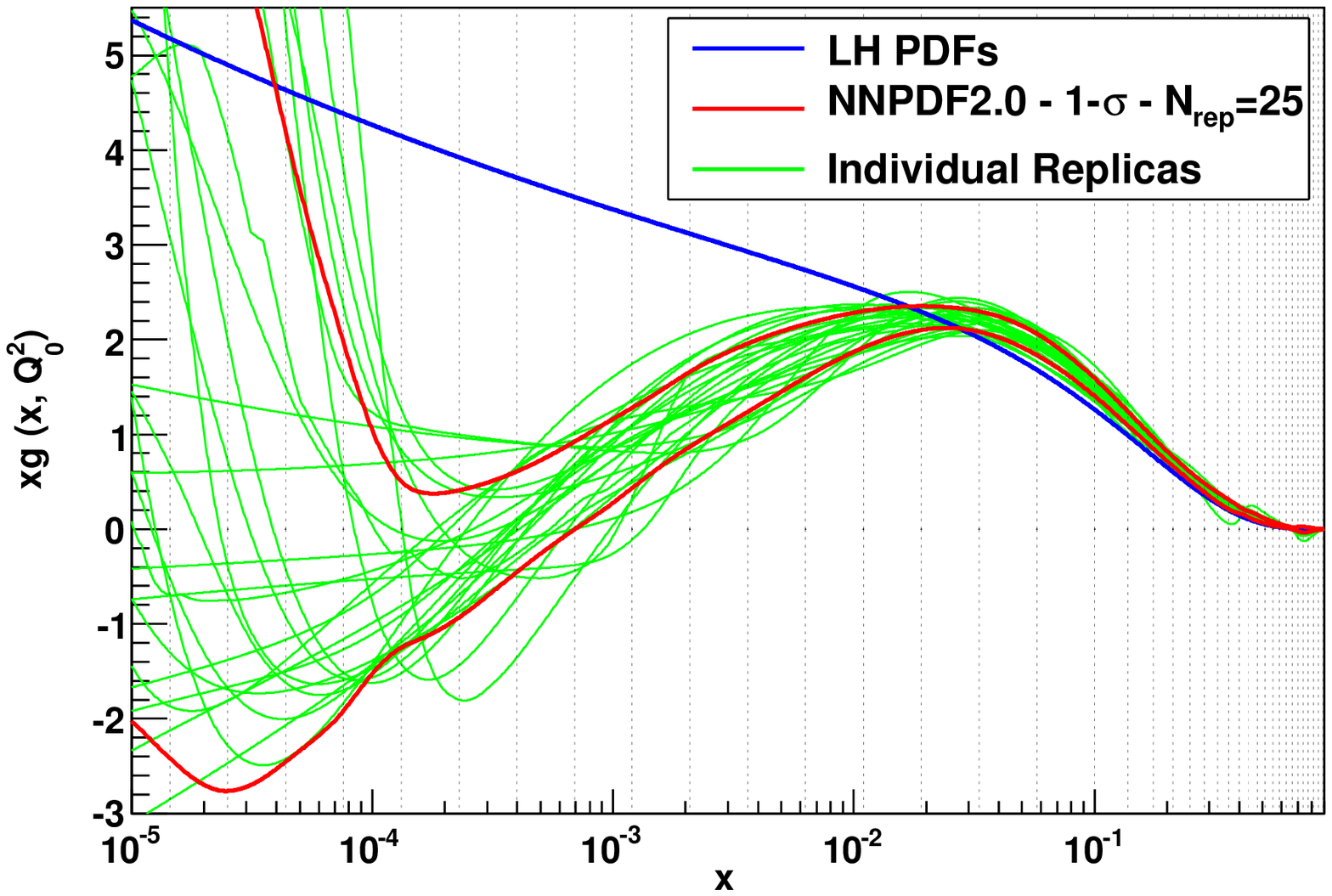}
\includegraphics[width=.7\linewidth]{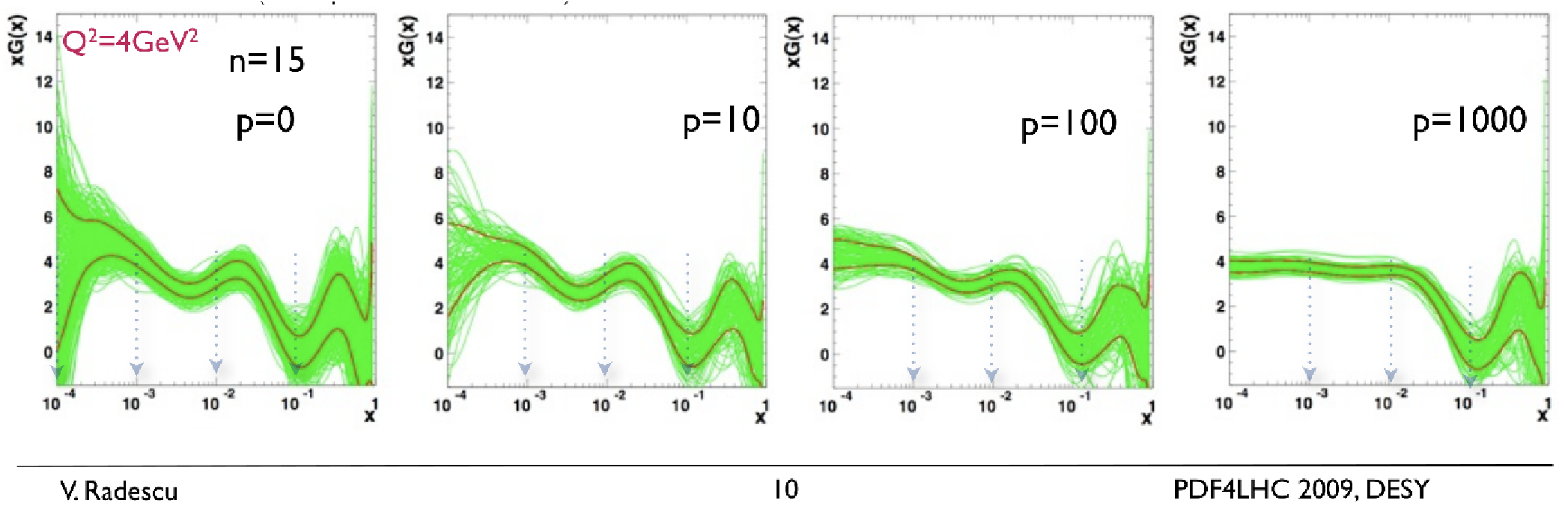}
\caption{Left: 25 gluon replicas based on a neural network
  parametrization, together with their one-$\sigma$ range, compared to
  a toy gluon distribution of the form Eq.~(\ref{pdfparmgen}) with
  $g(x)=1$ and typical values of the parameters $\alpha$ and $\beta$
  (from Ref.~\cite{Ball:2010de}). Right: gluon replicas  based on a
  parametrization on a basis of Chebyshev polynomials together with
  their one-$\sigma$ range; subsequent plots correspond to an
  increasingly high penalty proportional to the length of the fitted curve.
(from Ref.~\cite{Glazov:2010bw}).}
\label{nnvsgenparm}
\end{center}
\end{figure}
A very simple example of neural network is the
function
\be
f(x)=
\frac{1}{1+e^{{\theta_1^{(3)}}-\xuno-\xdue}}
\label{nnex}
\ee
where $\theta^{(i)}_n$ and $\omega^{(i)}_{n m}$ are free parameters.
 This is a $1-2-1$ neural network, parametrized by six free
parameters: 1-2-1 refers to the way the neural network is constructed,
by iterating recursively the response function
$g(x)=\frac{1}{1+e^{\theta -\beta x}} $ on nodes arranged in layers which
feed forward to the next layer, with the first (last) layer containing the
input (output) variables. 

In Refs.~\cite{Ball:2009mk}-\cite{Ball:2010de} PDFs
are parametrized using 2-5-3-1 neural networks, with 37 free
parameters (the input has two variables because $x$ and $\ln x$ are
treated as two independent inputs, thereby increasing the redundancy
of the parametrization). The six light flavours and antiflavours are
parametrized and the gluon are parametrized in this way, so that the
total number of parameters is $37\times7=259$, thus rather larger than
the typical numbers used when dealing with parametrizations of the
form Eq.~(\ref{pdfparmgen}). Such a large number of parameters
clearly reduces considerably the risk of a parametrization bias, but
it poses the problem that if the best fit is  determined as the
absolute minimum of the $\chi^2$ one may end up fitting data
fluctuations, which is clearly not desirable. Even if these fluctuations average
out when averaging over Monte Carlo replicas this would be a very
inefficient way of proceeding.

The advantage of a neural network parametrization can be understood
from Fig.~\ref{nnvsgenparm}, where a gluon distribution determined using neural
networks is compared to the simplest version of
parametrization Eq.~(\ref{pdfparmgen}), and also to a very flexible
parametrization based on orthogonal polynomials. The 
neural network gluon distribution shown in Fig.~\ref{nnvsgenparm}
corresponds to $N_{\rm rep}=25$ replicas from the Monte Carlo
set of Ref.~\cite{Ball:2010de}, and it is displayed along with the
average and one-$\sigma$ contour computed from the set. On the same
plot 
a parametrization of the form
Eq.~(\ref{pdfparmgen}) is also shown, 
with typical values of the parameters $\alpha$
and $\beta$, and with $g(x)=1$. 
It is compared to a set of Monte
Carlo replicas of the gluon which were constructed in
Ref.~\cite{Glazov:2010bw}
by expanding the
gluon distribution on a basis of 15 independent Chebyshev
polynomials, while also imposing an increasing penalty $p$ to fits
with large arc-length (and thus more oscillations). The fits based 
on orthogonal polynomials display large uncontrolled oscillations
which are only tamed by appropriately
tuning the length penalty. The fits based on
neural networks, despite having a  number of free parameters which
is more than double than those using orthogonal polynomials, do not
display a similarly unstable behaviour, even though they do show
considerable flexibility, and in fact the ensuing one-$\sigma$ band,
though accounting well with its width for the functional freedom  is
actually quite stable.

\begin{figure}\begin{center}
\includegraphics[width=.4\linewidth]{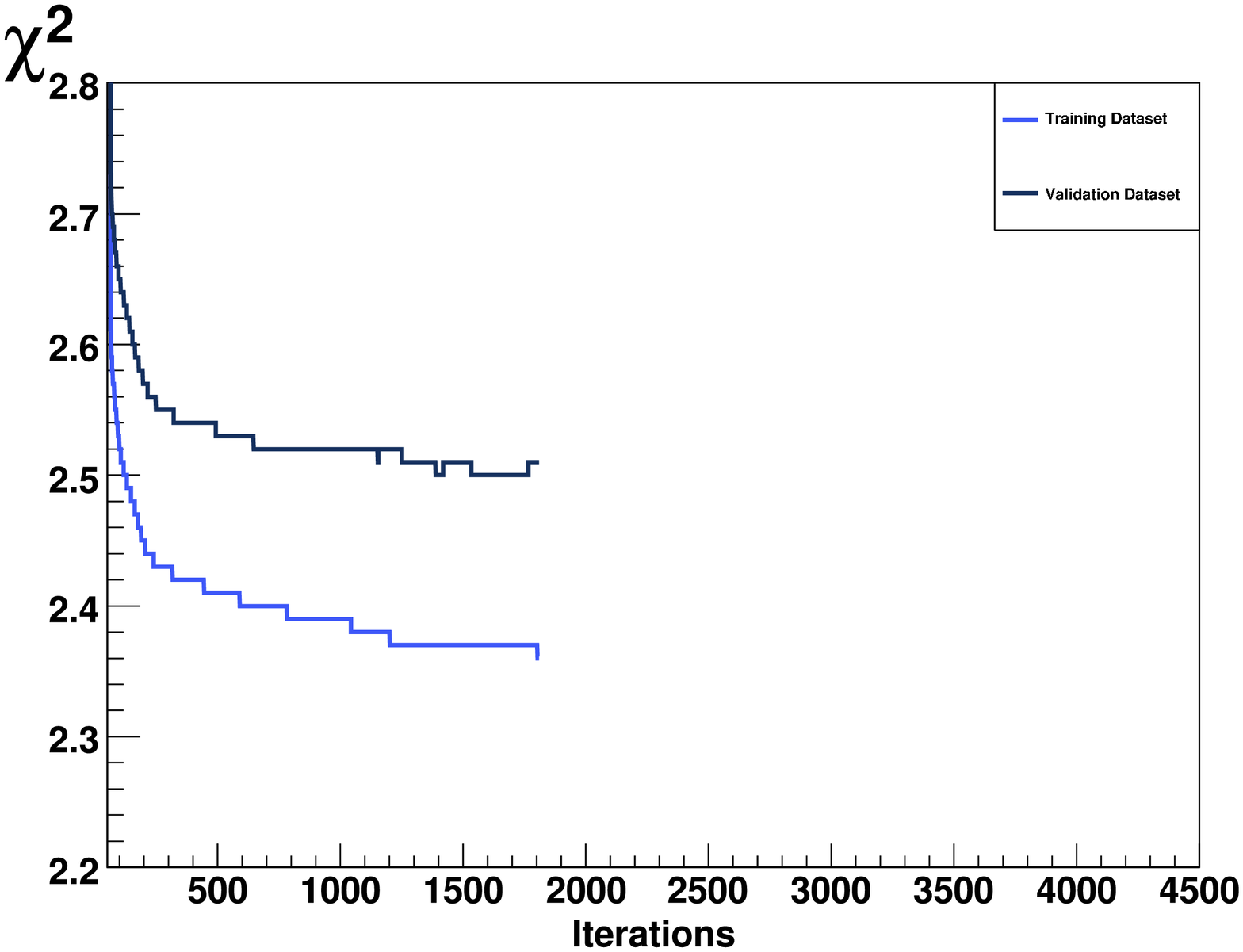}
\includegraphics[width=.4\linewidth]{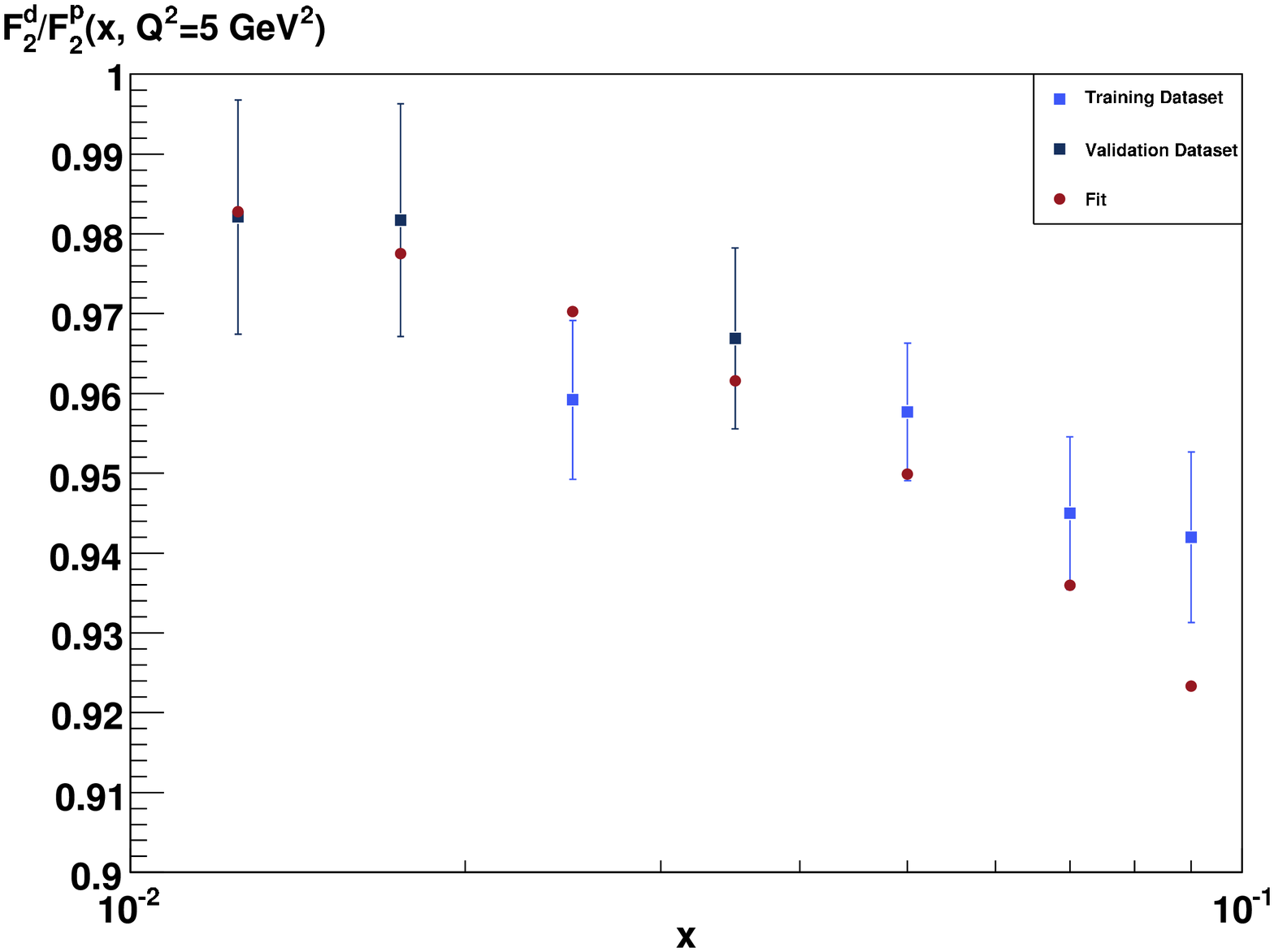}\\
\includegraphics[width=.4\linewidth]{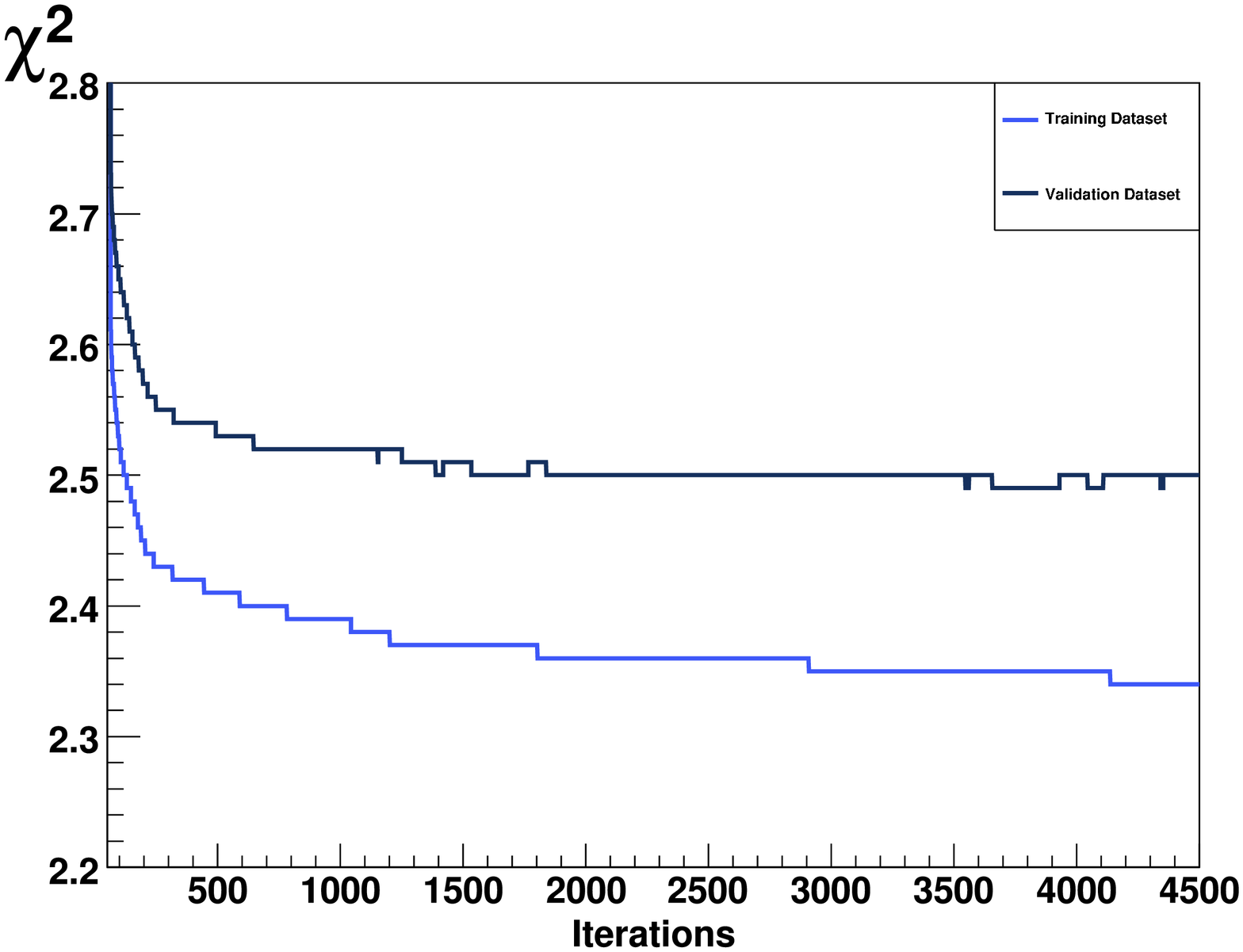}
\includegraphics[width=.4\linewidth]{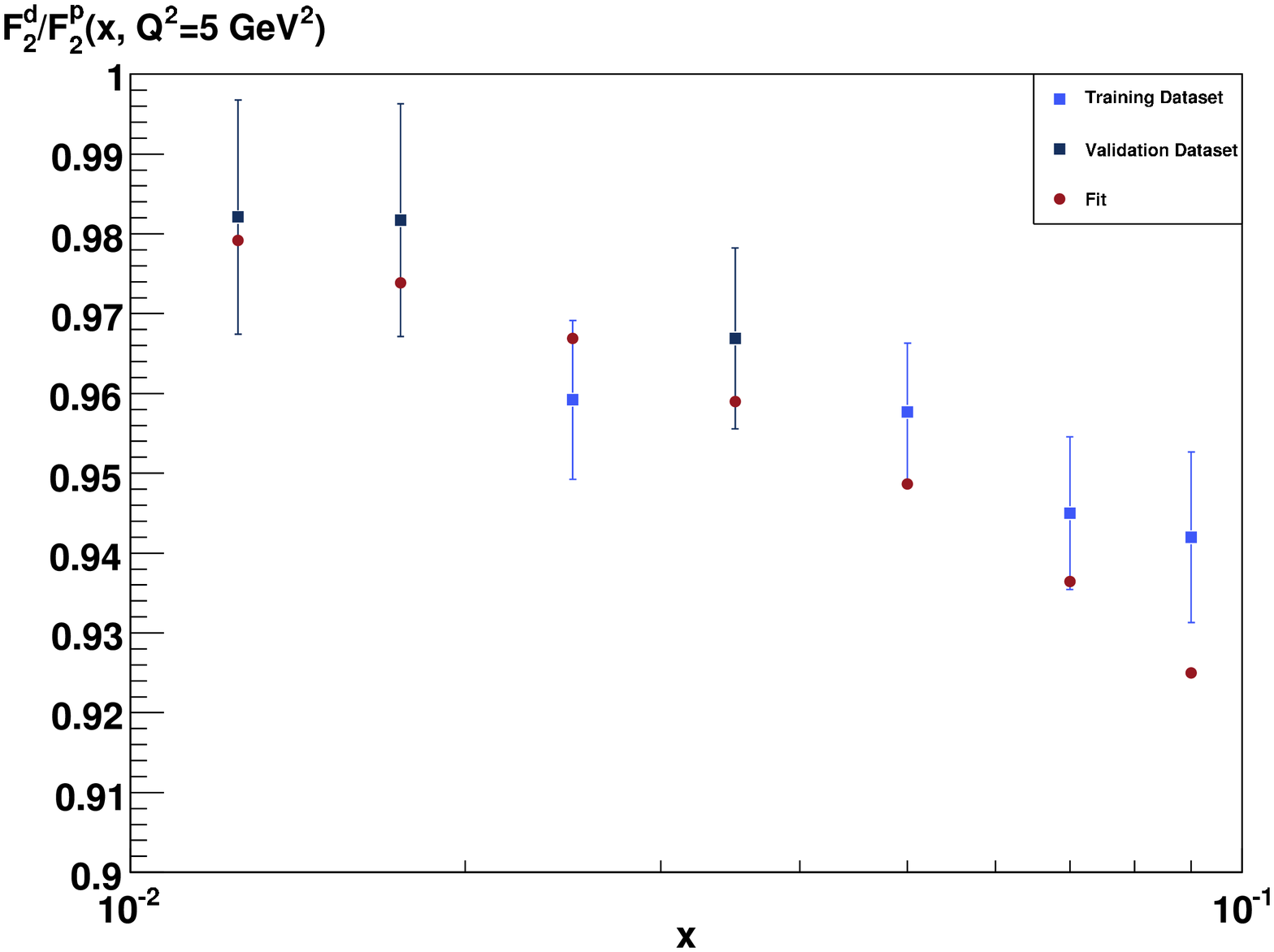}
\caption{Determination of the best-fit by cross-validation: 
the  $\chi^2$ of the fit to 
   points in the training set
  (light blue)  and the validation set (dark blue) are shown as a
   function of the number of iterations of the minimization algorithm 
   (left plots), but only the $\chi^2$ for training data is
   minimized.  The data are shown in the right plot, along with the
   best fit (red points without error band).  The upper plots show the
best fit at stopping point (optimal fit) and the $\chi^2$ up to this
point, the lower plots show the $\chi^2$ up to and the best fit at an
``overlearning'' point.} 
\label{crossvalid}
\end{center}
\end{figure}

The best fit is instead determined using a cross-validation
method (see Fig.~\ref{crossvalid}). Namely, the data are randomly divided into a
training and a validation sample. The $\chi^2$ is
computed both for the data in the training sample and those in the
validation sample. Only the training $\chi^2$ is minimized, but the
 validation $\chi^2$ is also monitored as the minimization
 proceeds. The best fit is defined as the point at which the
 validation $\chi^2$ stops improving even though the training
 $\chi^2$ may keep improving: this is the point at which one is starting
 to fit the statistical noise of the training sample. In order to
 ensure lack of bias, the partitioning of the data is done randomly
 in a different way for each data replica. Also, in practice, in
 order to minimize the effect of random fluctuations in the data (or
 of the minimization algorithm) the
 stopping criterion must be imposed after a suitable averaging, such
 as for instance the moving average of  values of the
 $\chi^2$ found in the last $N_s$ iterations of the minimization
 algorithm.

\section{From data to PDFs}
\label{data}

Once a parton parametrization and a methodology have been chosen, the
determination of a PDF set relies on the choice of a set of physical
observables. The problem is that, even after projecting the problem on
a finite-dimensional parameter space, we must still determine seven
independent PDFs, which means that we need seven linearly independent
pieces of information at fixed scale for each value of  $x$. For instance,  
a determination of deep-inelastic structure functions $F_1$ and
$F_3$  for charged-current  deep-inelastic scattering
provides, according to Eq.~(\ref{strfun}) four
independent linear combinations of quark
distributions (if $W^\pm$ can be distinguished),
with two more linear combinations  provided by neutral current
structure functions. All individual light quark and antiquark flavours
 then can be determined by linear combination. This situation
would be realistic at a neutrino factory with both neutrino and
antineutrino beams and the possibility of identifying the charge of the
final state
lepton on an event-by-event basis~\cite{Mangano:2001mj,Forte:2001ph}.

Unfortunately, this theoretically and phenomenologically very clean
option is at best far in the future, so at present the information on
individual PDFs can only be achieved by combining information from
different processes into so-called ``global'' fits. The idea is that,
even though each electroproduction or hadroproduction observable 
depends on all PDFs through the
factorization formulae Eqs.~(\ref{hadrfact},\ref{disfac}), inclusion
of specific processes or combination of processes may give a specific
handle on individual PDFs or combinations of PDFs on which it depends
most strongly (typically, through its leading order form). We will now
review one at a time each of these individual handles on PDF, then
briefly discuss how they are combined in modern more or less global fits. 

\subsection{Isospin singlet and triplet}
\label{isospinsingt}

Neutral current deep-inelastic (DIS) structure function data only provide a
determination of the charge-conjugation even combination $q_i+\bar q_i$ of
quarks and antiquarks, for each quark flavor $i$.  Specifically,
photon DIS data only determine the fixed combination in which each
flavor is weighted by the square of the electric charge, see
Eq.~(\ref{strfun}). However, one may separate off the isospin triplet
and singlet components by considering DIS on both proton and deuteron
targets, assuming that the deuterium structure function is simply the
incoherent sum of the proton and neutron ones
$F_2^d=\half(F_2^p+F_2^n)$
(up to small nuclear corrections which can be accounted for through
models, such as that of Ref.~\cite{Kulagin:2004ie}), and then 
 using isospin symmetry to relate the quark and antiquark
 distributions of the proton and neutron:
\beq
{ u}^{ p}(x,Q^2)= {d}^{ n}(x,Q^2);\quad
  { d}^{ p}(x,Q^2)= {u}^{ n}(x,Q^2).\label{isospin}
\eeq
One then has 
\beq
{ F_2^p(x,Q2)-F_2^d(x,Q^2)= \frac{1}{3}\left[\left(u^p+\bar
    u^p\right)-\left(d^p+\bar d^p\right)\right]\left[1+
    O(\alpha_s)\right]}\label{nonsingf2}
\eeq
so that the difference of proton and deuteron structure functions
provides a leading-order handle on the isospin triplet combination
\beq
T_3(x,Q^2)\equiv u(x,Q^2)+\bar
    u(x,Q^2) -\left[d(x,Q^2)+\bar d(x,Q^2)\right].
\label{tripdef}
\eeq
 Note that even beyond leading order $F_2^p-F_2^d$ only
depends on $T_3$, which can thus be determined 
without further assumptions: a theoretically very clean, though
necessarily not especially accurate determination~\cite{DelDebbio:2007ee}.

\subsection{Light quarks and antiquarks}
\label{lightqqbar}

Modern DIS data are available over a wide range of values of $Q^2$,
extending well into the region where the CC contributions are sizable:
in fact HERA-I data are available both for CC and NC scattering, both
with electron and positron beams. Unfortunately, collider data only
provide a fixed combination Eq.~(\ref{eq:ncxsect})
of the structure functions $F_1$ and $F_3$,
because for given $x$ and $Q^2$ Eq.~(\ref{eq:ncxsect}) implies that
$y$ can be varied only by changing the center-of-mass energy of the
hadron-lepton collision.  Hence, HERA data only provide three
independent combinations of structure functions and thus of parton
distributions (NC and CC with positively or negatively charged
leptons). However, a fourth combination is provided because the
$Q^2$ dependence of the $\gamma^*$ and $Z$ contributions to NC
scattering is different (see Eq.~(\ref{bdcoup}). It follows that the
very precise HERA data can determine four independent linear
combinations of PDFs, which can be chosen as the two lightest flavours
and antiflavours, with strangeness then determined by assumption.

Even without a neutrino factory, data on neutrino deep-inelastic
scattering are available, but typically on approximately isoscalar
nuclear targets. Because the energy of the neutrino beam typically has
a (more or less broad) spectrum, the value of $y$ Eq.~(\ref{ydef}) is
not fixed, and the contributions of $F_1$ and $F_3$ to the cross
section can be disentangled. On an isoscalar target at leading order
\begin{eqnarray}
  F_2^{\nu}&=& x( u+\bar{u}+d+\bar{d}+2s
  +2\bar{c} )+O(\alpha_s),\qquad
  F_2^{\bar{\nu}}= x( u+\bar{u}+d+\bar{d}
+2\bar{s}+2c)+O(\alpha_s),\nonumber \\
\label{f23nu}
  F_3^{\nu}&=&  u-\bar{u}+d-\bar{d}+2s-2\bar{c}+O(\alpha_s),\qquad
F_3^{\bar{\nu}}= u-\bar{u}+ d-\bar{d}-2\bar{s}+2c+O(\alpha_s),
\eea
so neutrino data provide an accurate handle on the total valence
component 
\be
V(x,Q^2)=\sum_{i=}^{n_f} (q(x,Q^2)-\bar q_i(x,Q^2)).
\label{valdef}
\ee

A more direct determination of the light flavour decomposition can be
obtained by exploiting the fact that the Drell-Yan cross section
probes various parton combinations, which can be selected by looking
at different final states. In particular one can notice~\cite{Ellis:1990ti}
 that for neutral-current Drell-Yan if both data on
 proton and neutron (or deuteron) targets are available, using isospin
 Eq.~(\ref{isospin}) one gets at leading order 
\be
\frac{\sigma^{pn}}{\sigma^{pp}}\sim \frac{\frac{4}{9}u^p\bar
  d^p+\frac{1}{9}d^p\bar u^p}{\frac{4}{9}u^p\bar
  u^p+\frac{1}{9}d^p\bar d^p}+ O(\alpha_s)+\hbox{heavier quarks},
\label{dyncasym}
\ee
where we have omitted the dependence on the kinematic variables, which
at leading order is as in Eq.~(\ref{lody}). As discussed there,
if the rapidity distribution is measured, the leading order partonic
kinematic is completely fixed: for given $y$ and $Q^2$ only partons
with $x_1$, $x_2$ given by Eq.~(\ref{lowerlim}) contribute. Here
``heavier quarks'' denote strange and heavier flavours, which give a smaller
contribution at least in the region of $x\gsim 0.1$ in which most of
the contribution to the sum rule
integrals Eq.~(\ref{momsr},\ref{momsr}) is concentrated. 
\begin{figure}\begin{center}
\includegraphics[width=.45\linewidth]{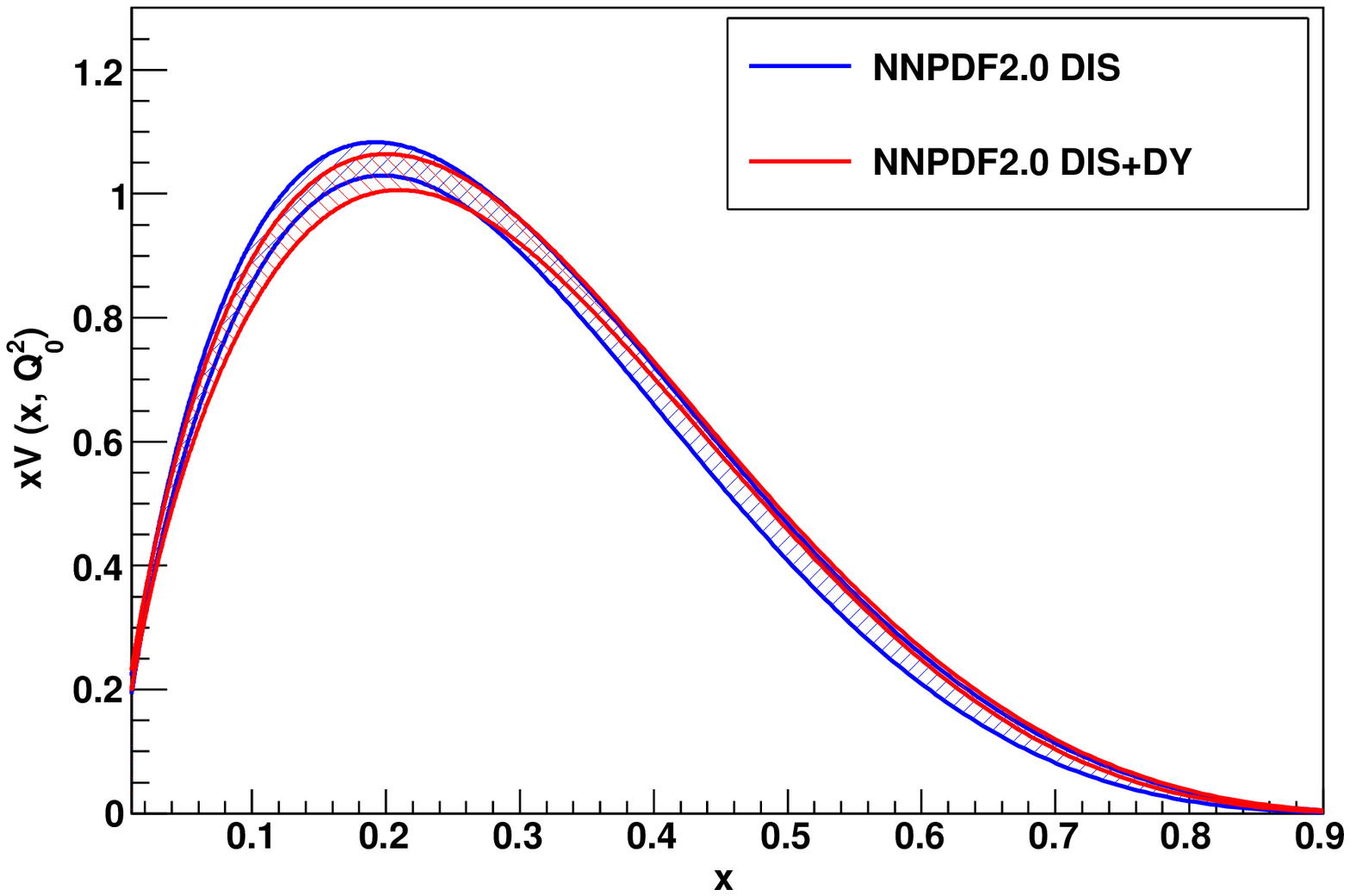}
\includegraphics[width=.45\linewidth]{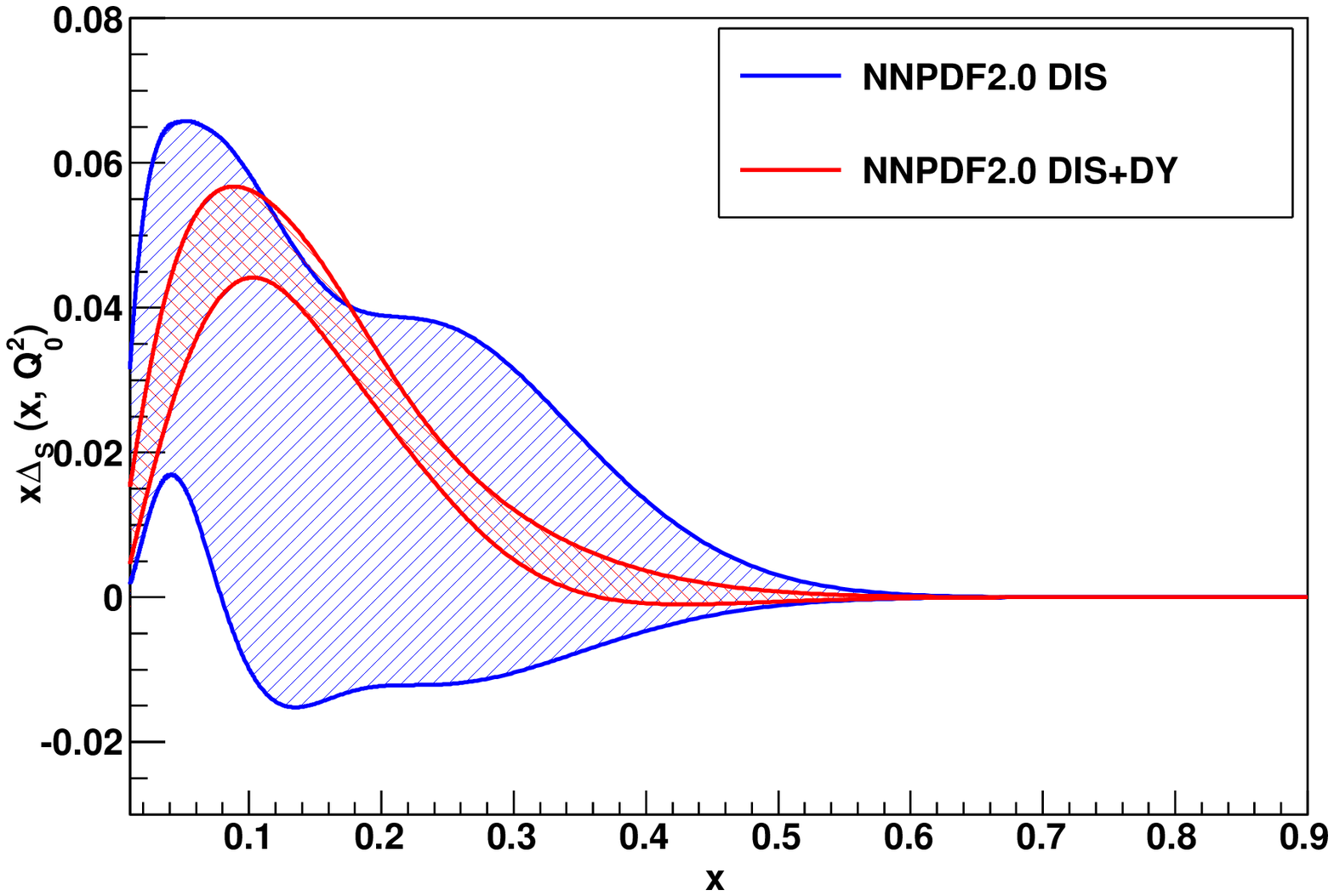}
\caption{Impact of Drell-Yan data on the NNPDF2.0~\cite{Ball:2010de}
  parton fit. Left: the total valence distribution; right: the light
  antiquark asymmetry.} 
\label{dyimp}
\end{center}
\end{figure}

In particular, because of the sum rule Eq.~(\ref{baryonsr}), in the
the region which gives the dominant
contribution to the integral (the ``valence'' region $x\gsim 0.1$) the
up distribution is roughly twice as large as the down distribution
(assuming $\bar u\sim\bar d$) so the
first term in both the numerator and the denominator of
Eq.~(\ref{dyncasym}) gives the dominant contribution, and the ratio
reduces to $ \frac{\sigma^{pn}}{\sigma^{pp}}\approx \frac{\bar
  d^p}{\bar u^p}$.
Hence this particular combination of cross sections provides a sensitive
probe of the $\bar u/\bar d$ ratio: indeed, it has been used to
provide first evidence that this ratio, though of order one, 
deviates from unity~\cite{Baldit:1994jk,Hawker:1998ty}.

In the charged current case, one may exploit the fact that
using charge-conjugation symmetry to relate the $p$ and $\bar p$ PDFs 
\be
q_i^{p}=\bar q_i^{\bar p}
\label{ccon}
\ee
at leading order one gets 
\be
\frac{\sigma^{p\bar p}_{W^+}}{\sigma^{p\bar p}_{W-}}=\frac{ u^p(x_1)
  d^p(x_2)+ \bar d^p(x_1)\bar u^p(x_2)}{d^p(x_1)
  u^p(x_2)+ \bar u^p(x_1)\bar d^p(x_2)}+ O(\alpha_s)+\hbox{ Cabibbo
  suppressed}+\hbox{ heavy quarks}
\label{dywasym}
\ee
where heavy quarks denotes charm and heavier flavours. 
In writing Eq.~(\ref{dywasym}) we have assumed that cross sections are
differential in rapidity. If the kinematics is chosen in  such  such a 
way that $x_i$ are in the ``valence''
region, in which quark distributions are sizably larger than antiquark
ones, the ratio Eq.~(\ref{dywasym}) is mostly sensitive to the light
quark ratio $u/d$~\cite{Berger:1988tu,Martin:1988aj} 
and indeed it has been used to provide the
first accurate determinations of it~\cite{Abe:1994rj}.

The sizable impact of Drell-Yan data on a PDF fit is demonstrated in
Fig.~\ref{dyimp}, where we compare  the value and uncertainty of PDF
combinations which are sensitive to the light flavour decomposition
before and after inclusion of Drell-Yan data in a DIS fit, 
namely the total valence
Eq.~(\ref{valdef}) and the light sea asymmetry
\be
\Delta_s(x,Q^2)\equiv \bar d(x,Q^2)-\bar u(x,Q^2).
\label{asymdef}
\ee
The DIS fit includes both the fixed-target proton and neutron data
(which thus determine well the isotriplet component and give a handle
on the singlet-triplet separation), the precise HERA
data (which give a handle on each individual light flavour and
antiflavour), and several neutrino data (which give a handle on the
total valence component). The Drell-Yan data included contain both proton
and deuteron fixed target $\gamma^*$ production data, and $W$ and $Z$ production.
It is apparent that the accuracy on the valence, which
is already quite good in the DIS-only fit, is reduced by a large
factor by the inclusion of Drell-Yan data, and the effect is even more
impressive on the light antiquark asymmetry which, despite the accuracy
of the HERA data, is only determined with large uncertainties by DIS data.

\begin{figure}\begin{center}
\includegraphics[width=.45\linewidth]{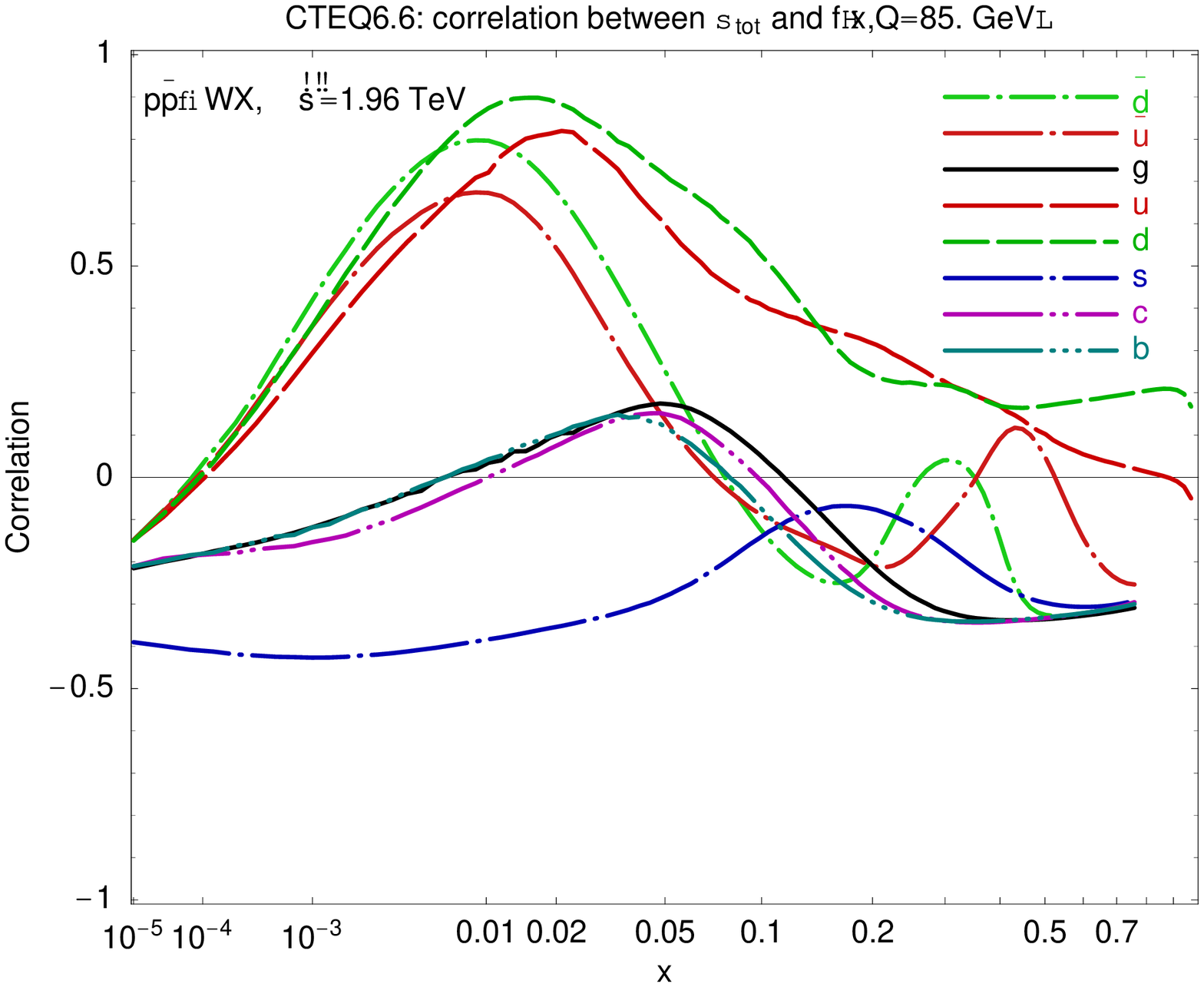}
\includegraphics[width=.45\linewidth]{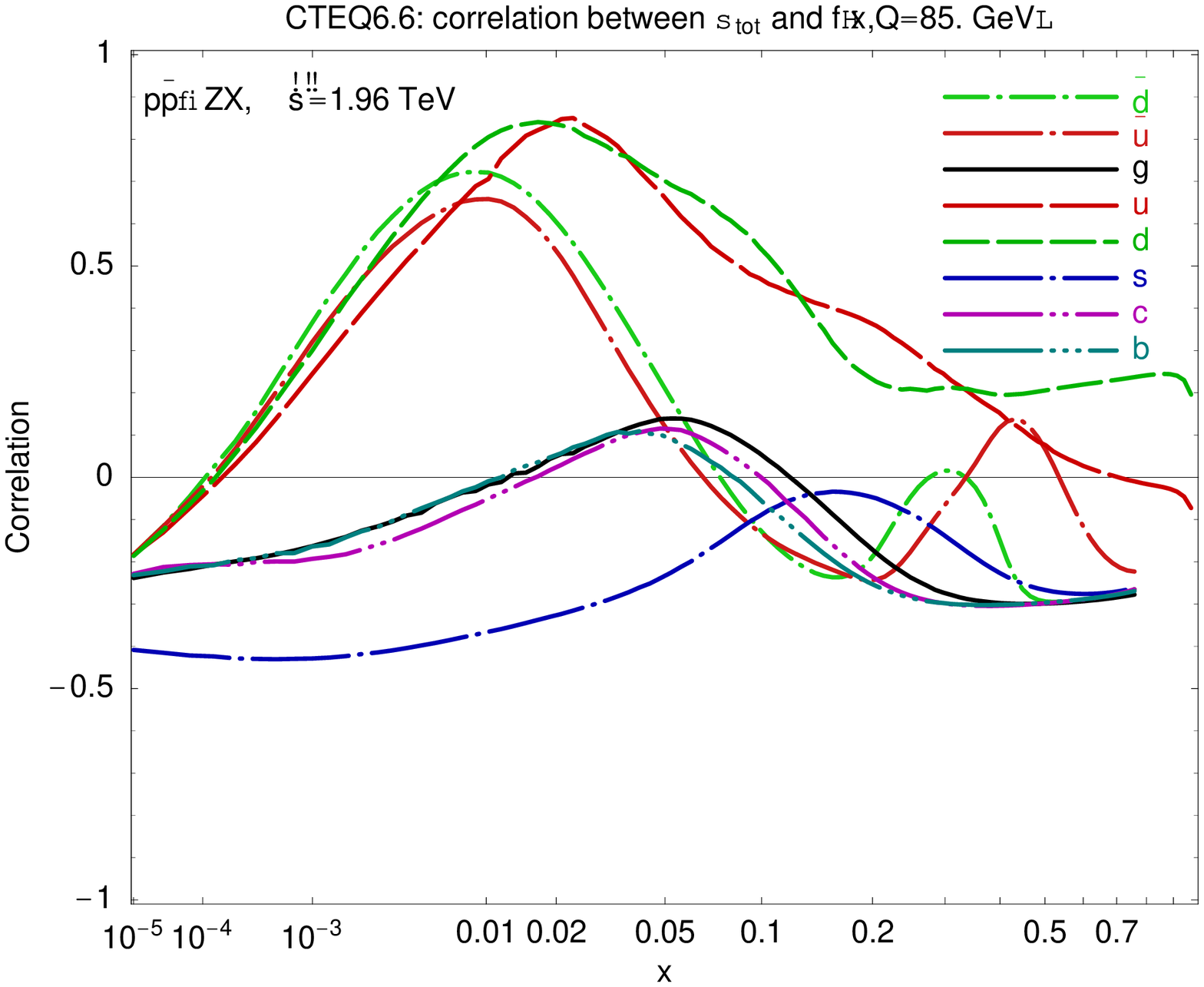}
\caption{Correlation between the total $W$ and $Z$ cross section at
  the Tevatron and individual parton distributions (from Ref.~\cite{Nadolsky:2008zw}).} 
\label{correln}
\end{center}
\end{figure}
The strong impact of $W$ and $Z$ production data can be seen
quantitatively by computing the correlation coefficient between the
$W$ and $Z$ cross section and individual parton distributions, which
can be computed both in a Hessian approach using standard error
propagation, or in a Monte Carlo approach from the covariance
of the cross section and the parton distribution over the Monte Carlo
sample. Results obtained in the Hessian approach using 
CTEQ6.6~\cite{Nadolsky:2008zw}
are shown in Fig~\ref{correln}: correlations are quite large, even
though results shown here are obtained using the total cross section,
which is a much less sensitive probe of PDFs than the rapidity
distributions discussed above.

\subsection{Strangeness}
\label{strangesec}

\begin{figure}\begin{center}
\includegraphics[width=.45\linewidth]{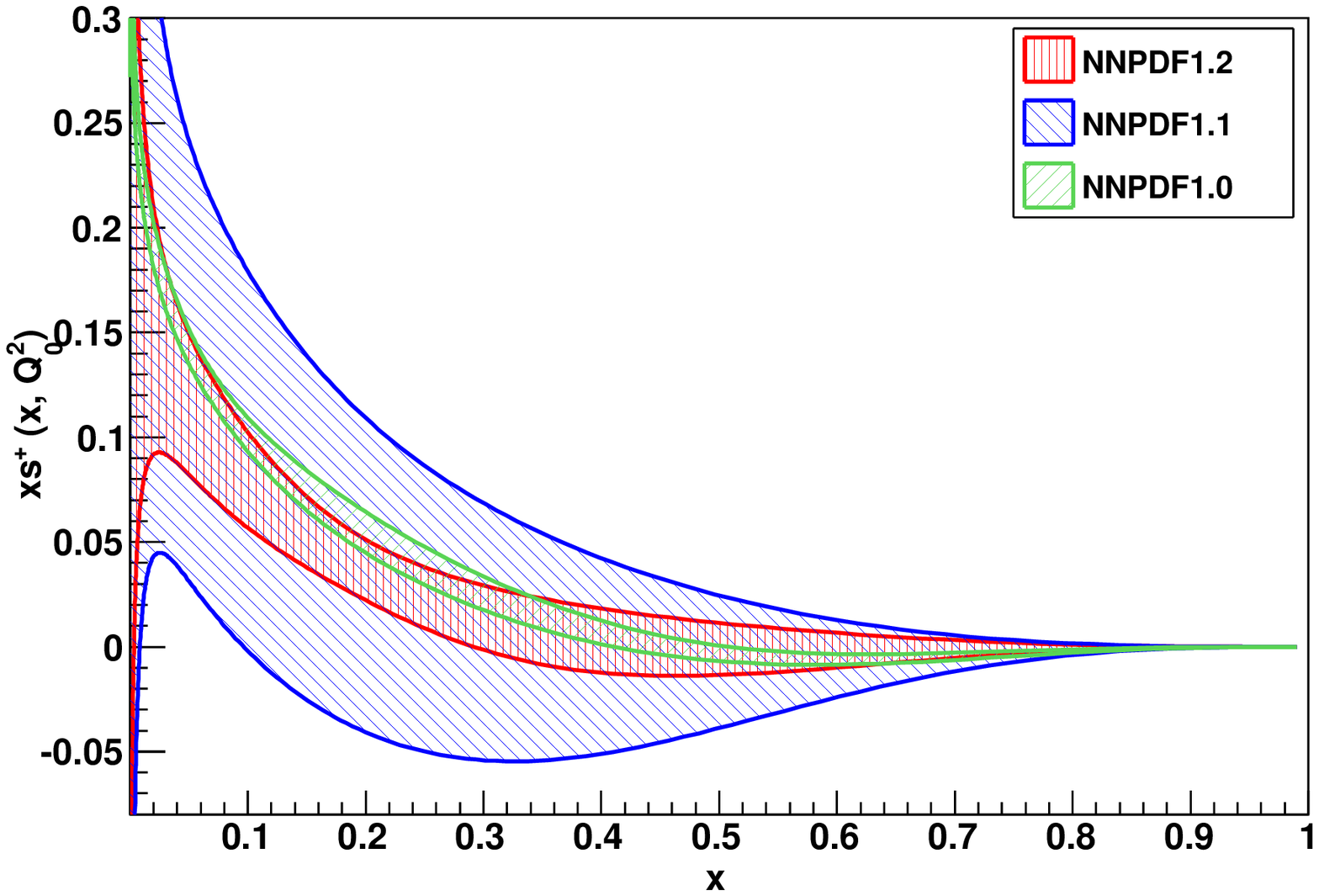}
\includegraphics[width=.45\linewidth]{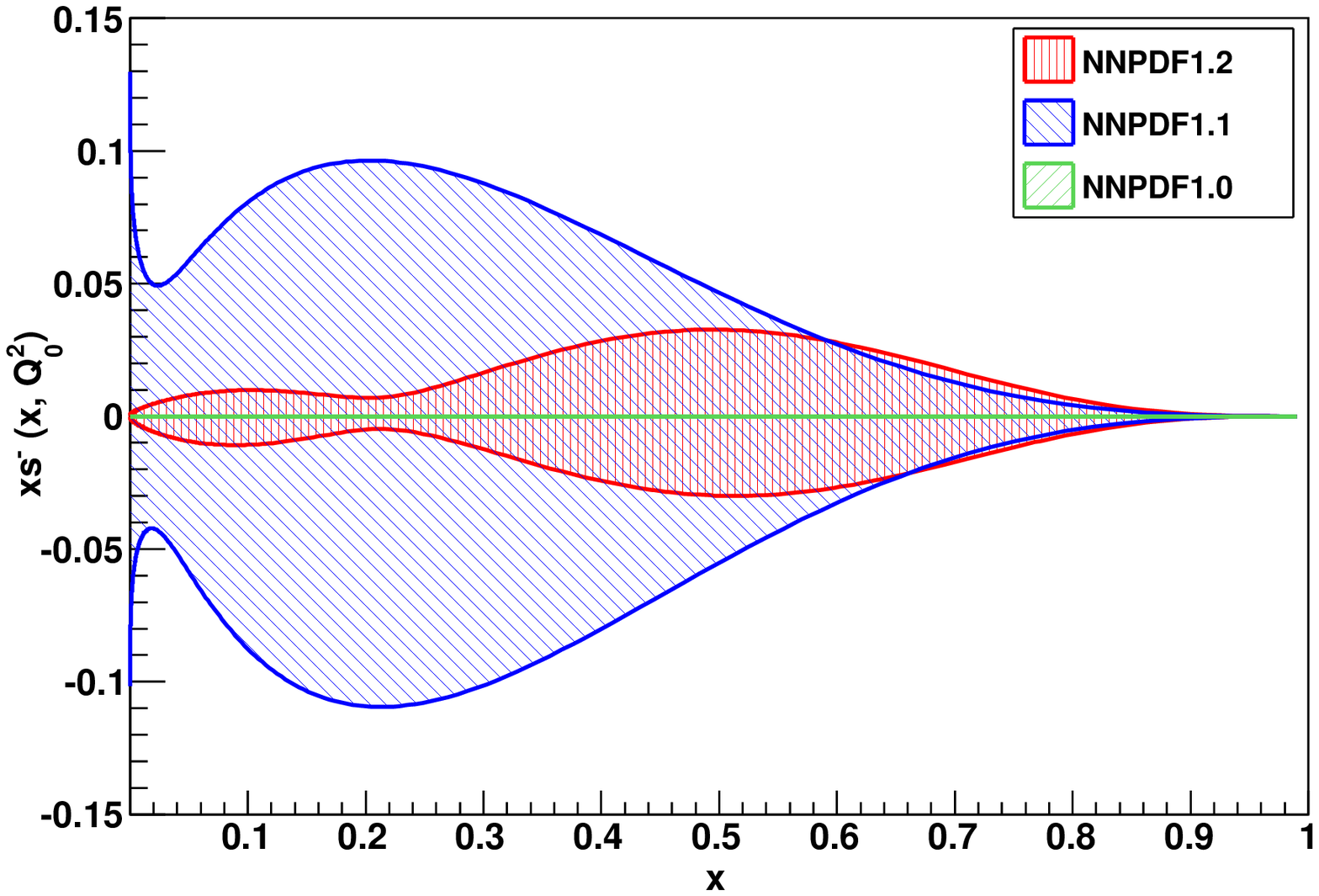}
\caption{The strange distributions $s^+=s+\bar s$ (left) and
$s^-=s-\bar s$ (right) determined in
  a fit to DIS including dimuon data (NNPDF1.2~\cite{Ball:2009mk}) and
  not including them 
  (NNPDF1.1~\cite{Rojo:2008ke}). A fit without dimuon data where strangeness is
  fixed by assumption ((NNPDF1.0~\cite{Ball:2008by}) is also shown.} 
\label{strangdim}
\end{center}
\end{figure}
The determination of strangeness is nontrivial because, of course, 
it has the same
electroweak couplings as the down distribution, while it is typically
smaller than it (except at small $x$ where all PDFs are the same size,
as discussed in Sect.~\ref{pdfconstr}). The only way of determining it
accurately from deep-inelastic scattering data is to include
semi-inclusive information. A simple way of doing this is to use data
for neutrino deep-inelastic charm production (known as dimuon
production, because charm is tagged by the muon from its decay
together with the muon due to the charged current neutrino
interaction). 
At leading order the structure functions are then just
\bea
  &&F_2^{\nu,p,c}(x,Q^2)=x F_3^{\nu,p,c}(x,Q^2)= 2x\,
  \big(|V_{cd}|^2\,d(x)\,+|V_{cs}|^2\,s(x)+|V_{cb}|^2\,b(x)\big)+O(\alpha_s^2),
  \nonumber \\
\label{fnbexpr}
  &&F_2^{\bar\nu,p,c}(x,Q^2)=-xF_3^{\bar\nu,p,c}(x,Q^2)=
2x\,
\big(|V_{cd}|^2\,\bar d(x)\,+|V_{cs}|^2\,\bar s(x)
+|V_{cb}|^2\,\bar b(x)\big)+O(\alpha_s^2)  ,
\eea
so up to CKM suppressed terms they measure strangeness directly. 

In
Fig.~\ref{strangdim} the behaviour upon inclusion of dimuon data 
of a  fit to a set of DIS data which
includes both neutrino and HERA data is shown: it is clear
that before inclusion of the dimuon data the fit
(NNPDF1.1~\cite{Rojo:2008ke}) cannot determine either of the two
strange combinations
\be
s^\pm(x,Q^2)\equiv s^+(x,Q^2)\pm s^-(x,Q^2),
\label{splusmdef}
\ee
but after their inclusion it determines both, though with limited
accuracy due to the limited accuracy and kinematic coverage of the
available dimuon data. In this plot, we also show the result one
obtains for strangeness if one simply assumes it to be proportional to
the light quark sea, i.e. if by assumption one sets  $s^-=0$ and
$s^+=\half(\bar u + \bar d)$. This is often done in PDF determinations
based on DIS
data only: the result is then misleadingly accurate. This comparison
should thus be taken as a warning that, when using PDF sets in which some
PDFs are fixed by assumption, some uncertainties may be significantly
underestimated. 

\begin{figure}\begin{center}
\includegraphics[width=.45\linewidth]{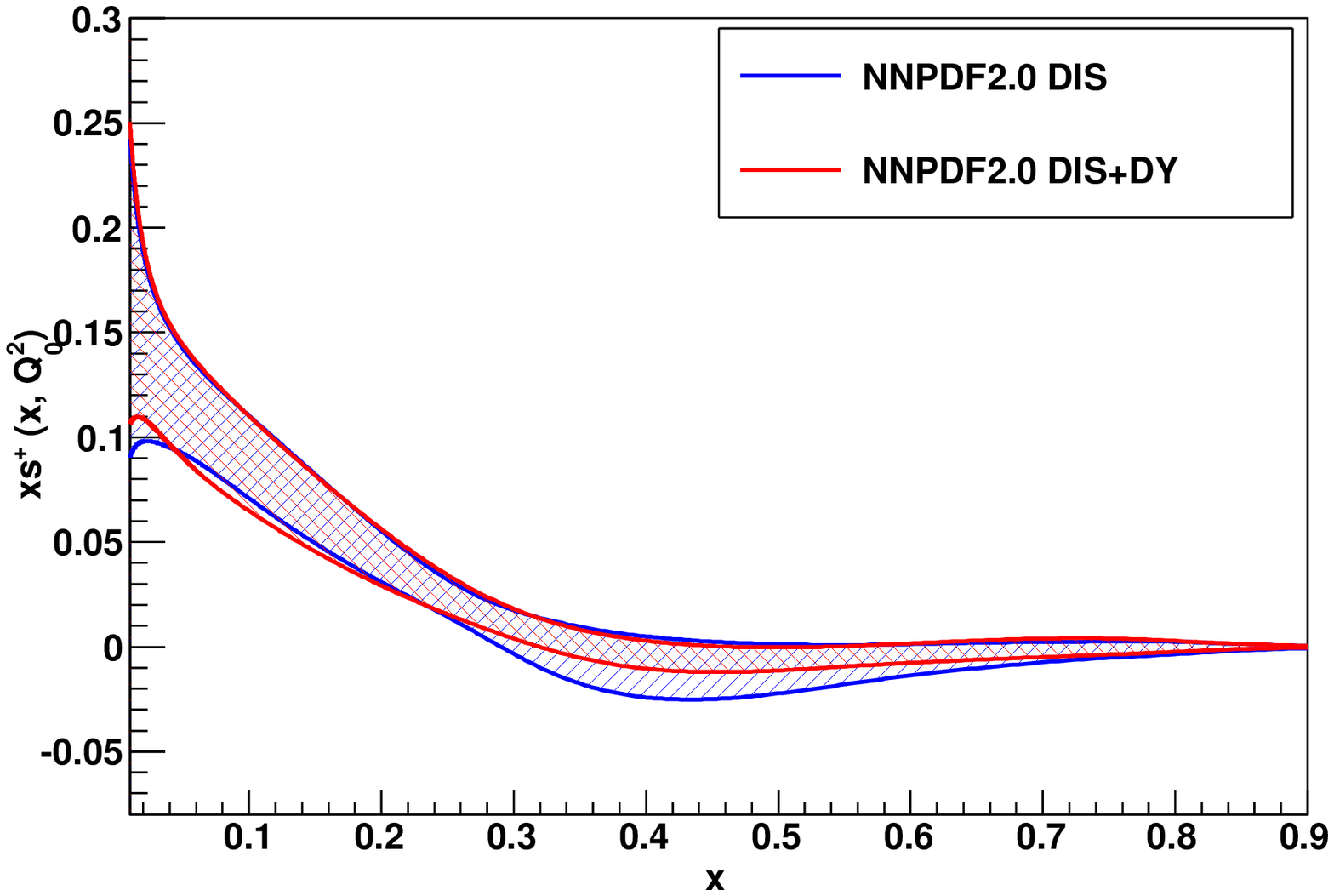}
\includegraphics[width=.45\linewidth]{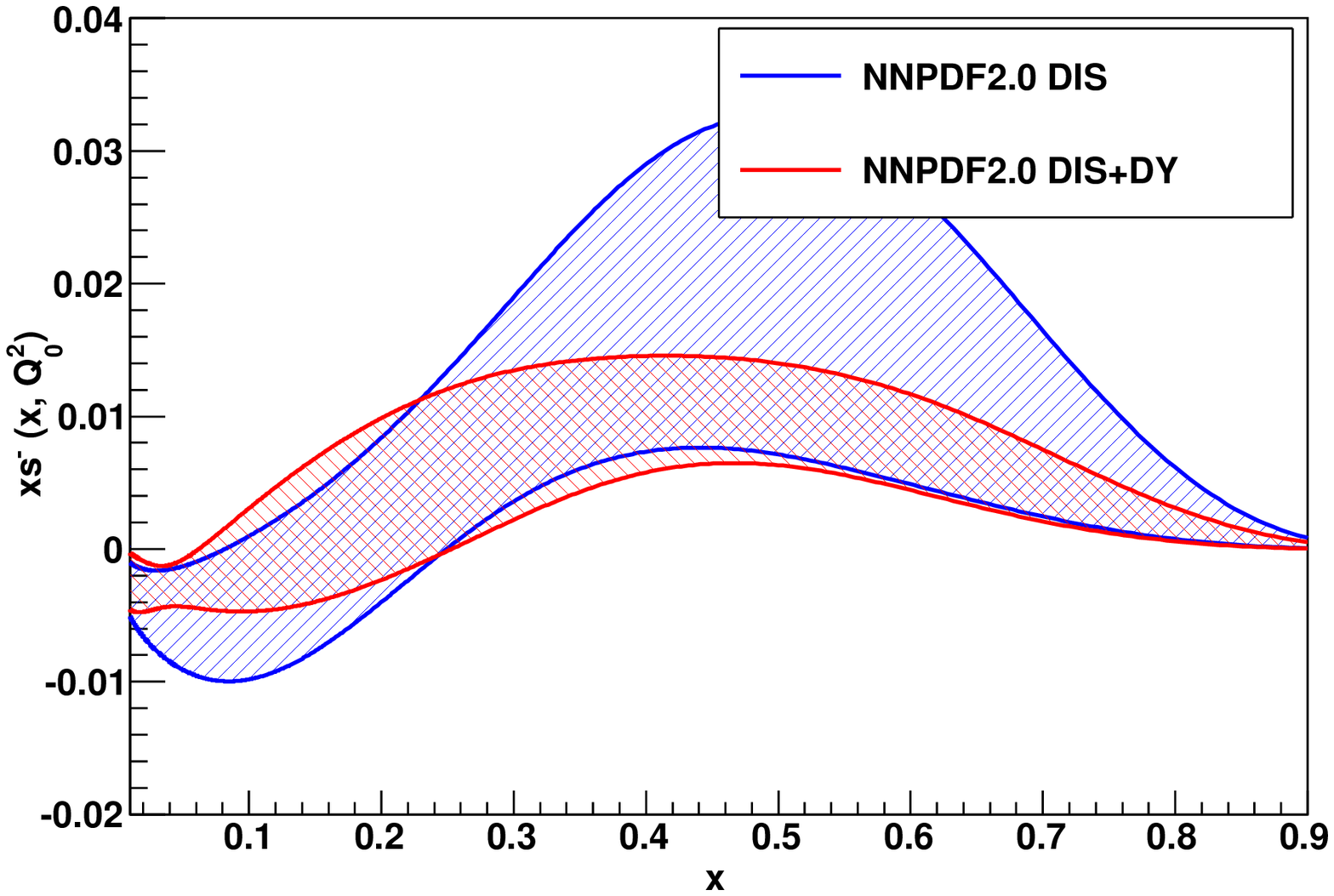}
\caption{Impact of Drell-Yan data on the
strange distributions $s^+=s+\bar s$ (left) and
$s^-=s-\bar s$ (right) using
the NNPDF2.0~\cite{Ball:2010de}
  parton fit.} 
\label{strangdy}
\end{center}
\end{figure}
Of course the Drell-Yan data discussed above also constrain
strangeness. Specifically, the cross section ratio Eq.~(\ref{dywasym})
receives a contribution from strange and charm quarks which, up to CKM
matrix elements, is identical to the contribution from down and up
quarks respectively. Well above charm threshold this contribution is
sizable, so comparing Drell-Yan data above and below charm threshold
potentially leads to a rather accurate determination of strangeness.
Indeed, in Fig.~\ref{strangdy} we show the impact of including
Drell-Yan data in a fit with DIS data only (same pair of fits already
shown in Fig.~\ref{dyimp}). The DIS dataset contains
dimuon data, and it is similar to the dataset on which the fit of 
Fig.~\ref{strangdim} is based, from which it mostly differs because of
improvements in the HERA data and in fit methodology; however, the
Drell-Yan data have a visible impact on the total strangeness $s^+$,
and lead to a very striking improvement in the determination of the
strangeness asymmetry $s^-$.

\subsection{Gluons}
\label{glusec}

The determination of the gluon distribution is nontrivial because the
gluon does not couple to electroweak final states. It does, however,
mix at leading order through perturbative evolution: so, even using
parton-model (i.e. O($\alpha_s^0$)) expressions for cross sections and
structure functions, the gluon does determine their scale
dependence. Indeed
\be
{ {d\over dt} F_2^s (N,Q^2)}=
{\alpha_s(Q^2)\over
2\pi} \left[{ \gamma_{qq}(N)} { F_2^{s}}+2  \, 
n_f{ \gamma_{qg}(N)g(N,Q^2)}\right]+O(\alpha_s^2),
\label{f2loscaldep}
\ee
where by $F_2(N,Q^2)$ we denote the Mellin moments
Eq.~(\ref{mellindef}) of the singlet 
component (defined as in Eq.~(\ref{singdef})) of the $F_2$ structure
function.

\begin{figure}\begin{center}
\includegraphics[width=.45\linewidth]{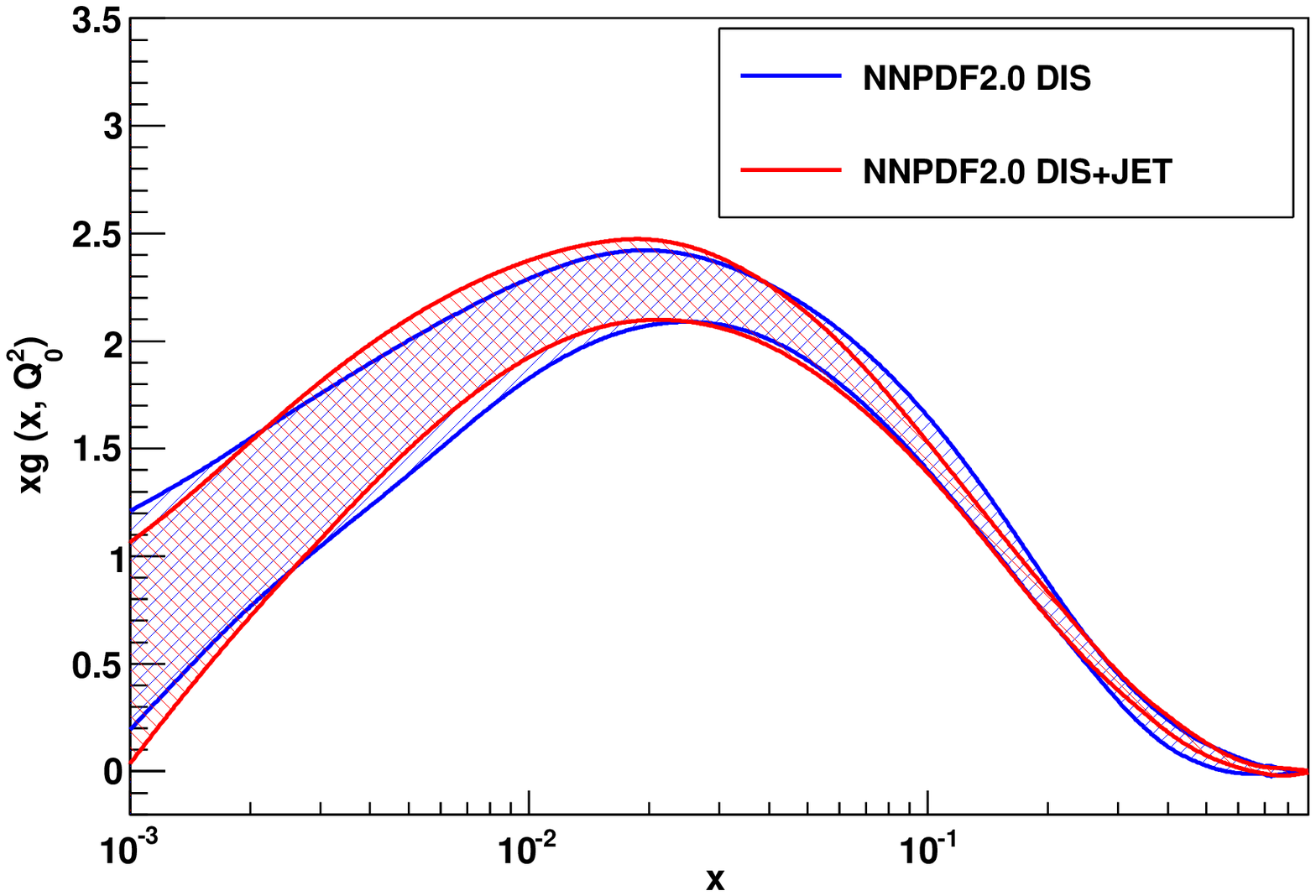}
\includegraphics[width=.45\linewidth]{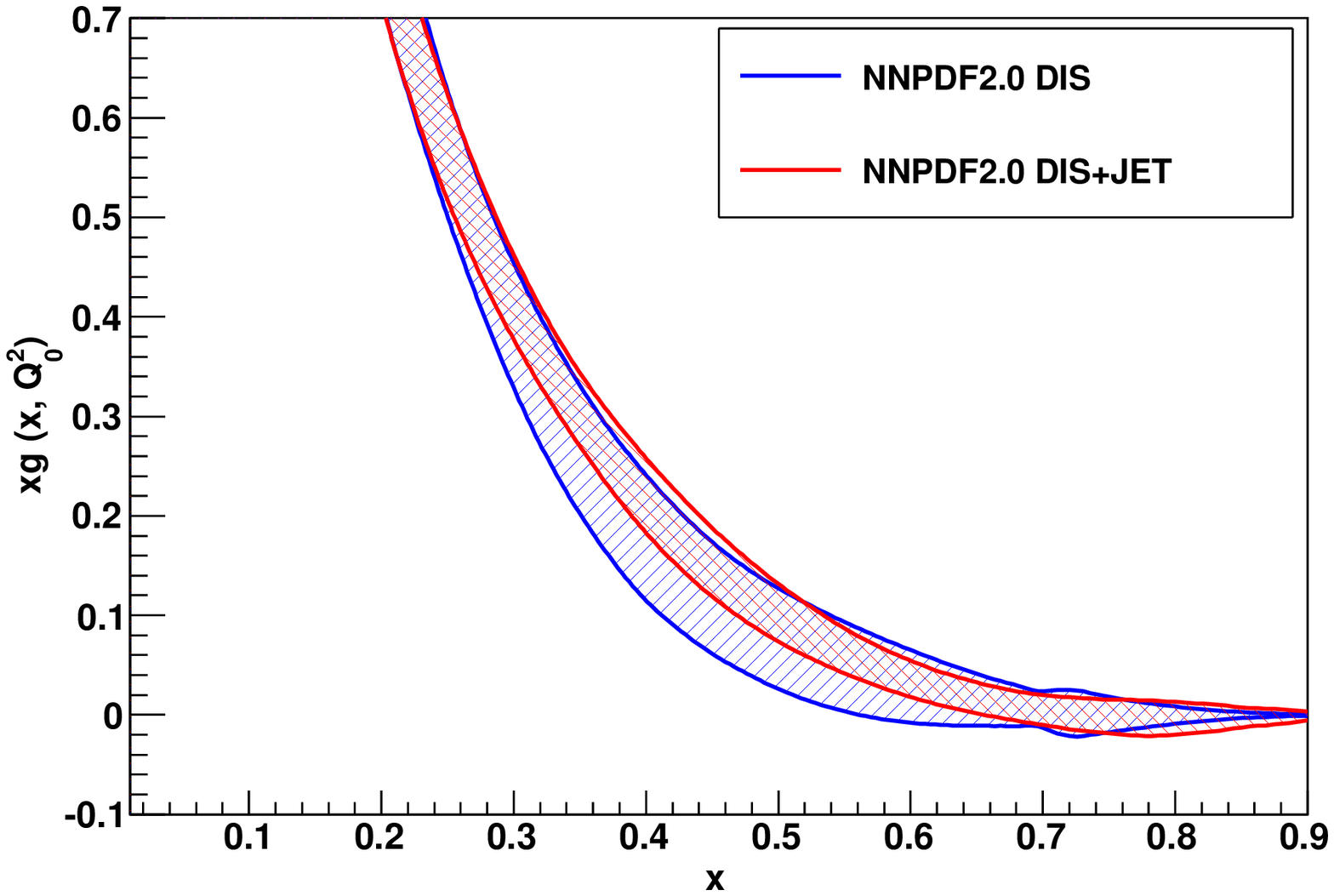}
\caption{Impact of jet  data on the
gluon distributions at small $x$ (left) and large $x$ (right)
using the NNPDF2.0~\cite{Ball:2010de}
  parton fit.} 
\label{glujet}
\end{center}
\end{figure}
 It follows that the gluon is mostly determined by scaling
violations, or by its coupling to strongly--interacting final-states,
i.e. jets. The main shortcoming of the determination from scaling
violations is that, as already pointed out in Sect.~\ref{pdfconstr},
the gluon only couples strongly to other PDFs for sufficiently small
$x$: for instance, Fig.~\ref{lodglap} shows clearly that  for $N>2$
the $\gamma_{qg}$ term rapidly becomes negligible in comparison to the
$\gamma_{gg}$ term.  On the other hand, the gluon distribution is
expected to be quite small at large $x$, and, 
as also discussed in
Sect.~\ref{pdfconstr}, to further shift towards smaller $x$  as the
scale increases. Hence, the large $x$ gluon is likely to be small and
affected by large uncertainties, which can only be reduced by looking
at hadronic (jet) final states.

Indeed, in Fig.~\ref{glujet} we show the effect of the inclusion of
jet data in a PDF fit based on DIS data. At small $x$ there is
essentially no effect: scaling violations are sufficient to determine
the gluon quite accurately. At large $x$, even though the
determination of the gluon from scaling violations is reasonably
accurate, its accuracy  is still quite significantly improved by the
inclusion of jet data. A feature of this plot which is worth noting
is the beautiful consistency of these two determinations. This is an extremely
strong consistency check for the perturbative QCD framework: the gluon
determined from scaling violation and evolved up to the much higher
jet scale is in perfect agreement with  the jet data, and indeed the
best accuracy is obtained 
combining the two determinations.

\subsection{Global fits}
\label{globfit}
\begin{figure}\begin{center}
\includegraphics[width=.7\linewidth]{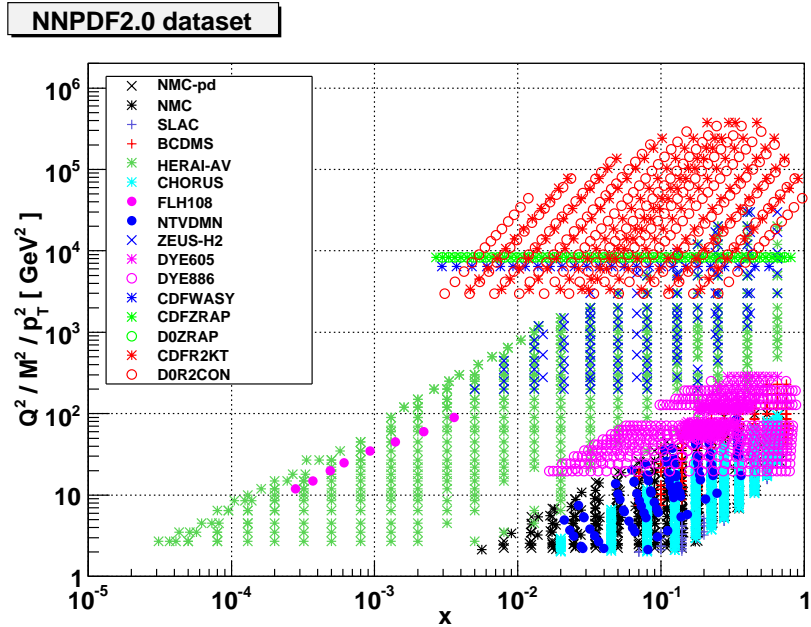}
\caption{The data set used in a typical global PDF determination
  (NNPDF2.0~\cite{Ball:2010de}).} 
\label{nnpdftdata}
\end{center}
\end{figure}
It is clear that the wider the number of different processes, the
greater the amount of information which is being used in the
determination of PDFs. The price to pay for this, as we will discuss
in Sect.~\ref{PDFunc}, is that the determination of PDFs and especially
their uncertainties from diverse and possibly inconsistent data might
be nontrivial --- at the very least, it is going to be computationally
intensive. Current global fits use all the processed discussed so far
in order to control as much as possible different aspects of
PDFs. 

The dataset used in one such fit (NNPDF2.0~\cite{Ball:2010de})
is shown in Fig.~\ref{nnpdftdata}. Different data in this set
constrain different aspects of PDFs, along the lines of the preceding
discussion, in a way which, referring to this specific dataset, can be
summarized as follows:

\begin{itemize}
\item information on the overall shape of quarks and gluons at medium
  $x$ as well as on the isosinglet-isotriplet separation come from
  fixed-target DIS data on proton and deuterium targets 
 (dominated by $\gamma^*$ exchange), denoted in
  the plot as NMC~\cite{Arneodo:1996qe}, NMCpd~\cite{Arneodo:1996kd},
  SLAC~\cite{Whitlow:1991uw} and BCDMS~\cite{Benvenuti:1989rh}; 
\item an accurate determination of the behaviour of the  gluon and
  quark at small $x$  (where it is dominated by the singlet) 
and by individual light flavours at medium $x$ (where CC and
  NC data play a role in separating individual flavours)
is found
  from the very precise HERA CC and NC data denoted in the plot
  as HERAI-AV~\cite{:2008tx}, which were obtained by combining the
  ZEUS and H1 data from the HERA-I run. More recent HERA-II
  ZEUS NC~\cite{Chekanov:2009gm}
  and CC~\cite{Chekanov:2008aa}
  data (ZEUS-H2) are also used.
\item information on the flavour separation at small $x$ comes from
  Tevatron Drell-Yan data  (in particular the $W$ asymmetry, as
  discussed above) denoted in
  the plot as CDFWASY~\cite{Aaltonen:2009ta},
  CDFZRAP~\cite{Aaltonen:2010zza}, D0ZRAP~\cite{Abazov:2007jy};
\item the flavour separation at  medium $x$ is mostly controlled by
  the   Tevatron Drell-Yan data on fixed proton and nucleus target, 
DYE605~\cite{Moreno:1990sf} and
  DYE866~\cite{Webb:2003ps,Webb:2003bj,Towell:2001nh} in the figure.
\item the total valence component is constrained by the neutrino
  inclusive DIS data, denoted as CHORUS~\cite{Onengut:2005kv} in the plot;
\item strangeness is controlled by neutrino dimuon data
  (NTVDMV~\cite{Goncharov:2001qe,Mason:2006qa}) , as well as by
  the interplay of the $W$ and $Z$ production data with lower scale
  DIS and Drell-Yan data;
\item the large $x$ gluon, already determined by DIS scaling
  violations, is further constrained by   Tevatron jet data
  (CDFR2KT~\cite{Abulencia:2007ez}, D0R2CON~\cite{:2008hua}).
\end{itemize}

Other global fits may differ in some detail, such as the specific
choice of experiments or the addition or subtraction of some set of
data, but are mostly based on datasets constructed on the basis of a
similar logic. Smaller datasets, typically a subset of the above, are
also considered.

Future improvements on some of these processes, in particular
Drell-Yan (including $W$ and $Z$) production and jet production will
certainly come from the LHC, both because of the higher available
center-of-mass energy (compare Fig.~\ref{lhckinxs}), and because of the
higher statistics which will be accumulated once higher or design
luminosity are reached. Some other processes which are likely to
become important at the LHC are prompt photon and heavy quark
production, as well as  Higgs production (if the Higgs is found and
understood), all of which are sensitive probes of the gluon distribution.
We will briefly come back on these issues in Sect.~\ref{lhcpd}, after
discussing the current main difficulty in the understanding of PDFs,
namely, the treatment of PDF uncertainties.

\section{PDF uncertainties}
\label{PDFunc}
\begin{figure}\begin{center}
\includegraphics[width=.4\linewidth]{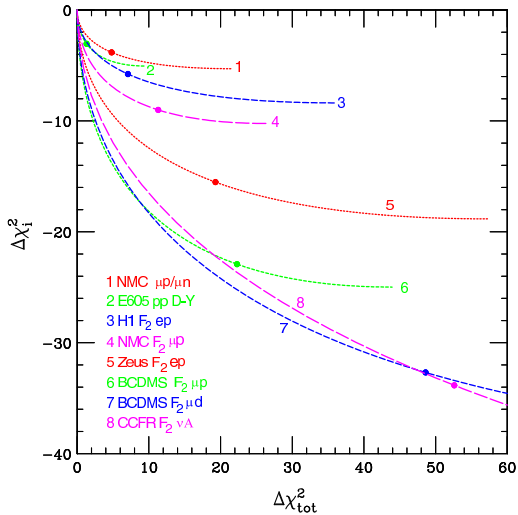}
\caption{Decrease of the $\chi^2$ for various dataset entering a
  global fit plotted as a function of the increase in $\chi^2$ of the
  global fit when moving away from the global minimum.
  (from Ref.~\cite{Collins:2001es}).} 
\label{dataincons}
\end{center}
\end{figure}
The accurate determination of PDF uncertainties is clearly necessary
if one wants to be able to obtain meaningful predictions from the
factorized QCD expressions of Sect.~\ref{fact}. 
Because PDFs are determined by comparing QCD predictions to the data,
as discussed in Sect.~\ref{statsec}, any uncertainty in the theory
used to obtain these predictions will
propagate onto the PDFs themselves. Such uncertainties include genuine
theoretical uncertainties, such as lack of knowledge of higher-order
perturbative corrections: these, generally,   do not have a simple
statistical interpretation (and in particular they are generally not
gaussian). They also include lack of knowledge of parameters in the
theory, in particular the value of the strong coupling constant
$\alpha_s$ and the heavy quark masses $m_c$ and $m_b$, which generally
do follow gaussian statistics. The treatment of these uncertainties is in
principle straightforward, in the sense that all one has to do is
propagate them onto the PDFs --- their effect on PDFs is no different
from their effect on the calculation of a physical observable, and PDFs
do not entail any new problem.
 For
example, if it is agreed that higher order corrections on
cross sections can be conventionally estimated by varying
renormalization and factorization scales in a certain range, to be
interpreted, say, as a 90\% confidence level with flat distribution,
the associate PDF uncertainty is simply found by repeating the PDF
determination while performing this variation. We will refer to these as
``theoretical uncertainties'', and come back to them in Sect.~\ref{th}.

On top of these, however, PDFs are
affected by statistical uncertainties which are related to the way
the information contained in the data is propagated onto a PDF
determination following the process summarized in
Fig.~\ref{nnpdfscheme}. The determination of these uncertainties is
highly nontrivial because, as discussed in Sect.~\ref{statsec}, the
desired final outcome of this process is the determination of a probability
distribution in a space of functions: these uncertainties are supposed
to behave as genuine statistical uncertainties, with a well-defined
probability distribution, and it is not obvious how to make sure, and
then verify, that this is the case. These 
will be referred to as ``PDF uncertainties'' for short. 

First attempts to determine PDF sets which include PDF uncertainties
are only quite
recent~\cite{Alekhin:1996za,Barone:1999yv,Botje:1999dj}; they immediately
met with the difficulty that as soon as wide enough datasets (such as those
discussed in Sect.~\ref{globfit}) are fitted, a standard statistical
approach does 
not seem to be adequate~\cite{Pumplin:2002vw,Martin:2002aw}. Furthermore, 
results obtained for relevant LHC processes such as Higgs production
using various different sets~\cite{Djouadi:2003jg} do not always agree well
with each other.  On both of these issues there has been considerable
progress over the last several years. On the one hand, the
understanding of statistical issues related to PDF uncertainties has
advanced considerably, and it will be reviewed in the remainder of
this section. On the other hand, existing PDF determination show a
distinct convergence as various phenomenological and theoretical
issues are addressed and understood, as we will see in Sect.~\ref{lhcpd}

\subsection{Tolerance}
\label{tol}

\begin{figure}\begin{center}
\includegraphics[width=.5\linewidth]{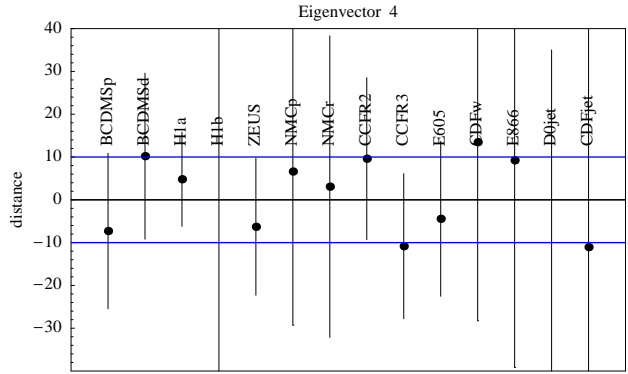}
\caption{Determination of the tolerance for a global fit (from
  Ref.~\cite{Pumplin:2002vw}). The distance on the $y$ axis is given
  in units of $\sqrt{\bar\chi^2}$ Eq.~(\ref{unnormchisqdef})
of the global fit. The interval shown
  for each experiment corresponds to a $90\%$ confidence level. The band
shown, corresponding to $\Delta \bar\chi^2=100$ is just wide enough to
accommodate the ZEUS (upper variation) and CCFR2 (lower variation)
experiments.}  
\label{toldet}
\end{center}
\end{figure}
Available fits to wide enough sets of data based on the Hessian
approach and the ``standard'' parton parametrization
Eq.~(\ref{pdfparmgen}) 
 discussed in
Sect.~\ref{sechessian} run into the difficulty that the best-fit is not
simultaneously a best-fit for individual datasets. Specifically, 
one can test for the possibility that the $\chi^2$
of the fit to individual datasets entering the
global fit may be improved  by moving away from the global minimum by
introducing Lagrange multipliers to select which dataset to
minimize~\cite{Collins:2001es}. Results, shown in
Fig.~\ref{dataincons}, are disquieting: not only  the minima of
individual experiments do not coincide with the global minimum  but
some of these minima seem to deviate much more than one might expect
on the basis of statistical fluctuations, and there even seem to be
runaway directions for some experiments.

This suggests that likelihood contours (for example one-$\sigma$) for the global fit
can only be determined while simultaneously testing for the degree of
agreement of individual experiments with it. The way this is done is
by introducing the concept of ``tolerance'', defined as follows~\cite{Pumplin:2002vw}. 
First, the Hessian matrix is diagonalized. Next, one moves the
value of each
eigenvector away from the minimum of the global fit in either
direction, and one computes  the $\chi^2$ of each experiment. Then,
for each experiment one determines both the position of the minimum of
the $\chi^2$ and the one-$\sigma$  interval
about it (corresponding to the $\Delta\bar \chi^2=1$ variation
about the minimum), or equivalently the 90\% confidence level
(obtained by rescaling the former interval by the  factor
$C_{90}=1.64485\dots$~\cite{PDGstat}). Finally, one takes the envelope
of the error bands for individual experiments at the desired
confidence level (c.l., henceforth). For example, at the 90\% c.l. one
determines the range of variation in parameter space along
this eigenvector about the minimum such that the 90\% c.l. interval
of each experiment overlaps with this range. This gives a tolerance
interval for the given eigenvector. The width of this interval can be
measured in units of  the variation of the $\chi^2$ of
the global fit. This defines a  tolerance: $T^2=\Delta
\chi^2$ is the width of the envelope (see Fig.~\ref{toldet}). 

The 90\% c.l.  is
finally taken to be  $\Delta
\chi^2=T^2$ instead of $\Delta \chi^2=c_{60}^2$ (equivalently, the one
$\sigma$ contour is $\Delta
\chi^2=T^2/c_{60}^2$).
  The logic behind this is
that PDFs should allow one to obtain predictions for new processes at
the desired confidence level: for instance, the actual result for a new measurement
should have a 68\% chance of actually falling into the predicted
one-$\sigma$ band. If new experiments behave as the
experiments which are already included in the fit do on average, then
this will happen for the one-$\sigma$ band defined in this way, while
if the one-$\sigma$ band were defined on the basis of standard
statistics the chances of the measurements falling outside the band
would be much higher. It should be stressed that therefore a tolerance
analysis is required in order for a fit based on this methodology to
be reliable (unless the dataset adopted is very small and/or consistent).
\begin{figure}\begin{center}
\includegraphics[width=.45\linewidth]{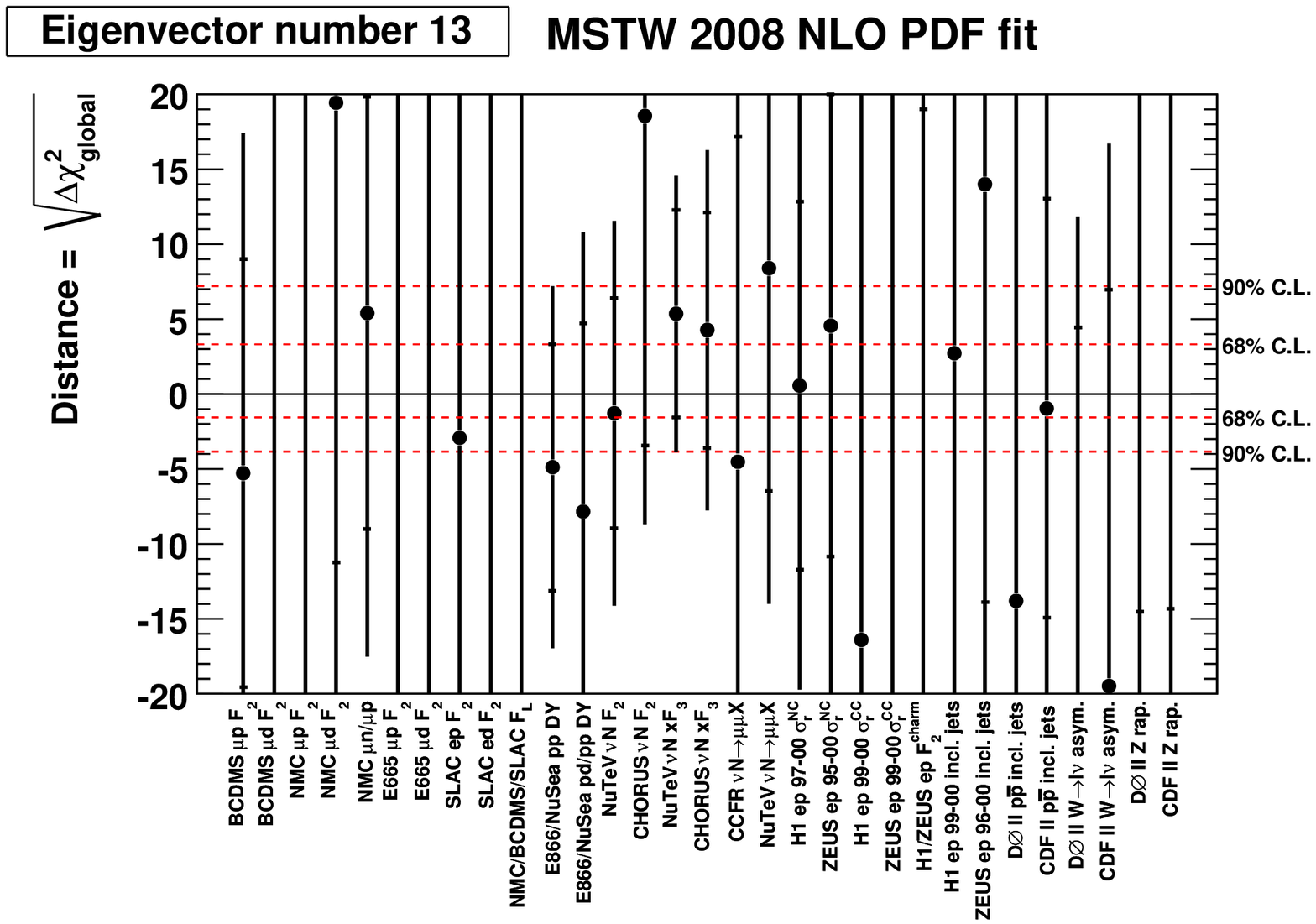}
\includegraphics[width=.45\linewidth]{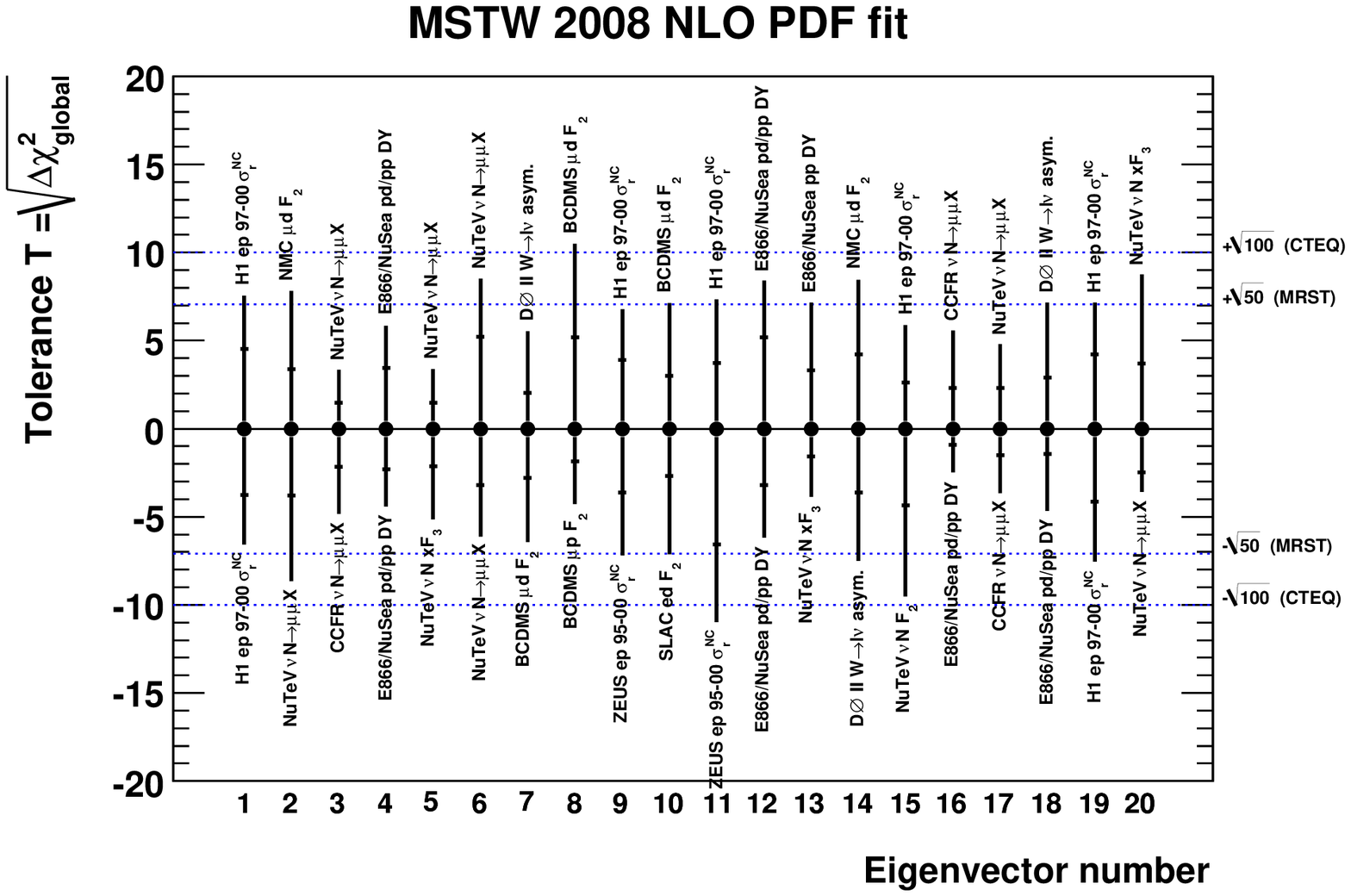}
\caption{Dynamical tolerance for the MSTW08 PDF fit. 
Left: the tolerance interval for the
  $13^{\rm th}$ eigenvector; the inner and outer uncertainty bands
  correspond for each experiment to the 68\%c.l. and
  90\%c.l. ranges. Right: the tolerance interval for each eigenvector,
  along with the experiments which determine it; the $T^2=50$ and
  $T^2=100$ previously~\cite{Pumplin:2002vw,Martin:2002aw} adopted are also shown (from
  Ref.~\cite{Martin:2009iq}).} 
\label{dyntol}
\end{center}
\end{figure}

In Ref.~\cite{Pumplin:2002vw} it was found that in practice $T^2=100$
worked for all eigenvalues and experiments at 90\% c.l. for the
dataset and fit considered there, corresponding to $\Delta
\chi^2=T^2/c_{60}^2\approx 37$ at one sigma. A similar analysis in
Ref.~\cite{Martin:2002aw} found instead $T^2=50$.
Taken at face value,
this would imply that all experimental uncertainties have been
underestimated by a factor of about $T/c_{60}\approx 6$ (for
$T^2=100$) or $T/c_{60}\approx 4$ (for
$T^2=50$) . While 
some uncertainty underestimation is
possible, such a large factor is at best puzzling, and thus its origin
deserves further investigation. 

The concept of tolerance was subsequently refined,
by suggesting that instead of a global tolerance value for all
eigenvalues, a different tolerance value, determined as above, 
 be adopted along each eigenvector direction. This is called
``dynamical'' tolerance~\cite{Martin:2009iq}. Proceeding in this way, one
finds a tolerance $T \lsim 6.5$, with most
values being in the range $2<T<5$, so the large tolerance problem is somewhat
mitigated. Also, in this approach it is possible to trace which
individual experiment is controlling the tolerance range for each
eigenvalue. This, together with the expression of the eigenvector in
terms of the original parameters,  provides insight on the
relation between data and PDF parameters and their mutual consistency.
Such an analysis is displayed in Fig.~\ref{dyntol}, where both the
tolerance analysis for one specific eigenvector, and then the
experiments and corresponding band which control the tolerance
interval for each eigenvector. A related but different refinement
was suggested in Ref.~\cite{Lai:2010vv}, based on the idea of
determining the band such that each experiment agrees with the global
fit to say 90\% c.l. by means of  a sharply rising
penalty term in the global $\chi^2$ based on the $\chi^2$ of each experiment.

It is interesting to contrast this treatment of uncertainties in the
Hessian approach with ``standard'' parametrization with the  
Monte Carlo approach together
with neural network parametrization discussed in Sect.~\ref{mcnnpdf}. 
In that approach, uncertainty bands corresponding to any given
confidence level can be computed directly from the Monte Carlo sample:
the one-$\sigma$ interval is just the standard deviation of the sample, and
one may even check whether it indeed corresponds to the central 68\% of the
distributions of PDF replicas. This is shown in
Fig.~\ref{onesigmannpdf} for the gluon distribution (from
Ref.~\cite{Ball:2009mk}): in this case (and in fact~\cite{Ball:2010de} in
most cases) the one-$\sigma$ and 68\% c.l. intervals coincide. In a
Monte Carlo approach, whether or not the fits behave consistently when
comparing fit results to new data, and then including these new data
into the fit, can be verified a posteriori by performing statistical
tests on the fit results. These tests were performed successfully for
the fits of Refs.~\cite{Ball:2009mk,Ball:2008by,Ball:2010de}.

\begin{figure}\begin{center}
\includegraphics[width=.45\linewidth]{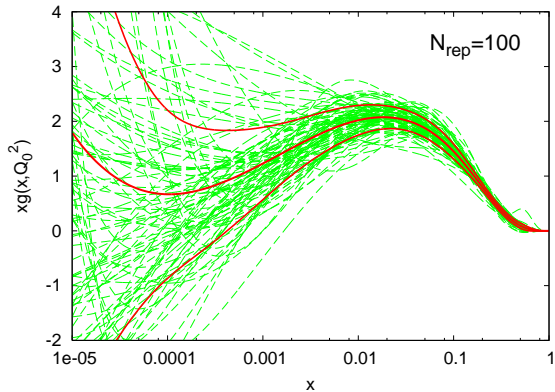}
\caption{One-$\sigma$ interval computed from a distributions of 100
  replicas of the gluon distribution.} 
\label{onesigmannpdf}
\end{center}
\end{figure}
The question of the
appropriate range in global $\chi^2$ which corresponds to one sigma
is thus side-stepped. In principle, it can be
answered a posteriori: in a Monte Carlo approach, the $\chi^2$ of the
mean is a
property of the Monte Carlo sample, so one could compute the one-$\sigma$
interval  from the sample itself. In practice, it is nontrivial to do
this accurately because, as
explained in Sect.~\ref{hessian}, the $\bar\chi^2$ has fluctuations of
order $N_{\rm dat}$, and in a Monte Carlo approach these fluctuations
take place  replica by replica, so one needs a very large sample to
determine the $\chi^2$ accurately. 

However, the issues which may be responsible for the large
tolerances can be addressed both in a Hessian and in a Monte Carlo
approach as we will now discuss.

\subsection{Parametrization bias and data incompatibility}

The large tolerance values discussed in Sect.~\ref{tol} are, by
definition, a manifestation of
the poor mutual compatibility of the 
experiments that go into the global fit. One possible explanation for
this is that experiments are genuinely incompatible with each other
within their stated uncertainties, i.e. that their published
uncertainties are underestimated. We will refer to this possible
explanation as ``data incompatibility''.

Another possible explanation is that
the way uncertainties are propagated from experiments onto PDFs leads
to underestimating the uncertainty in the latter. For example, assume
that  experiment $A$ 
does not depend on some PDF parameter, and that one determines PDFs
from this experiment, but
instead of leaving the undetermined parameter  free, one fixes it in
some arbitrary way. If the ensuing PDF is then used to predict another
experiment B which happens to depend on the undetermined parameter the
likelihood of results being in agreement with the prediction will not
depend on statistics, but rather in the arbitrary way the parameter has
been fixed. We will refer to this as ``parametrization bias''. 

Of course other options are possible: for example, that the theory
which is being used is not adequate. In the latter case, however, one
would have to find a convincing argument why this theoretical
inadequacy has not been seen elsewhere.

\begin{figure}\begin{center}
\includegraphics[width=.35\linewidth]{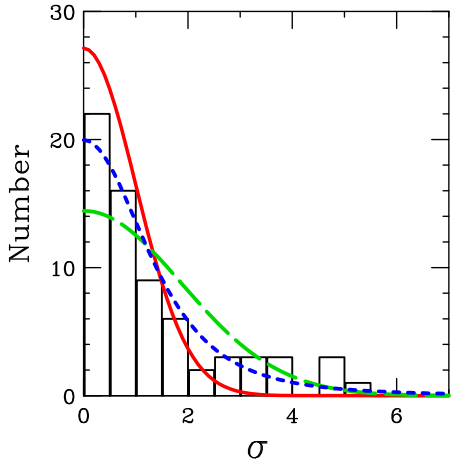}
\caption{Distribution of discrepancies between each experiment
  entering a global fit, and  the
  global best fit. The solid (red) curve is the standard
  gaussian distribution, the dashed (green) curve is a gaussian
  distribution with $\sigma$ rescaled by a factor 1.88, the dotted
  (blue) curve is a Lorentzian distribution (from Ref.~\cite{Pumplin:2009sc})} 
\label{pumpdata}
\end{center}
\end{figure}
Data incompatibility in the Hessian approach
was recently studied in a quantitative way in
Ref.~\cite{Pumplin:2009sc}, exploiting the
observation~\cite{Pumplin:2009nm} that once the $\chi^2$ has been
written in the form of Eq.~(\ref{diaghessian}) one can perform a
further linear transformation of the parameters which preserves this
form, while also diagonalizing the contribution to the $\chi^2$ from
some specific subset of data. After this simultaneous diagonalization,
the $\chi^2$ is written as the sum of a contribution from the data in
the given subset and the rest: the distance of the minima of these two
contributions to the $\chi^2$ in units of the corresponding standard
deviation measures the compatibility of the given subset of data with
the rest of the global dataset. 
The idea is then to study the distribution of such distances,
in all cases in which the experiment does contribute significantly to
the global minimum. If experimental uncertainties are correctly
estimated, they should be gaussianly distributed. The results of this
analysis, shown in Fig.~\ref{pumpdata}, suggest that the distribution
of discrepancies deviates significantly from a gaussian distribution,
and that if it is fitted to a gaussian its uncertainty should be
rescaled by about a factor 2. This suggests uncertainty
underestimation by a similar factor, 
which corresponds to a value of the tolerance for 90\% c.l. of order
of $T^2\sim 10$. 

\begin{figure}[t]\begin{center}
\includegraphics[width=.4\linewidth]{xg_Q_2_lin.eps}
\includegraphics[width=.4\linewidth]{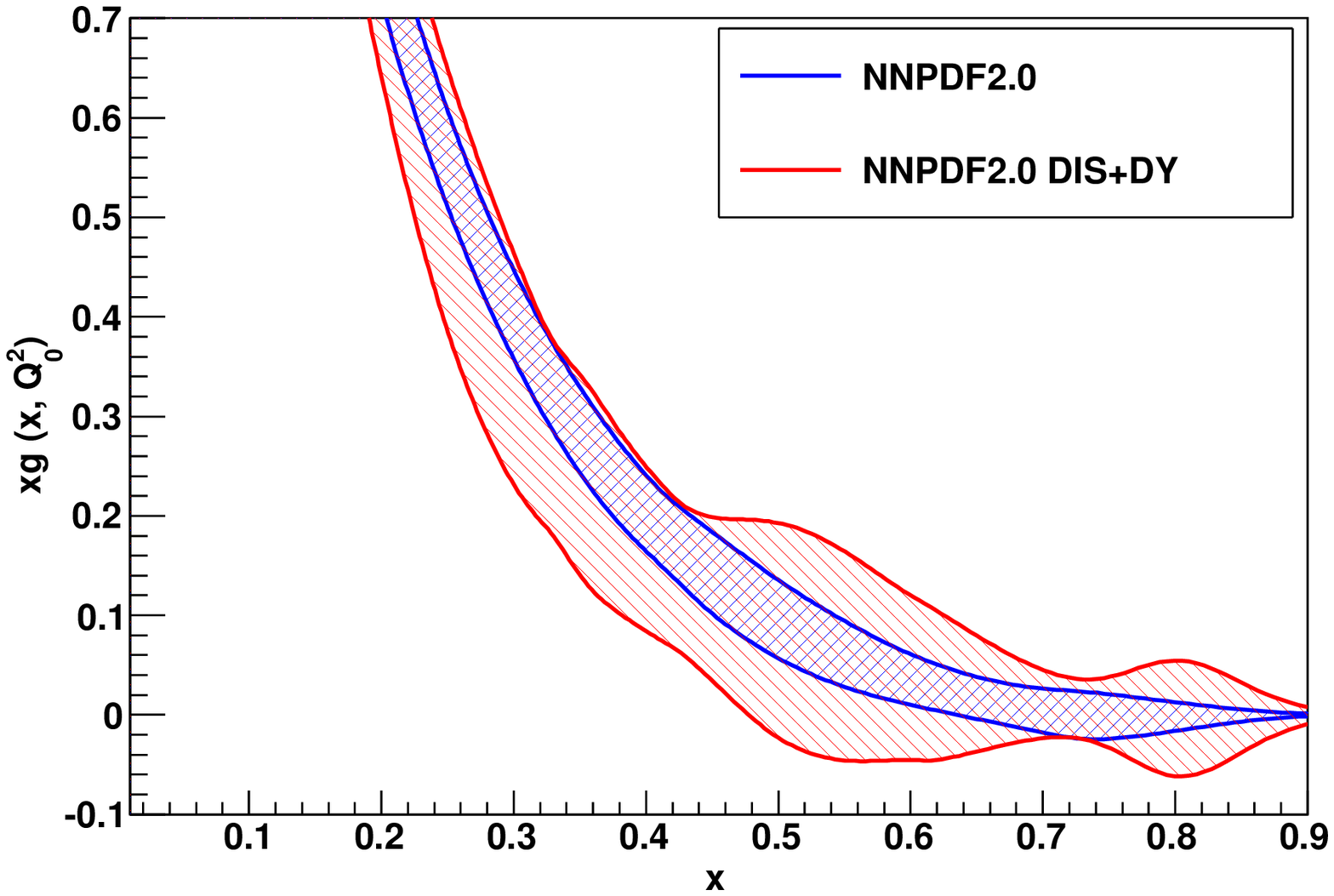}\\
\includegraphics[width=.4\linewidth]{xDeltaS_Q_2_lin2.eps}
\includegraphics[width=.4\linewidth]{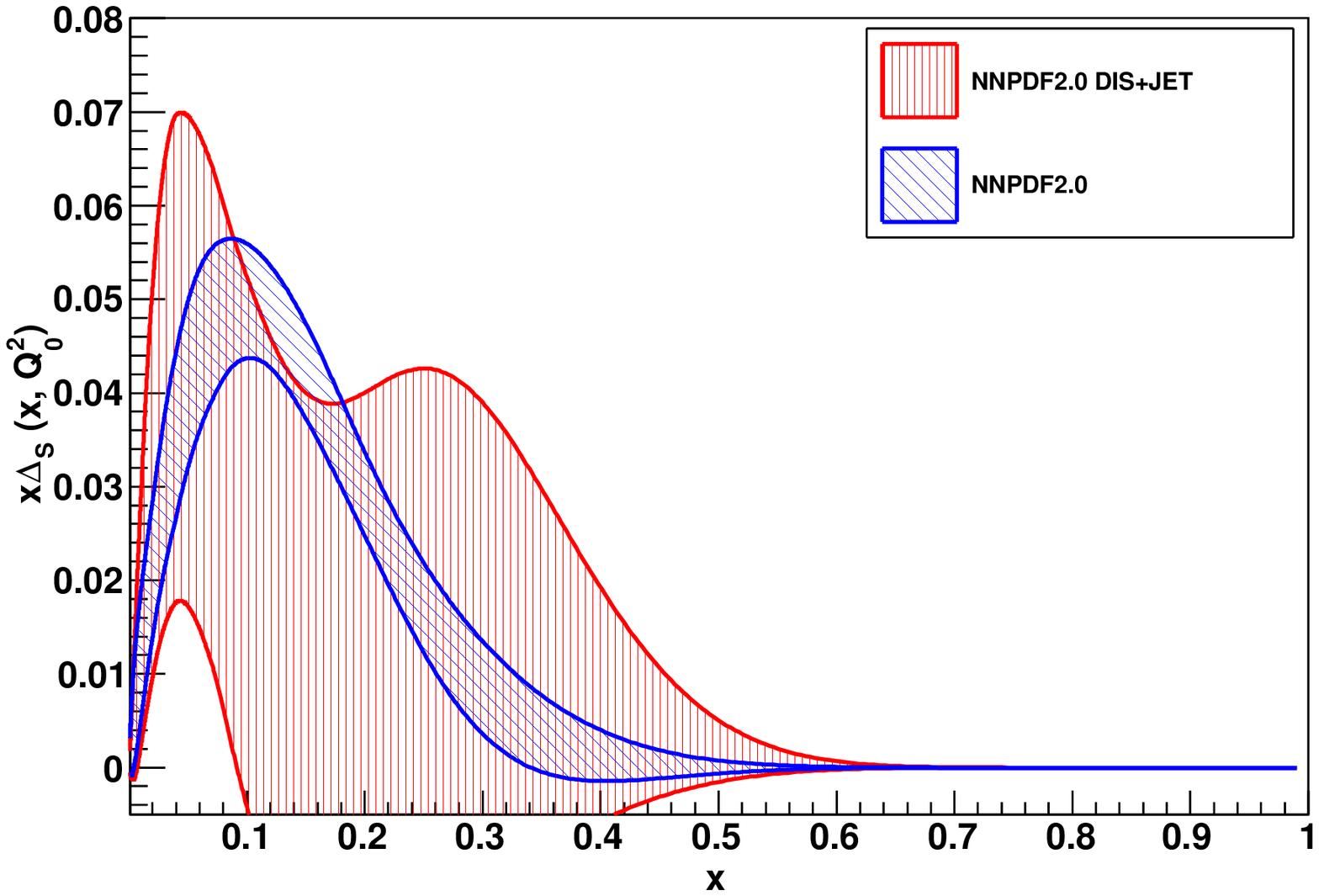}
\caption{Tests of data compatibility by changing the order of  
inclusion of data in fits with
different datasets, based on the NNPDF2.0~\cite{Ball:2010de} PDF
determination. 
Top: effect on the gluon distribution of the
inclusion of jet data in a fit to DIS data
only (left) or on a fit to DIS+Drell-Yan data (right). Bottom: effect
on the sea asymmetry Eq.~(\ref{asymdef})  of the inclusion of
Drell-Yan data in a fit to DIS data
only (left) or on a fit to DIS+jet data (right).}
\label{commutefits}
\end{center}
\end{figure}
This suggest that data incompatibility can explain only in part the
need for large tolerance. Further evidence that data incompatibility
is at most moderate can be obtained in a Monte Carlo approach, by
comparing the effect of the subsequent inclusion of different datasets
into a fit. Indeed, if some datasets were incompatible with others,
then the effect of their inclusion in the global fit would change
according to whether the global fit already includes
the data with which they are incompatible or
not. Assume for example that the gluon determined from jets is
compatible with that found exploiting scaling violations in DIS data,
but less compatible with that found from scaling violations in
Drell-Yan: then, inclusion of jet data in a pure DIS fit would have a
different effect than their inclusion in a fit which contains both DIS
and Drell-Yan. When such tests are performed~\cite{Ball:2010de}   no
evidence for data incompatibility is found, as demonstrated in
Fig.~\ref{commutefits}).

\begin{figure}\begin{center}
\includegraphics[width=.35\linewidth]{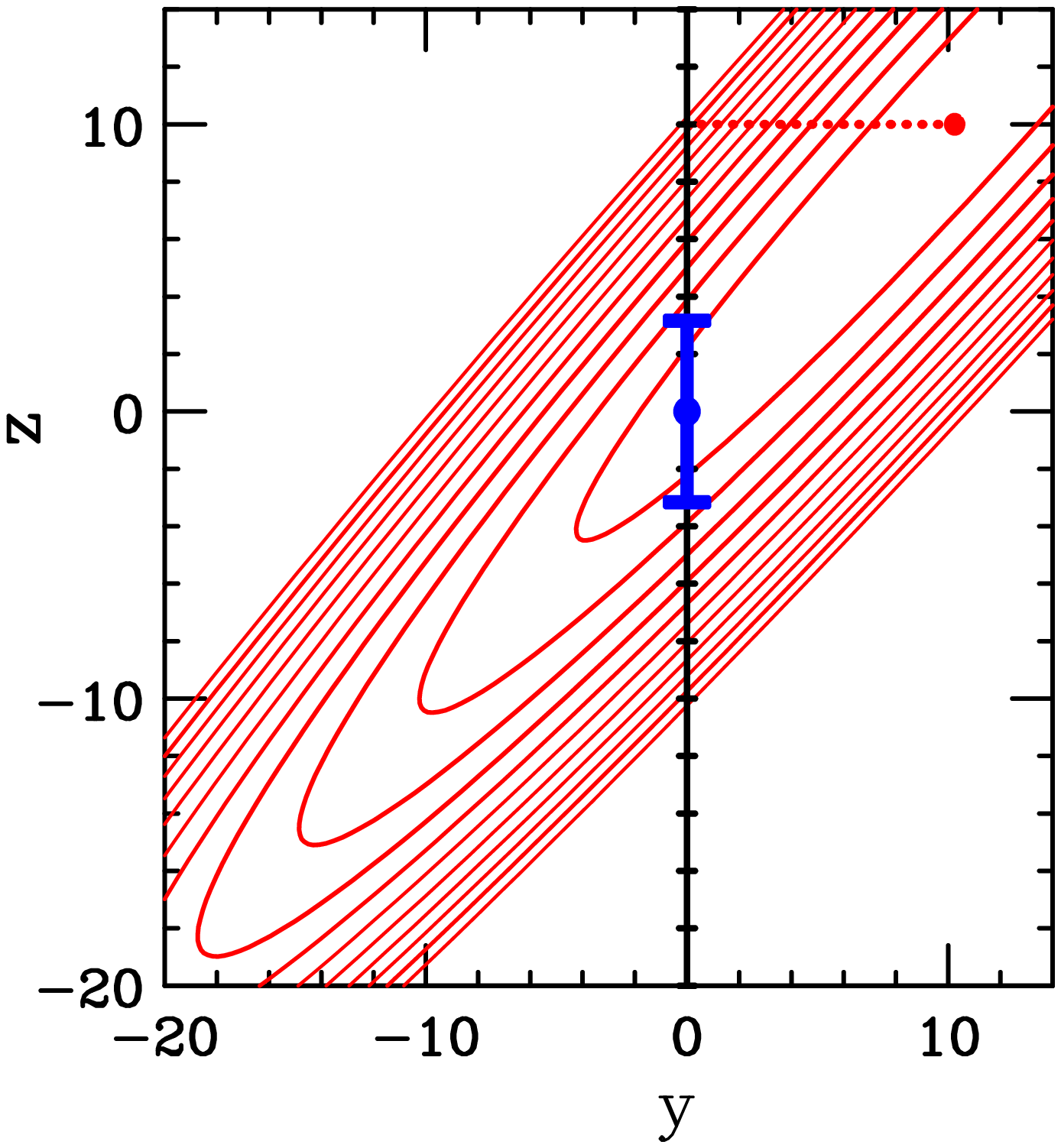}
\includegraphics[width=.27\linewidth]{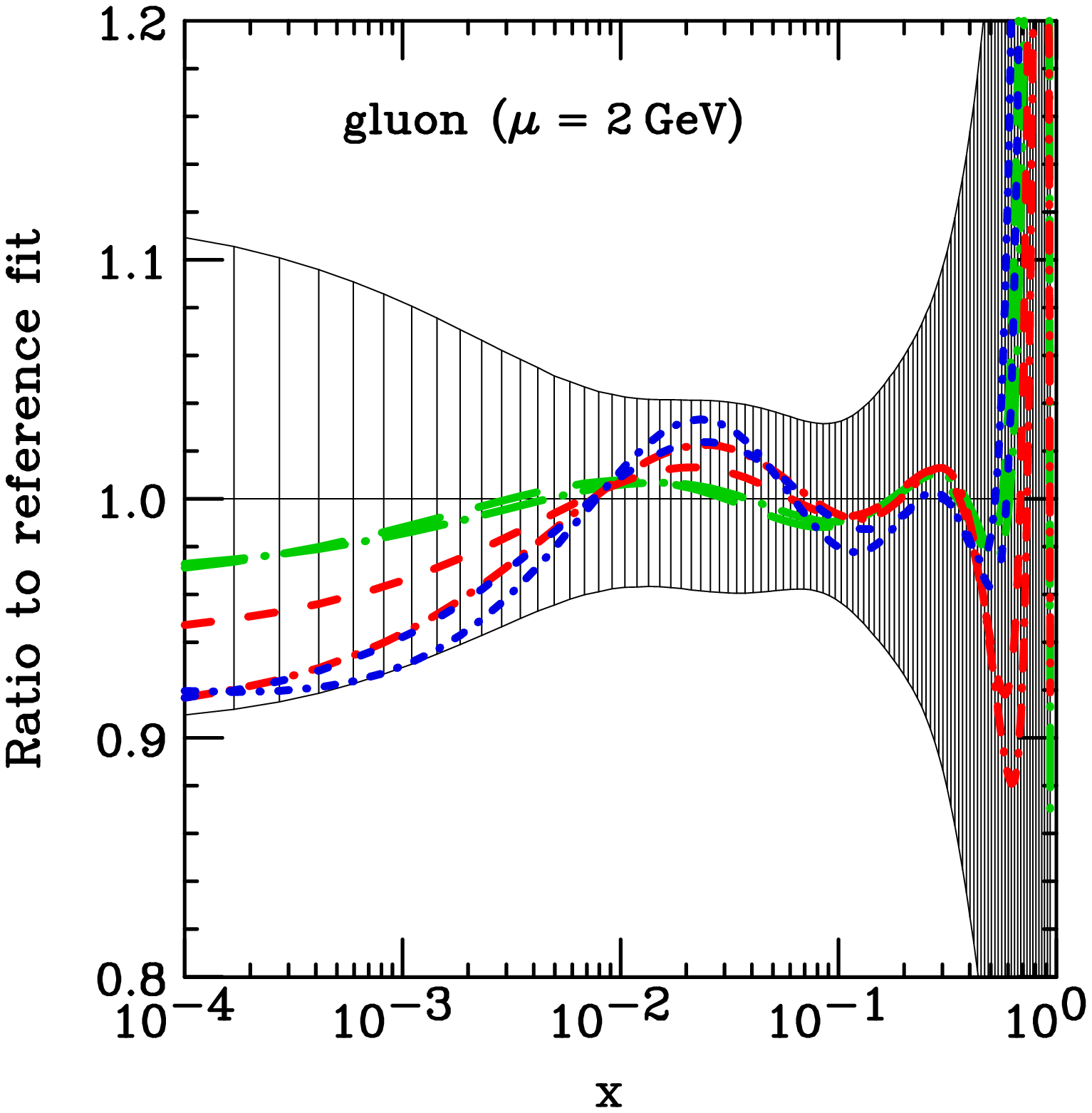}
\caption{Parametrization bias. Left: the one-$\sigma$ range for parameter $z$
  when parameter $y$ is kept fixed away from the best-fit value
  corresponds to a range which is larger than $\Delta
  \chi^2=1$. Right: various gluons (dashed curves) based on a general parametrization 
  which lead to the same or better fit quality as the best-fit gluon
  of Ref.~\cite{Nadolsky:2008zw}, compared to the $\Delta \chi^2=10$
  band about the latter gluon (from Ref.~\cite{Pumplin:2009bb}).} 
\label{pumpfunc}
\end{center}
\end{figure}
Let us now turn to the possibility that parameterization bias may be
responsible for the effect. The way this could happen in a Hessian
approach was recently exemplified  in Ref.~\cite{Pumplin:2009bb}.
Assume a relevant parameter on which PDFs depend is not fitted, but
rather fixed by assumption at a value which is away from its best-fit.
 Then, clearly (see Fig.~\ref{pumpfunc}) the one-$\sigma$ range for
 the other parameters when this parameter is kept fixed corresponds to
 a variation of $\chi^2$ which is greater than the standard $\Delta
 \chi^2$ found when moving away from the minimum. A first estimate of the
 possible size of this effect was also provided in
 Ref.~\cite{Pumplin:2009bb} by simply repeating the PDF fit of
 Ref.~\cite{Nadolsky:2008zw}, but with a much more general
 parametrization, based on expanding the gluon on a basis of
 orthogonal polynomial, analogous to that of Ref.~\cite{Glazov:2010bw} shown in
 Fig.~\ref{nnvsgenparm}. With this more
 general parametrization,  fits whose $\chi^2$ is similar to or better
 than  the best-fit $\chi^2$
 of the more restrictive parametrization are found to span a band
 which corresponds roughly to  the
 $\Delta \bar\chi^2=10$
 range for the restrictive parametrization. So this suggests a
 tolerance of at
 least $T^2=10$ just to account for the bias on the gluon shape
 imposed by the  parametrization of Ref.~\cite{Nadolsky:2008zw}


\begin{figure}\begin{center}
\includegraphics[width=.24\linewidth]{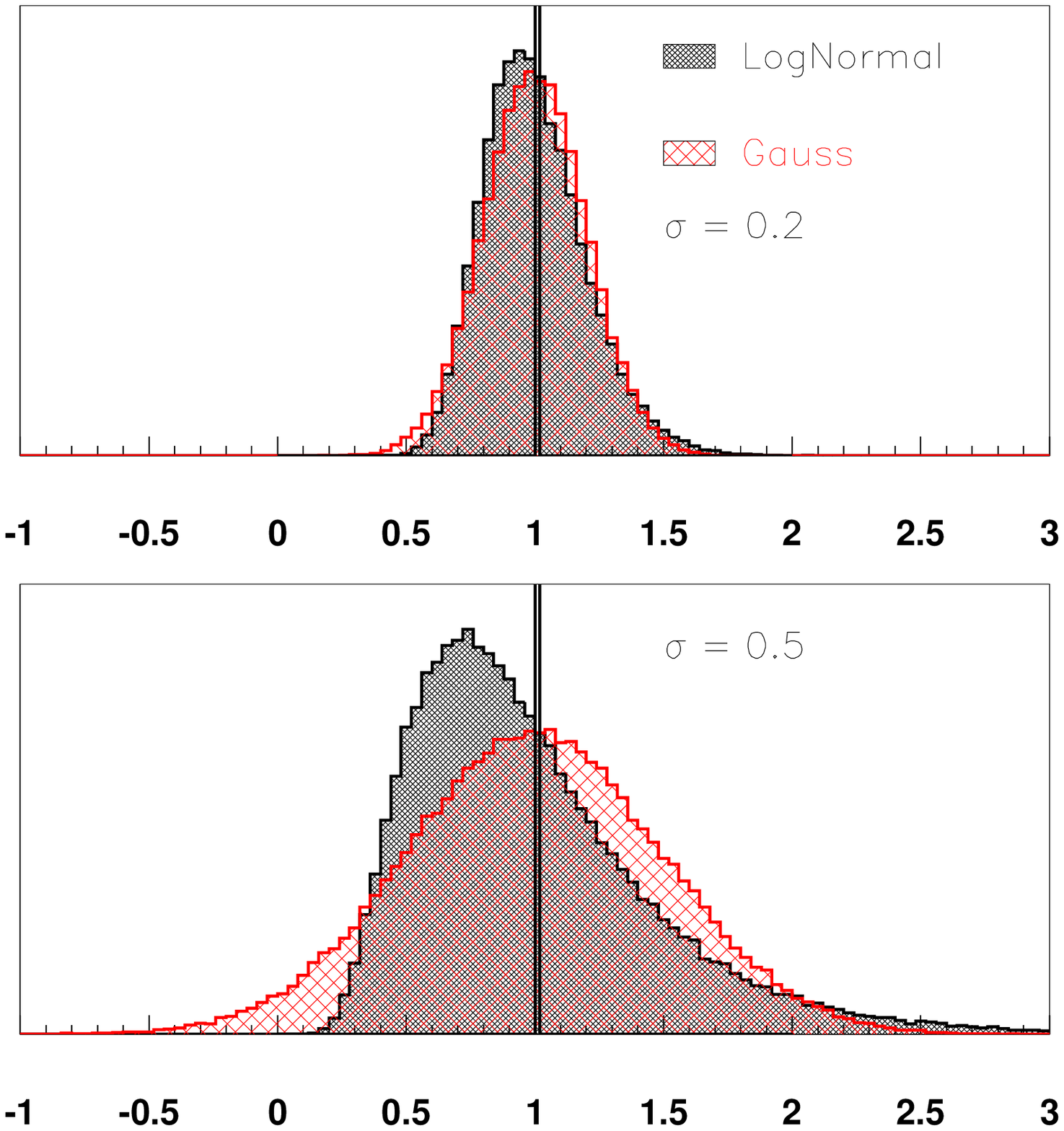}
\includegraphics[width=.24\linewidth]{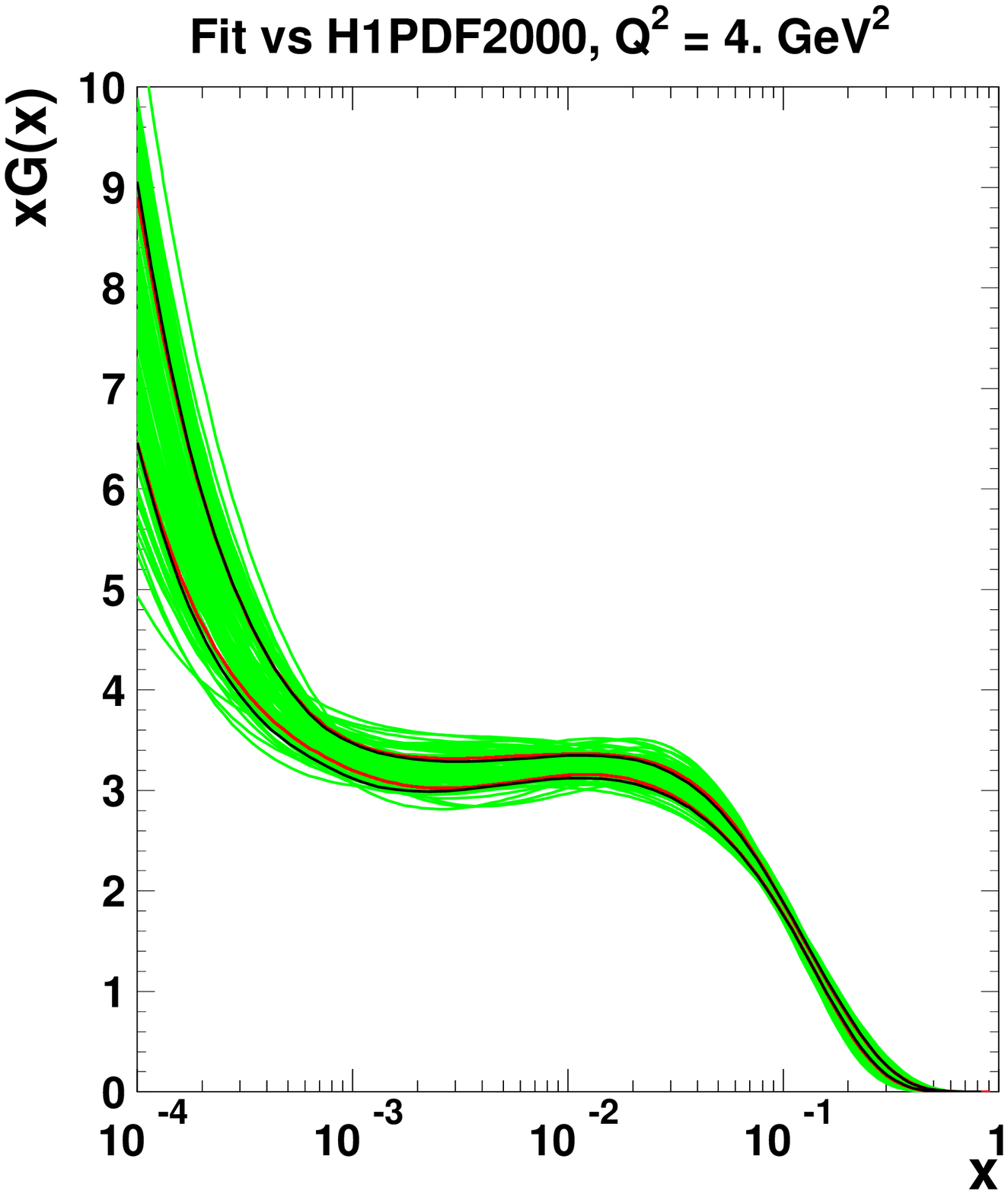}
\includegraphics[width=.24\linewidth]{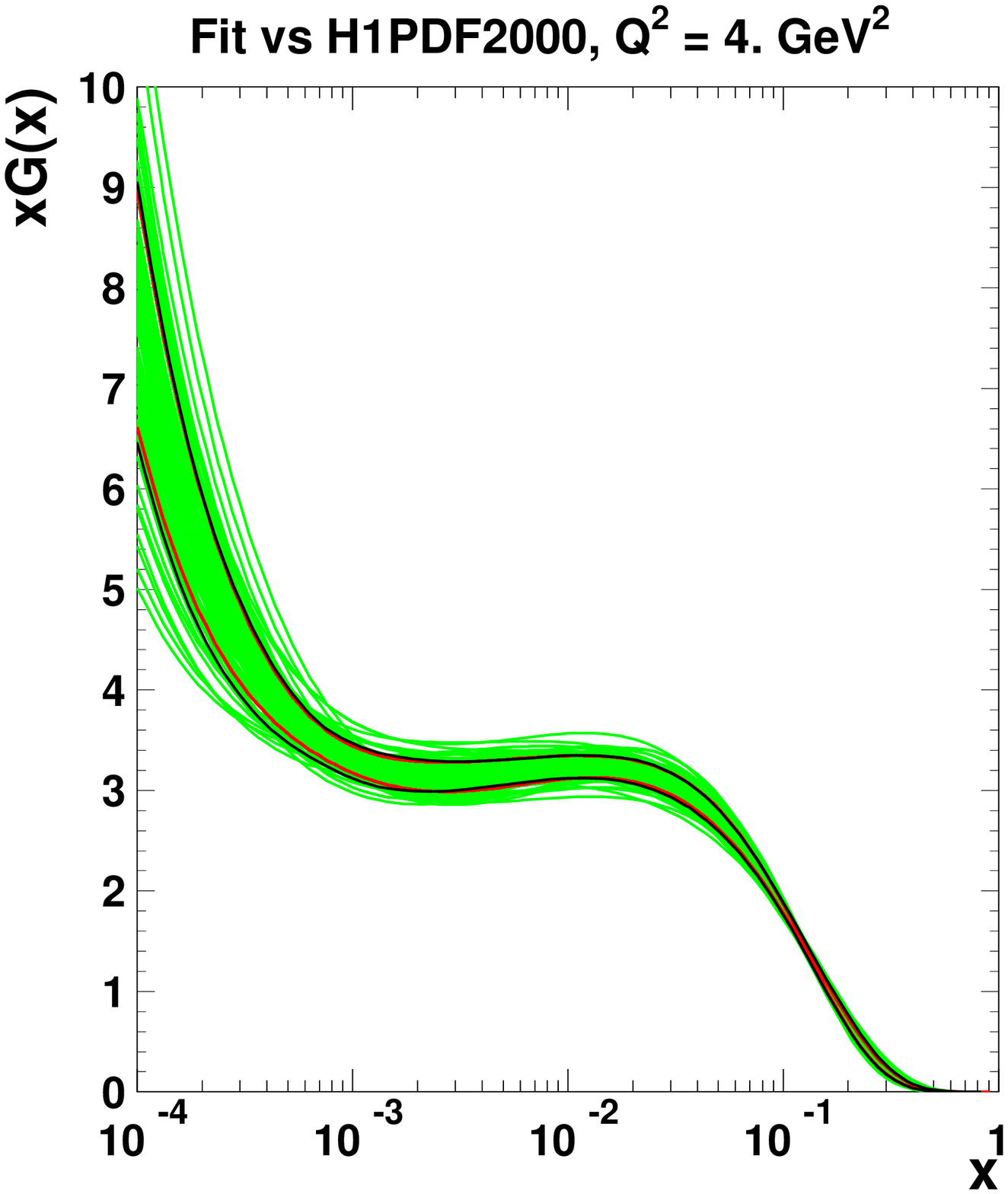}
\includegraphics[width=.24\linewidth]{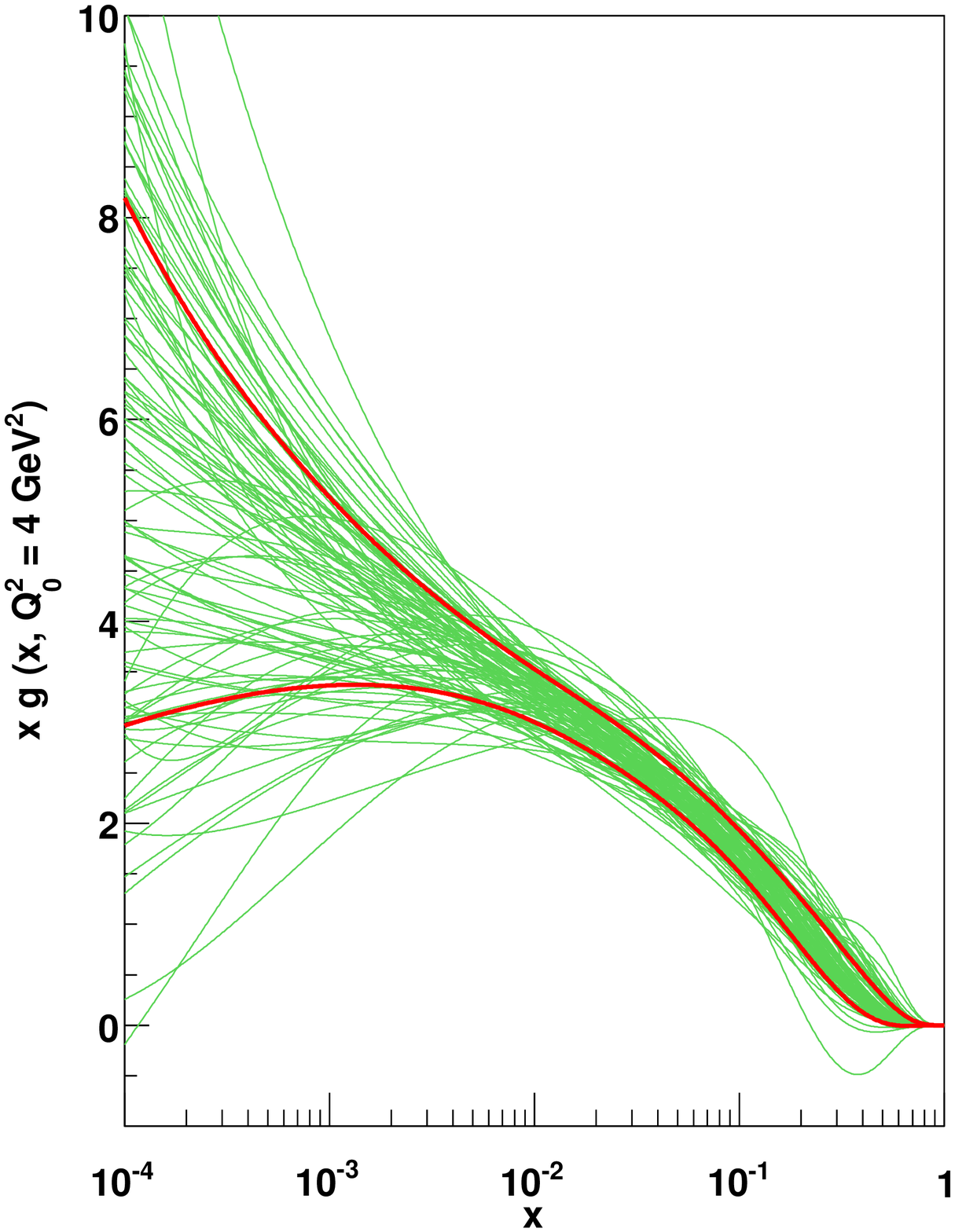}
\caption{Comparison of fits based on gaussianly or log-normally
  distributed data. From left to right: Comparison of the gaussian and lognormal
  distribution. Gluon determined from lognormal data. 
Gluon determined from gaussian data. In the latter two cases, the plot shows
individual replicas (green) Hessian uncertainty band (black) and Monte
Carlo uncertainty band (red) .(From
  Ref.~\cite{FGR}). Gluon determined from the same
  gaussian data, but using the neural network parametrization of
  Ref.~\cite{Ball:2008by} (from Ref.~\cite{Dittmar})} 
\label{gaussvsnormal}
\end{center}
\end{figure}
The issue of parton parametrization and bias thus deserves further
investigation. First, one may ask whether the gaussian assumption 
is by itself a source of bias. This has
been investigated in Ref.~\cite{FGR}, using a Monte Carlo approach
together with a standard parton parametrization. Data replicas based
on the HERA DIS data
($F_I(n)$ of Fig.~\ref{nnpdfscheme}) have been generated either using a
gaussian or a lognormal distribution (see
Fig.~\ref{gaussvsnormal}), 
and used for a  PDF fit based on the ``standard'' functional form
Eq.~(\ref{pdfparmgen}). In each case, both the Monte Carlo and Hessian
uncertainty are computed.
 Results are shown in
Fig.~\ref{gaussvsnormal}: the lognormal and gaussian results are essentially
indistinguishable, and so are the Hessian and Monte Carlo
uncertainties.  
The choice of the probability distribution 
of the data
does not seem to play any major role, as one might have expected from
the central limit theorem: with so many data, everything looks
gaussian. This also provide a nice visual demonstration of the
equivalence of Hessian and Monte Carlo uncertainty computation in the
Gaussian case.

On the other hand, in the same
figure we also show the gluon obtained in
a fit to exactly the same data, but using the neural network
functional form and  associate cross-validation methodology 
of Ref.~\cite{Ball:2008by}. It is clear
that the uncertainty is now much wider. This  suggests that it is the
form of parametrization which plays a dominant role, rather than the
form of the probability distribution.

\begin{figure}[t]\begin{center}
\includegraphics[width=.4\linewidth]{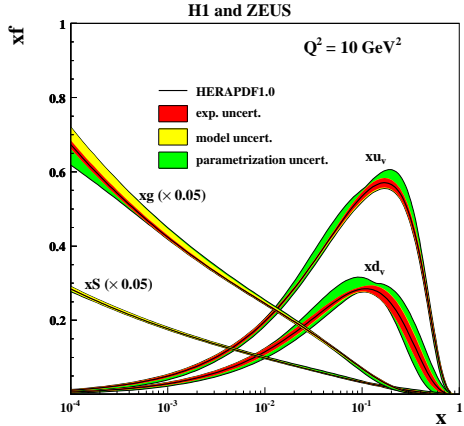}
\caption{Comparison of statistical and parametrization
  uncertainties in the HERAPDF fit (model uncertainties denote what
  we call theoretical uncertainties), from Ref.~\cite{:2009wt}.} 
\label{hessparm}
\end{center}
\end{figure}
The issue has been investigated further in the
HERAPDF~\cite{:2009wt} PDF fits, where the standard $\Delta
\chi^2=1$ PDF uncertainty based on a ``standard'' functional form 
Eq.~(\ref{pdfparmgen}) has been supplemented by a further
parametrization uncertainty, obtained by varying the assumed
functional form (in particular, the large $x$ behaviour, the number of
terms in the polynomial Eq.~(\ref{pdfparmct}), and the assumptions on
strangeness which is not fitted). It is clear from Fig.~\ref{hessparm}
that this leads to a sizable enlargement of the PDF uncertainty band.

The Monte Carlo approach together with neural network offers an
interesting way of searching for the origin of uncertainties, in that
different sources of uncertainty can be switched on and off one at a
time. In particular, one may perform the following exercise. First,
one freezes the generation of data replicas, and one takes each
replica dataset equal to the central values of the data. Recall from 
Sect.~\ref{mcnnpdf}
that each replica is fitted to a different, randomly chosen subset of
the data. Hence, all datasets $F_I$ of Fig.~\ref{nnpdfscheme} are now
the same, and only the way they are partitioned in validation and
training sets changes between replicas. 
Each PDF replica is thus obtained as the
fit to a different partition of the central experimental data. The 
(square) fluctuation of the data are reduced by a factor two:
instead of having replicas which fluctuate about experimental data
which in turn fluctuate about their ``true'' values, one only has
different subsets of central data fluctuating about their true values.
\begin{table}[h]
\begin{center}
{\begin{tabular}{|c|c|c|c|}
\hline 
& replicas &  central value & fixed partition\\
\hline 
 $\chi^2$ &  1.32 & 1.32     &  $\sim$1.3  \\
\hline 
$\langle\chi^2 \rangle_{\rm rep}$& $2.79\pm 0.24$&  $1.65\pm 0.20$ & $\sim1.6\pm 0.2$  \\
\hline
$\langle\sigma\rangle_{\rm dat}$ &0.039 &0.035 & $\sim$0.03 \\
\hline
\end{tabular}}\end{center}
\caption{Values of the $\chi^2$ for the best fit (first row),  the average and
  standard deviation of the $\chi^2$ of individual replicas (second row), and the
  percentage uncertainty of the prediction averaged over all data
  points (third row) for the PDF determination of
  Ref.~\cite{Ball:2009mk} (first column); the same but with all PDF
  replicas fitted to different partitions of the experimental
central values (second column); the same but with all data replicas
fitted to the same partition of the experimental central values
(in the latter case the process has been repeated with 5 different
choice of fixed partition and averaged)}
\label{testfittab}
\end{table}

In Tab.~\ref{testfittab} we compare some indicators of fit quality and
results for a fit obtained in this way to those of the corresponding
standard fit (using the fit of Ref.~\cite{Ball:2009mk}, based on
DIS data). In particular, we compare the $\chi^2$ of the best fit in
either case: this is unchanged. However, the average $\chi^2$ of each
replica fit is smaller by a factor two. This is as it should be: in
both cases the best-fit is reproducing the same central best-fit
value, but the fluctuation of replicas about it are now suppressed by
a factor two, and thus the average $\chi^2$ per replica is reduced by
approximately the same factor. The surprising result however is found
when one computes the average percentage uncertainty in the prediction
obtained in either case. This is determined 
as the percentage uncertainty  of 
the prediction obtained from the final
replica PDF set, for all datapoints included in the fit, averaged
over datapoints. One might expect that having halved  the
fluctuations, the average uncertainty
should be reduced by a factor $\sqrt{2}$;  in actual
fact, it is reduced by a much smaller amount.

\begin{figure}\begin{center}
\includegraphics[width=.45\linewidth]{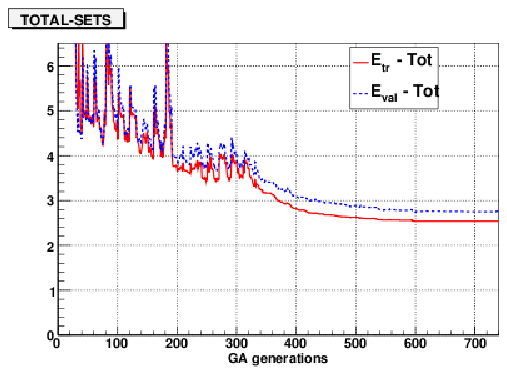}
\includegraphics[width=.45\linewidth]{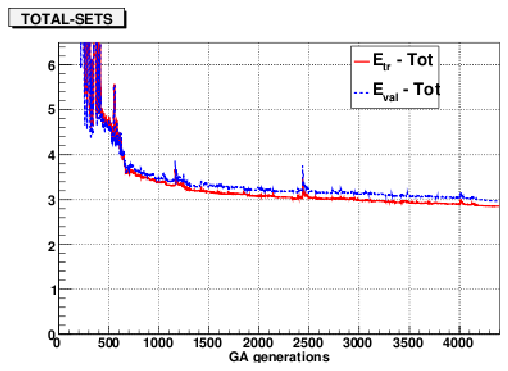}
\caption{The $\chi^2$ for the training dataset and total dataset for
  two different replicas fitted to exactly the same training subset of
  the central values of the data of Ref.~\cite{Ball:2009mk}, shown
as a function of the number of iterations of the
minimization algorithm.} 
\label{profilecomp}
\end{center}
\end{figure}
The origin of this state of affairs can be understood by performing
an even more extreme text: one simply produces 100 replicas fitted to
exactly the same partition of the central data. In this case, all
$F_I$ contain the same data, partitioned in the same way into training
and validation sets.
Naively one may think
that this may lead to simply repeating
100 times the same fit. This is not
necessarily the case because each replica is determined by
initializing the neural networks at random, and then minimizing by means
of an (equally random) genetic algorithm. Hence one starts each time
from a different point in the very wide parameter space, and then
the minimum is approached along a different path. Indeed, in
Fig~\ref{profilecomp} we show the $\chi^2$ profiles along the
minimization, as a function of the number of iterations of the
minimization algorithm, for two individual replicas: 
it is clear that even though the final $\chi^2$ values are quite
similar, the number of iterations and profiles that take there are
quite different, thereby showing that the minimum is approached along
different paths.

In this
case, in order to make sure that results do not depend on 
the particular partition that has been
picked in the first place, the whole procedure is repeated
five times with five different choice of starting partition and
results are then averaged. Results are shown in Tab.~\ref{testfittab}
(they should be taken as indicative, because one should use rather
more than five fixed partitions for accurate results). These results 
are quite surprizing. 
 The $\chi^2$ of the best fit and the average over replicas are
 unchanged, and this is to be expected: it shows that indeed the five
 replicas chosen are not special. However, very surprizingly, the
 average uncertainty, which one might expect to be tiny, 
is more than 50\% of that of the original
 fit. This is also seen by comparing results for PDFs (see
 Fig.~\ref{fixedpartpdf}: 
the uncertainty band is smaller, but of the same order of magnitude as
that of the full fit. The inevitable conclusion is that a large
fraction of the uncertainty band, probably more than half, does not
depend on the fluctuations in the data. Rather, it is a consequence of
the fact that there is an infinity of functions that provide fits of
comparable quantity to the data. Different minimization profiles such
as those shown in Fig.~\ref{profilecomp} land on somewhat different
minima; the uncertainty of this fit is then a measure of the spread of
this space of minima. 
Once understood that a sizable
fraction is simply due to this ``functional'' uncertainty, it is clear
why it is more difficult to capture with a fixed parametrization,
which then requires a suitable tolerance in order to mimic it. 
\begin{figure}[h]\begin{center}
\includegraphics[width=.4\linewidth]{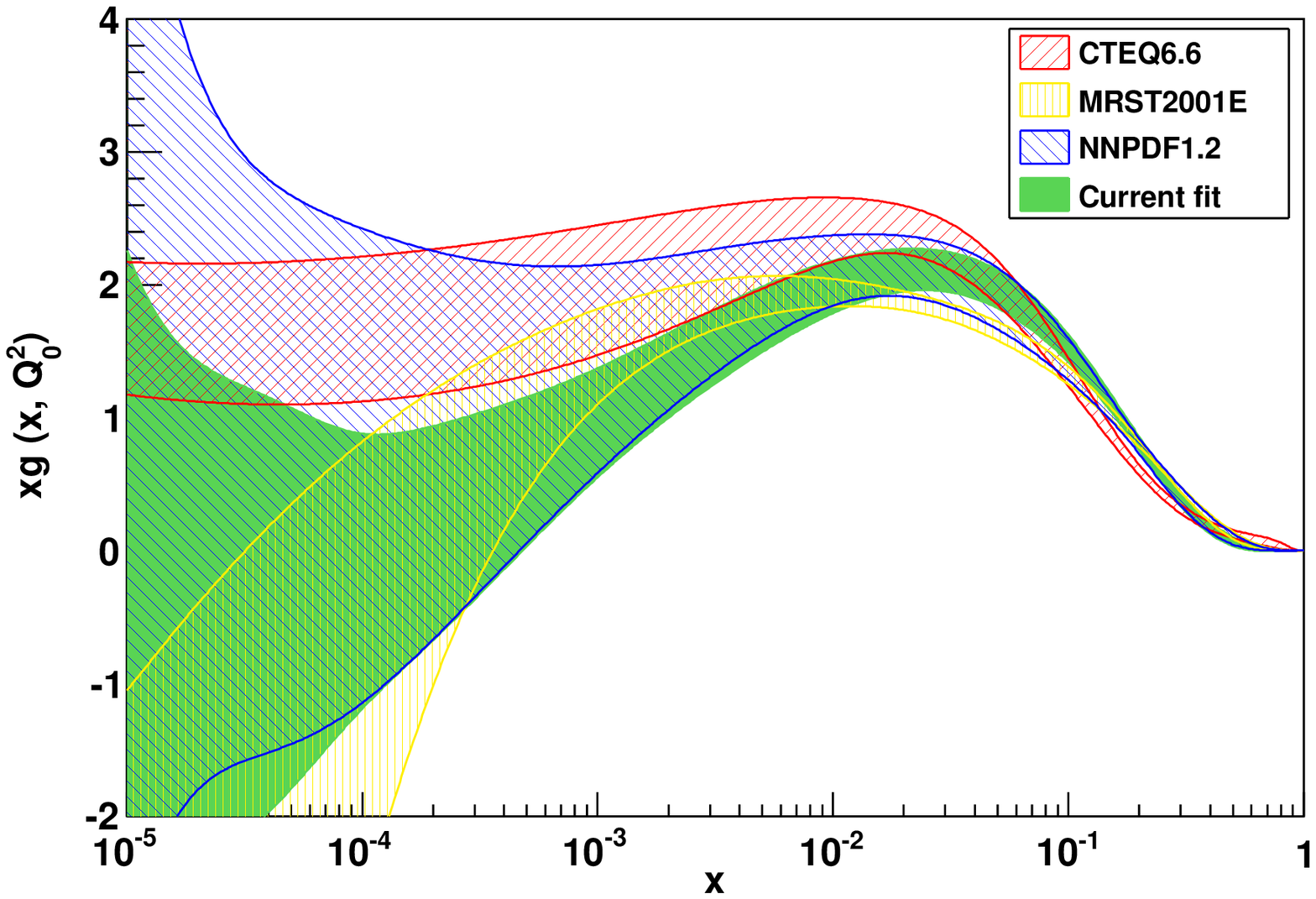}
\includegraphics[width=.4\linewidth]{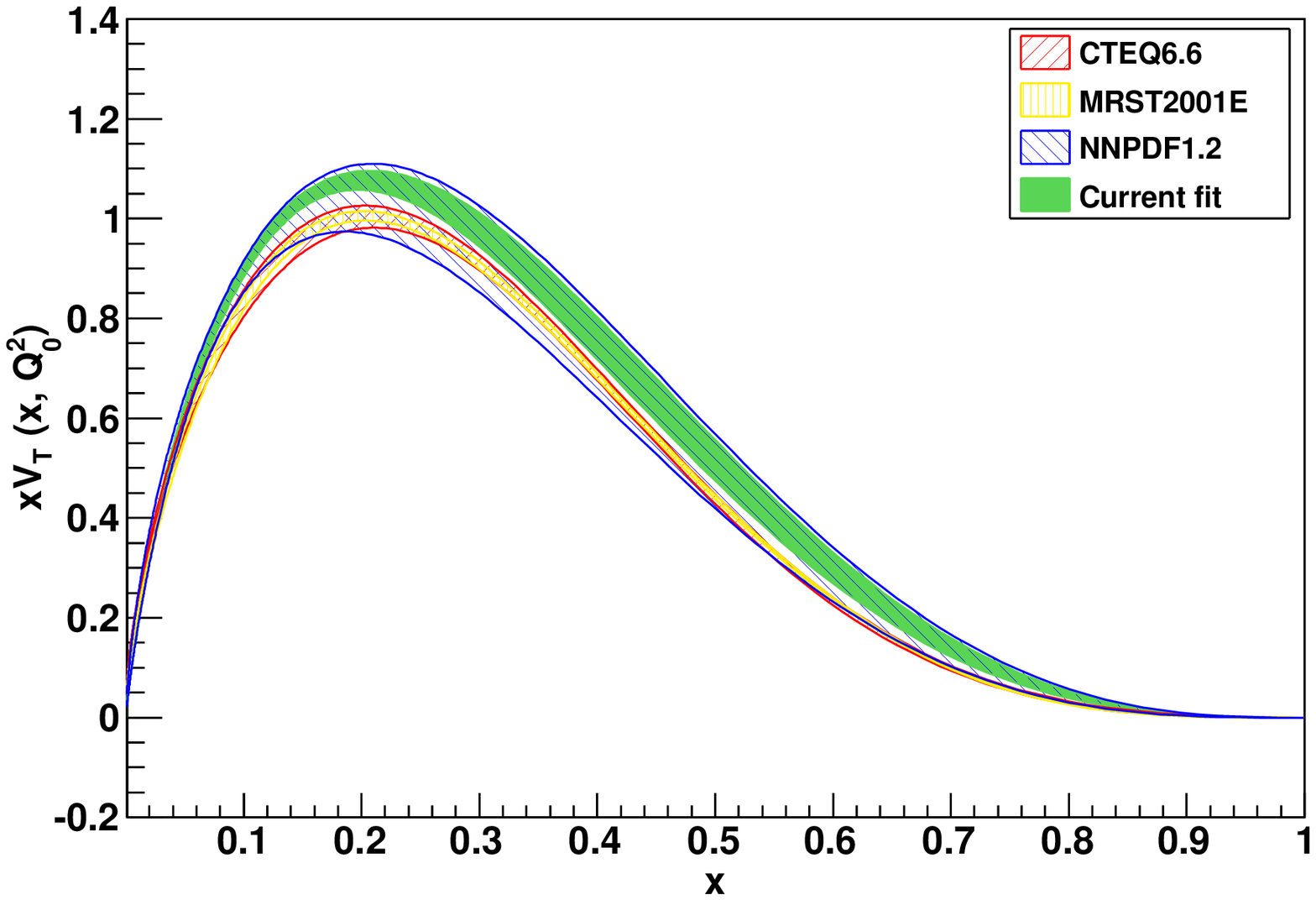}\\
\includegraphics[width=.4\linewidth]{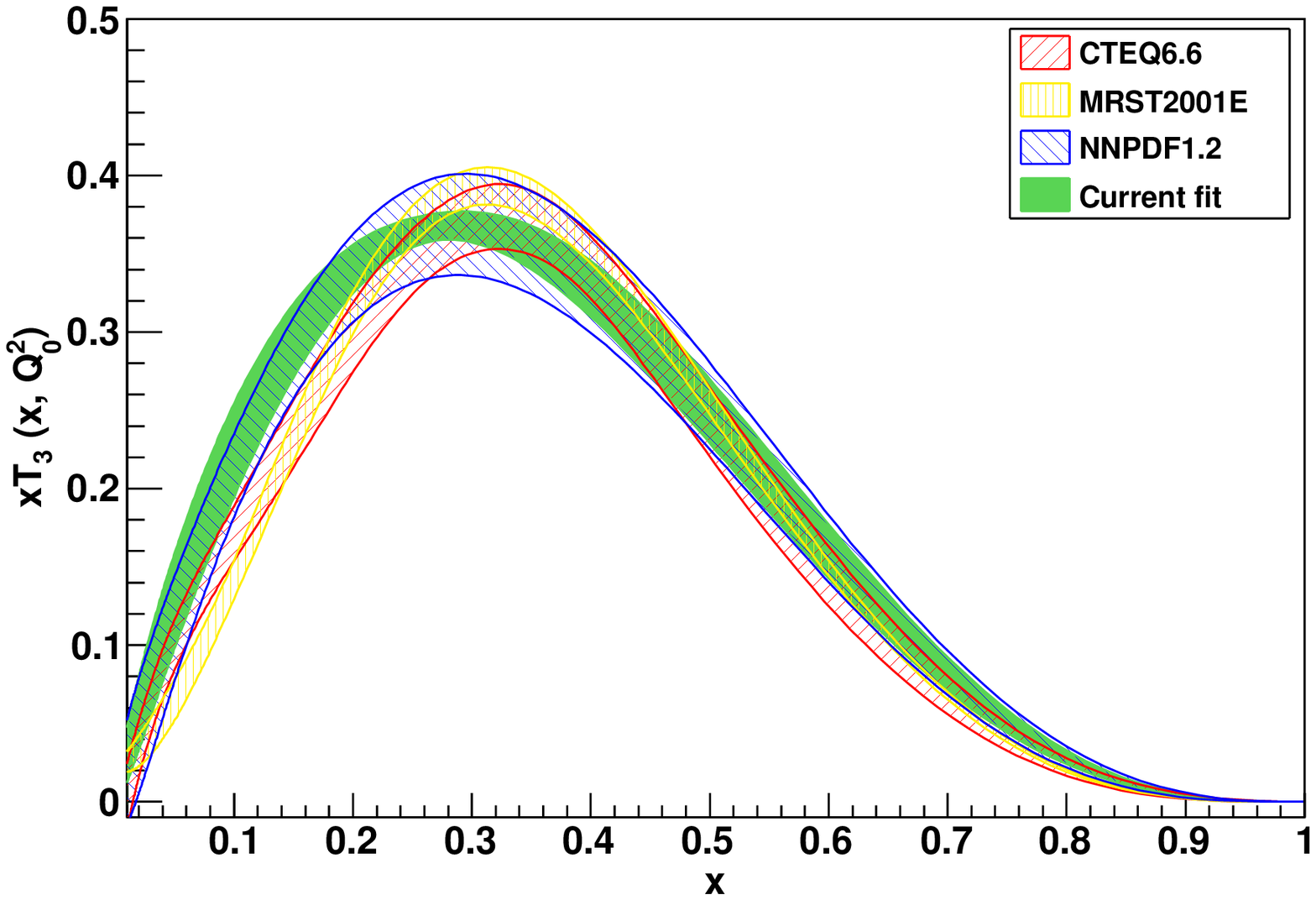}
\includegraphics[width=.4\linewidth]{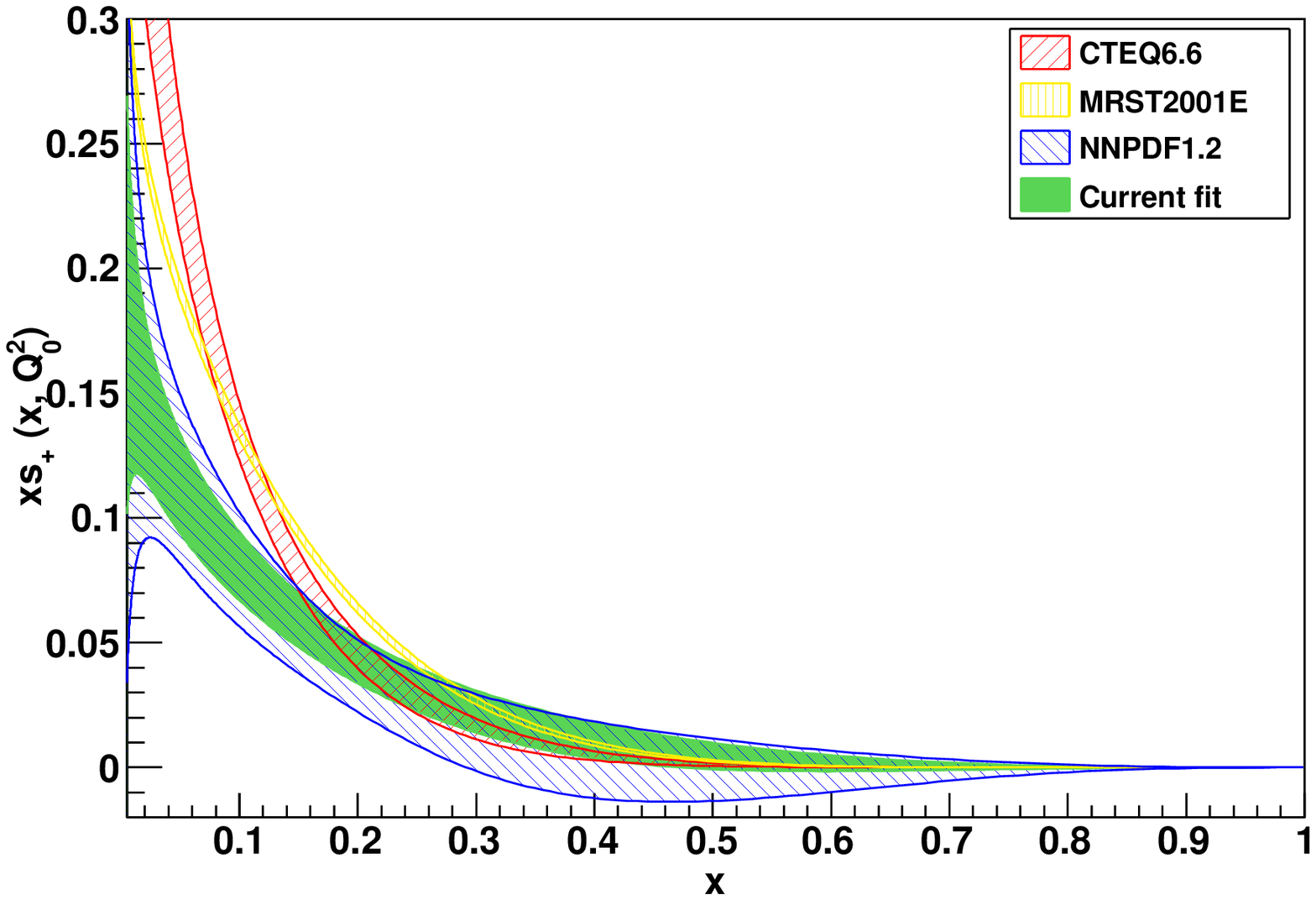}
\caption{Comparison of PDFs from the Monte Carlo set of
Ref.~\cite{Ball:2009mk} (NNPDF1.2,
  blue band) to PDFs determined using the same data and procedure, but
  fitting all replicas to the central data (green band). The PDFs from Ref.~\cite{Martin:2002aw}
  (MRST2001E, yellow band) and Ref.~\cite{Nadolsky:2008zw} (CTEQ6.6, red band) are
  also shown for comparison. The PDFs shown are the gluon (top left),
  total valence Eq.~(\ref{valdef}) (top right), triplet
  Eq.~(\ref{tripdef}) (bottom left) and total strangeness $s^+$
  Eq.~(\ref{splusmdef}) (bottom right).} 
\label{fixedpartpdf}
\end{center}
\end{figure}

\section{Recent developments}
\label{recent}

The state of the art in PDF determination is moving very fast and thus
any attempt to review it would necessarily become obsolete quite
rapidly:
a recent review of the status of the field
as of November 2010 is in Ref.~\cite{higgsYR}. Here, in
an attempt to discuss issues of somewhat less fleeting value, we will
first briefly review theoretical uncertainties, which are the current
frontier of PDF determination, then summarize what progress has been
made and what remains to be done in the determination of PDFs for
the LHC in the years to come.

\subsection{Theoretical uncertainties}
\label{th}
As already mentioned, the PDF
uncertainties discussed in Sect.~\ref{PDFunc} are the result of
propagating into the space of PDFs the uncertainty on the data on
which the PDF determination is based. 
Most of the effort has gone so far in their determination and understanding
because they
are likely to be at present the dominant uncertainty. However, it has
been recently realized that in many cases uncertainties related to the
theory used to extract PDFs from the data may be larger than one may
think. These, as already mentioned, include both uncertainties in the
theory itself (such as higher order corrections) but also
uncertainties in the knowledge of the free parameters on the theory.

\begin{figure}\begin{center}
\includegraphics[width=.32\linewidth]{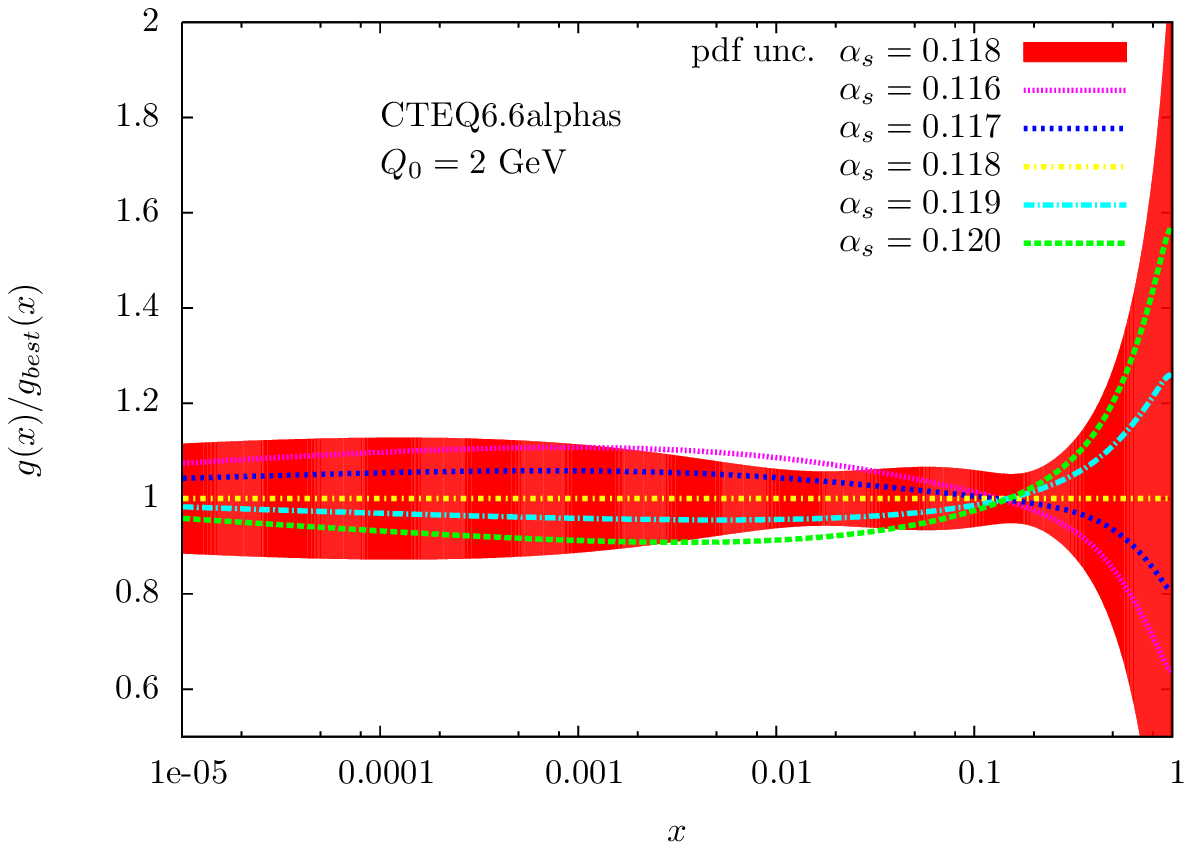}
\includegraphics[width=.32\linewidth]{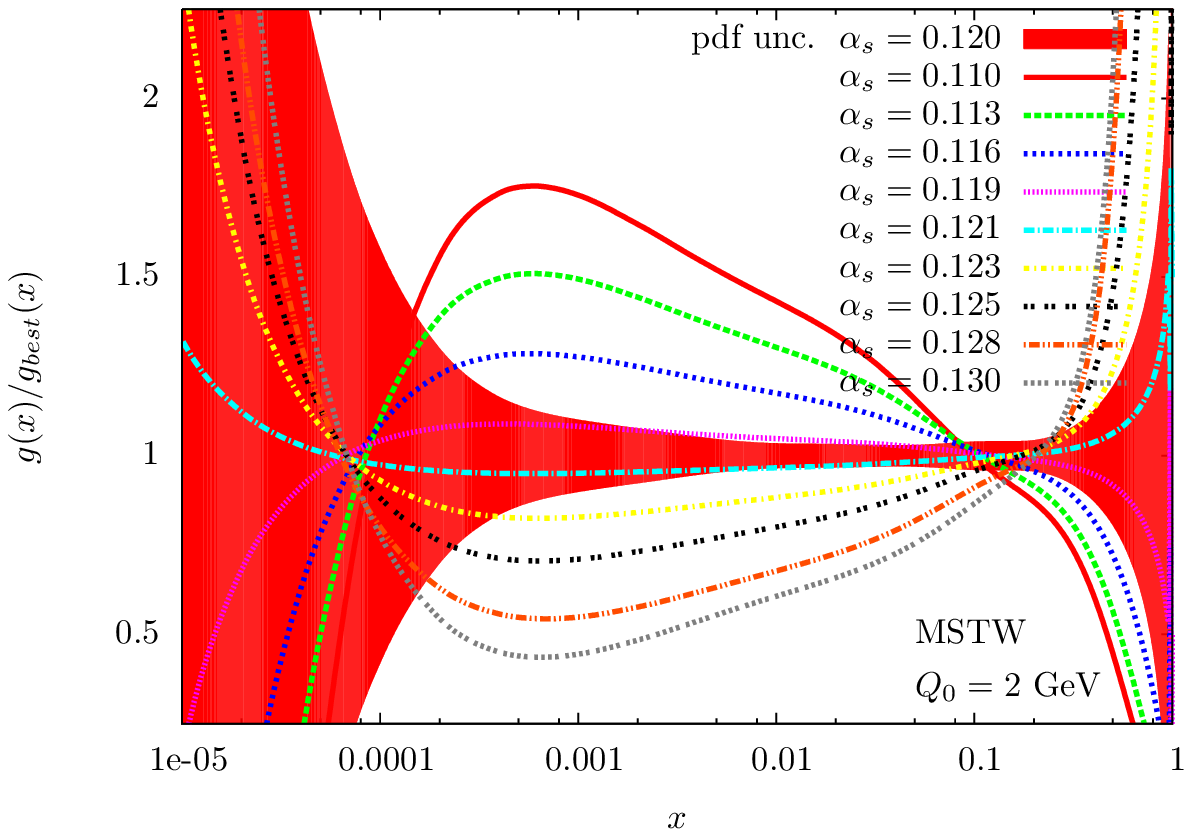}
\includegraphics[width=.32\linewidth]{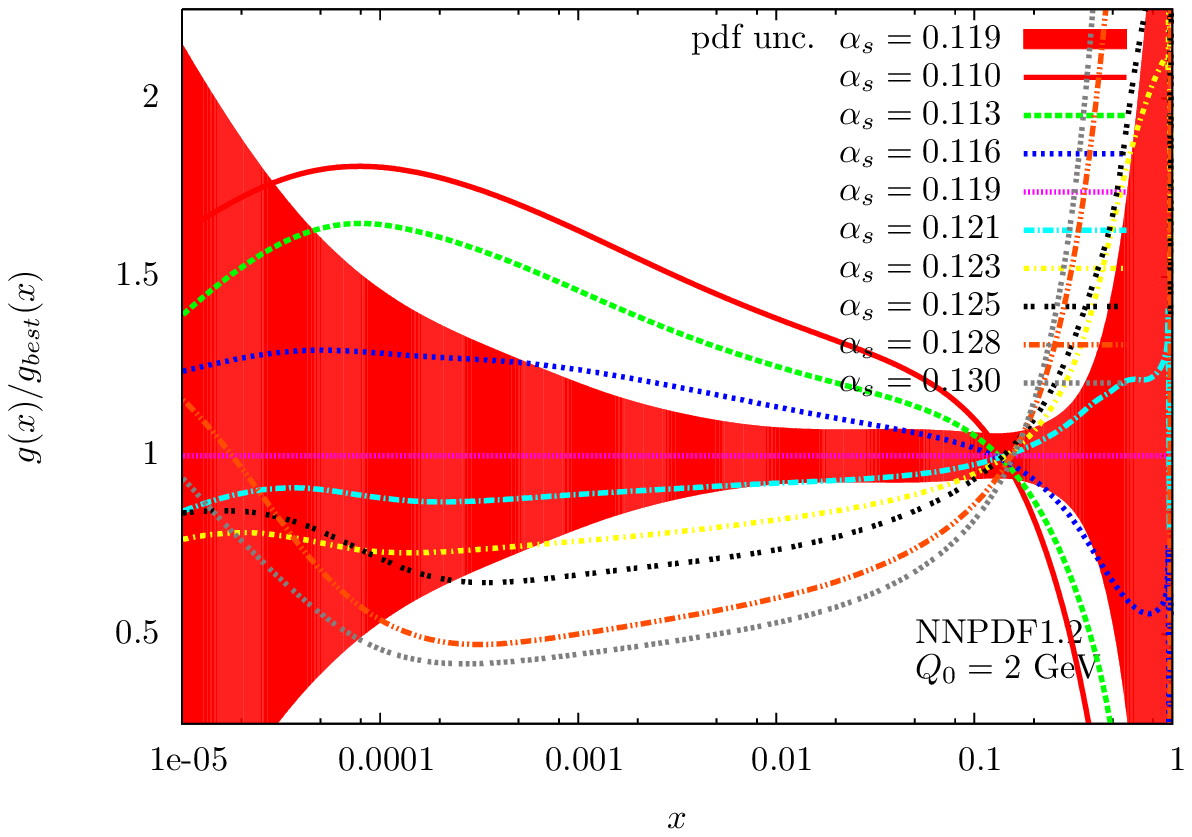}
\caption{Dependence on $\alpha_s$ of the gluon distribution determined
  in the fits of Ref.~\cite{Nadolsky:2008zw} (CTEQ6.6, left), Ref.~\cite{Martin:2009iq,Martin:2009bu} (MSTW08,
  center) and Ref.~\cite{Ball:2009mk} (NNPDF1.2, right) (from Ref.~\cite{Demartin:2010er}).} 
\label{glucalcorr}
\end{center}
\end{figure}
The most obvious source of theoretical uncertainty is the value of the
strong coupling $\alpha_s$. The PDF which depends most strongly on it
is the gluon distribution, which, as discussed in Sect.~\ref{glusec}
 is largely determined by scaling violations: the rather strong
 dependence of the gluon on the value of $\alpha_s$ is shown in 
 Fig.~\ref{glucalcorr} for various PDF sets. Note that even
 though, as discussed in Sect.~\ref{sechessian} the total
 PDF+$\alpha_s$ uncertainty can be obtained by determining these two
 uncertainties separately and adding results in
 quadrature~\cite{Lai:2010nw},  when determining the $\alpha_s$
 uncertainty,  the value of $\alpha_s$ in the
 factorization formula Eq.~(\ref{hadrfact})
must be varied both in the PDFs
 and in the partonic cross section $\hat\sigma$. This is especially
 important when dealing with processes, such as Higgs production in
 gluon-gluon fusion~\cite{Demartin:2010er,higgsYR} (or also top production)
which depend on the gluon PDF and start at a high order in
$\alpha_s$. For this purpose, PDF sets corresponding to different
values of $\alpha_s$ are necessary and have thus been produced by
several group (using sets with PDFs given 
as a continuous function of $\alpha_s$  is in principle also possible,
but practically more cumbersome and less accurate).

\begin{figure}\begin{center}
\includegraphics[width=.6\linewidth]{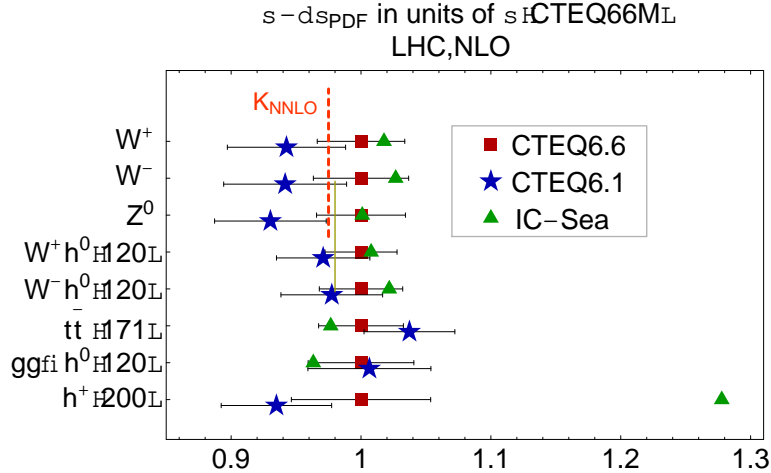}
\caption{The total $W$ and $Z$ cross section at the LHC with $\sqrt
  s=14$~TeV computed using PDFs determined including 
charm mass corrections (red squares~\cite{Nadolsky:2008zw})
  and neglecting them (blue stars~\cite{Stump:2003yu}). Other
  processes and PDF sets not discussed here  are also shown 
(from Ref.~\cite{Nadolsky:2008zw}). All uncertainty bands shown are at
90\% c.l..} 
\label{hqpheno}
\end{center}
\end{figure}
The only other free parameters in the QCD Lagrangian are the quark
masses, i.e., in the perturbative regime, the heavy quark masses. 
Dependence of PDFs on them are larger than one might naively expect,
and has two different origins which we will now discuss in turn.
The first, is simply the fact that even though all perturbative
computation are done up to power-suppressed terms in $Q^2$, terms of
order $\frac{m_c^2}{Q^2}$ and $\frac{m_b^2}{Q^2}$ may have a
non-negligible impact on PDF fits. This was brought to general attention
by the comparison~\cite{Nadolsky:2008zw}
of two calculations of the $W$ and $Z$ production cross-sections based on
PDF sets  which differ mostly because one does
include  $\frac{m_c^2}{Q^2}$ corrections
(CTEQ6.1~\cite{Stump:2003yu}) while the other (CTEQ6.6~\cite{Nadolsky:2008zw})
does not. It turns out (see
Fig.~\ref{hqpheno})  that
these corrections change the result
 by an amount which is almost twice the (statistical) PDF
uncertainty (as defined in Sect.~\ref{PDFunc}): note that all the
bands in Fig.~(\ref{hqpheno}) are at 90\%c.l., and the one-$\sigma$
intervals would be accordingly smaller.

In order to understand what is going on here, one must recall the way
heavy quark PDFs are defined, and heavy quarks are treated in
perturbative QCD computations. Usually, perturbative QCD computations
are performed in a decoupling renormalization
scheme~\cite{Collins:1978wz}, in which heavy quarks decouple from
perturbative Feynman diagrams for
scales much lower than the quark mass. In such a scheme, below
threshold the number of flavours in the evolution equation for the
strong coupling and for parton distributions Eq.~(\ref{dglap})
is equal to the number of light flavours. So in particular 
whereas there may still exist a
non-perturbative  nonvanishing ``intrinsic'' heavy quark PDF below
threshold~\cite{Brodsky:1980pb}, it will only start evolving and coupling to other PDFs
through evolution equations above threshold: for all practical
purposes, below charm threshold $n_f=3$.
However, for scales which are much
larger that the heavy quark mass, there is no reason to treat the
heavy quark on a different footing from any other light quark: for
scales $Q^2$ much larger than the charm mass, all terms of order
$\frac{m_c^2} {Q^2}$ can be neglected, and there are $n_f=4$ massless
flavours. The problem arises for scales which are higher but not much higher
than  the heavy quark mass: then, the heavy quark can be produced and it
does not decouple, yet it is not necessarily a good approximation to treat
it as massless, i.e. to neglect $\frac{m_c^2} {Q^2}$ corrections.

This situation is illustrated in Fig.~\ref{hqresum}, where we show the
neutral-current photon-induced DIS
$F_2^c$ structure function (i.e. the contribution to $F_2$ in which
the virtual photon couples to a charm quark) computed in various
approximations. The two curves labelled ZM-VFN (purple, highest
curves) are curves in which charm is simply treated as another
massless flavour. The charm PDF is assumed to vanish below threshold,
and above threshold a (massless)  charm component is generated by
perturbative evolution --- note that even if an intrinsic component did
exist, it should be small, and only non-negligible for very large
$x$~\cite{Brodsky:1980pb}.
The two NLO and NNLO ZM-VFN (zero mass-variable flavour number)
curves correspond to the case in which
anomalous dimensions are computed up to $O(\alpha_s^2)$ and
$O(\alpha_s^3)$ respectively: the solution to evolution equations
then includes all
contributions of order
$\alpha_s^k \ln^n\frac{Q^2}{\mu^2}$ to all orders in $\alpha_s$, with
$n\ge k-1$ at NLO and with $n\ge k-2$ at NNLO. They are called
``variable flavour number'' curves because the number of active flavours
is increased by one unit when each heavy quark threshold is crossed.
For
light quarks, $\mu^2=Q^2_0$ --- the starting scale of perturbative
evolution --- and for the
charm contribution $\mu^2=m_c^2$. Given that $Q_0\sim m_c$, at high
scale there is
no reason not to include  the charm contribution in evolution
equations along with
the light contributions. 
\begin{figure}\begin{center}
\includegraphics[width=.4\linewidth]{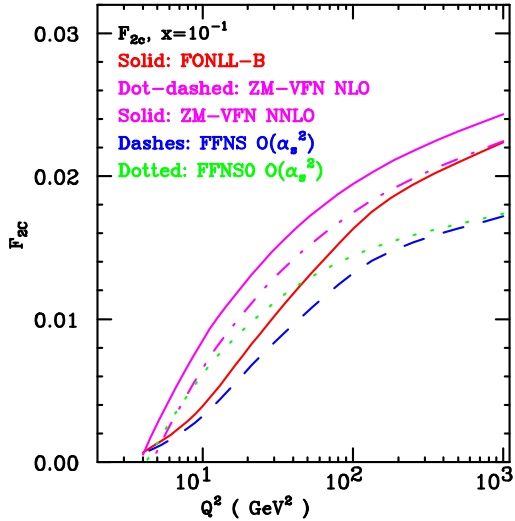}
\caption{The deep-inelastic charm structure function $F_{2c}$ computed
  in various approximations (see text), plotted as a function of scale for
  fixed $x=10^{-1}$. The same conventional PDFs are used for all
plots.} 
\label{hqresum}
\end{center}
\end{figure}

However, the evolution
equations neglect all quark mass effects. Indeed, the curve
labelled FFN (fixed flavour number) $O(\alpha^2_s)$ shows the result
of the computation
fixed order in $\alpha_s$, but now with the dependence on $m_c$ fully
included. This is a fixed-flavour number result because charm is
included as a massive quark in partonic cross sections, but only the
lighter flavours contribute to evolution equations and the running of
$\alpha_s$. 
Its limit for $Q^2\to\infty$, which  coincides with the
contributions to the VF-ZFN NNLO, but up to $O(\alpha^2_s)$ only, is
labelled as FFN$0$. The FFN$0$ and FFN results are different,  
and their difference, which is
sizable for $Q^2 \lsim 50$~GeV$^2$ is a measure of the size of mass
suppressed contributions (note however that all curves come together
at threshold, where $F_{2c}$ vanishes). In this region,
$ \ln\frac{Q^2}{m_c^2}$ is not large and indeed the ZM-VFN NNLO curve and
the FFN$0$ are quite close: the inclusion of higher order
powers of $\ln\frac{Q^2}{m_c^2}$ in the ZM-VFN NNLO has little effect. 
 On the other hand, for
$Q^2\gsim100$~GeV$^2$ the FFN$0$ and FFN curve become quite close ---
their difference are mass-suppressed contributions which here 
have little effect --- but the ZM-VFN curve is much higher: here
$\ln\frac{Q^2}{m_c^2}$ is large and its all-order inclusion in the ZM-VFN
result is important. In fact, the difference between  the FFN curve
and the ZM-VFN curve is much larger than the difference between the
NLO and NNLO ZM-VFN curves: it is important to resum
$\ln\frac{Q^2}{m_c^2}$ to all orders, but whether this resummation is done
to NLO or NNLO has a much smaller impact.

So summarizing: for lower $Q^2 \lsim 50$~GeV$^2$ it is important to
include mass corrections and not important to resum $\ln\frac{Q^2}{m_c^2}$
to all orders, so the most accurate result is the FFN one. For $Q^2
\lsim 50$~GeV$^2$ the converse is true and the most accurate result is
the ZM-VFN one. The question is whether the two can be combined. That
this is possible in principle to any perturbative order is a
consequence of a
factorization theorem for massive quarks proven to all orders
in Ref.~\cite{Collins:1998rz}. A practical implementation was
suggested and worked out up to NLO in Ref.~\cite{Aivazis:1993pi} - the
so-called ACOT method, which was used to produce the massive CTEQ6.6
result of Fig.~\ref{hqpheno}. Other implementations were suggested in
Refs.~\cite{Thorne:1997ga,Thorne:2006qt} (TR method) for
deep-inelastic scattering, and in Ref.~\cite{Cacciari:1998it} for
hadroproduction (FONLL method, recently generalized to DIS and worked
out up to NNLO in
Ref.~\cite{Forte:2010ta}). These various methods, which thus combine
both a massive fixed-order FFN computation and a massless all-order
ZM-VFN resummation, are often referred to as  ``GM-VFN'' (general
mass, variable flavour number) methods.

\begin{figure}
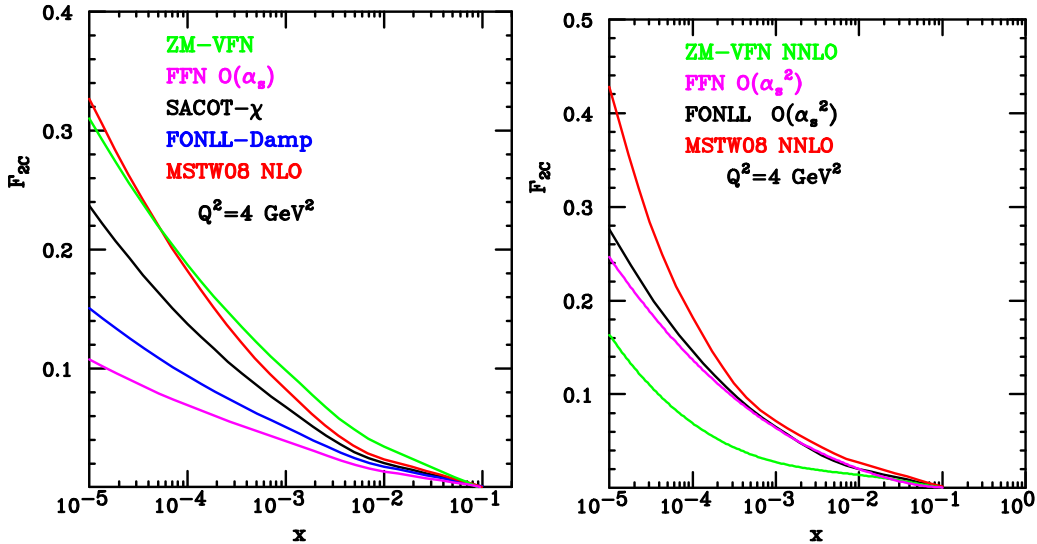
\begin{center}
\includegraphics[width=.4\linewidth]{f2c1.ps}
\includegraphics[width=.4\linewidth]{f2cnn1.ps}
\caption{Comparison of FFN, ZM-VFN and GM-VFN computations of the 
deep-inelastic charm structure function $F_{2c}$: left, FFN
$O(\alpha_s)$ and ZM-VFN NLO; right,  FFN
$O(\alpha_s)$ and ZM-VFN NNLO. In each case the various GM-VFN combine
all terms in the FFN and ZM-VFN results, and only differ by subleading
terms. Results are shown as a function of $x$, at a scale little
higher than the threshold. The same conventional PDFs are used for all
plots.} 
\label{hqschcomp}
\end{center}
\end{figure}
The FONLL curve is also shown in
Fig.~\ref{hqresum}, in a version (called FONLL-B) which includes all
terms which are contained in the ZM-VFN NLO and FFN $O(\alpha_s^2)$
calculations. It is clear that the FONLL-B curve nicely interpolates between
the FFN curve, more accurate at low scale, and the ZM-VFN one, more
accurate at high scale. Without entering a discussion of these various
prescriptions, which would be quite technical, it is important to
understand that (unless an error is made) all  GM-VFN prescriptions include
all the terms included in the ZM-VFN and FFN calculations:  they may
differ first, in
the orders at which either of the ZM-VFN and FFN terms are included,
and furthermore, because even with the same terms included,
subleading terms are generally different, and may in practice have a
non-negligible impact. The impact of these subleading terms can only
be reduced by pushing to higher orders both the ZM-VFN and FFN
contributions which are included in the GM result. This is illustrated
in Fig.~\ref{hqschcomp}, where the FFN, ZM-VFN and GM-VFN results are
compared. The GM-VFN are shown in the version adopted by CTEQ6.6
Ref.~\cite{Nadolsky:2008zw} 
of the ACOT method of Ref.~\cite{Aivazis:1993pi} (only available at
NLO),  in the version
adopted by MSTW08~\cite{Martin:2009iq} of the TR method of
Refs.~\cite{Thorne:1997ga,Thorne:2006qt}, and with the FONLL method of
Ref.~\cite{Forte:2010ta}. It is clear that the differences between the
various GM schemes are sizable, and only start decreasing at
NNLO-$O(\alpha_s^2)$. 

Considerable progress has been made recently in the benchmarking of
all these GM schemes (see Ref.~\cite{Binoth:2010ra}), and the use of a
GM-VFN scheme, which is highly desirable, has become more
widespread. However, a systematic estimate of the uncertainties related to the
choice of specific heavy quark scheme is not available in  existing
PDF sets.

\begin{figure}\begin{center}
\includegraphics[width=.45\linewidth]{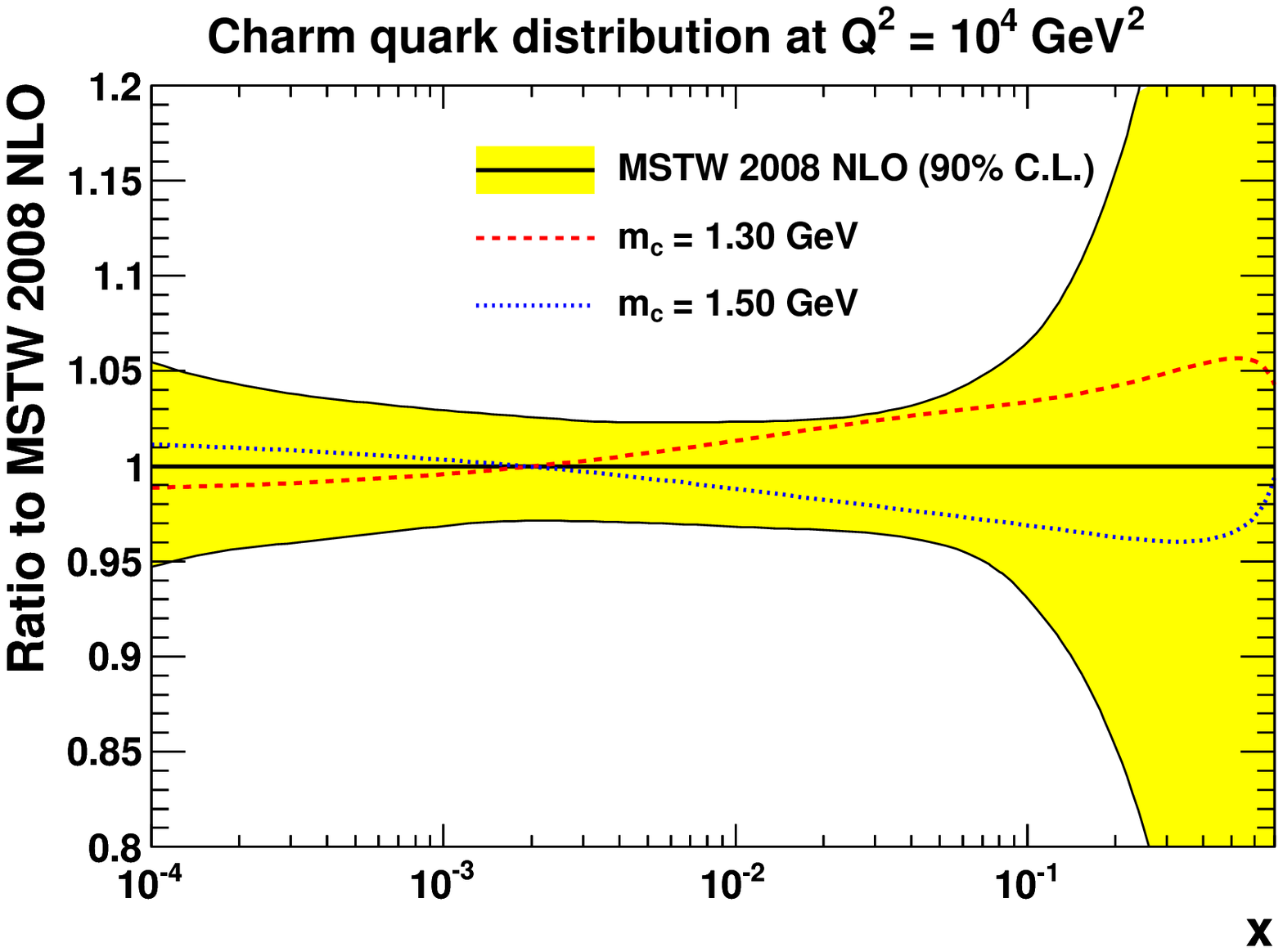}
\includegraphics[width=.45\linewidth]{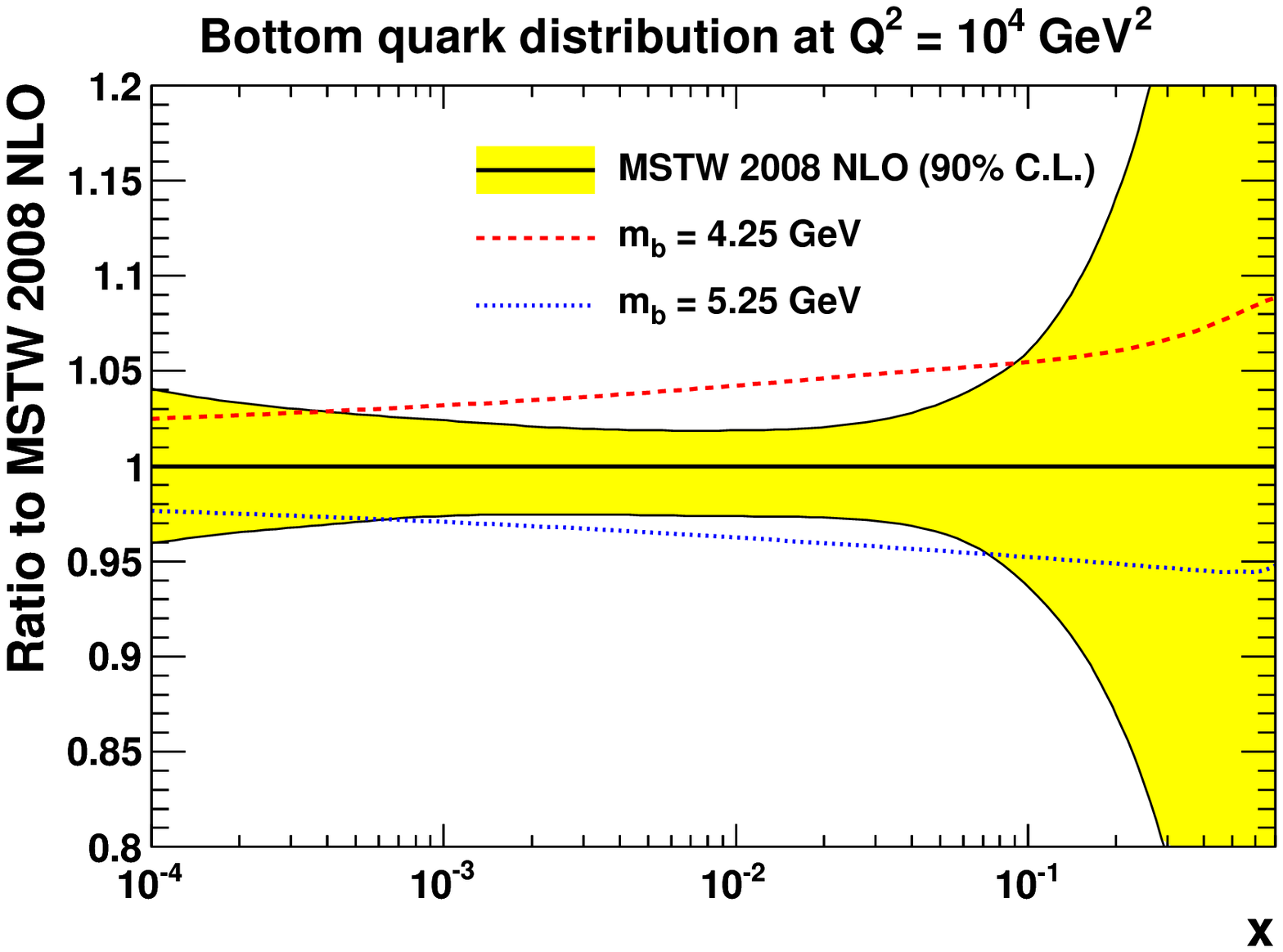}
\caption{Dependence of the heavy quark PDFs at a typical electroweak
  scale on the heavy quark mass: left, charm;
  right, bottom (from Ref.~\cite{Martin:2010db}).} 
\label{hqmdep}
\end{center}
\end{figure}
So far we have only discussed one of the two ambiguities related to
the treatment of heavy quarks. The other one has simply to do  with the
value of the heavy quark mass. This has a considerable impact because,
as we have seen, apart from possible intrinsic contributions, heavy
quark PDFs are obtained by assuming them to vanish at threshold, and
then to be generated by perturbative evolution. But changing the mass
changes the position of the threshold, and thus the amount of
evolution. This has been very recently argued to have a potentially
non-negligible impact on phenomenology~\cite{CooperSarkar:2010ik}. First
PDF sets with variable values of the heavy quark masses have been
presented in Ref.~\cite{Martin:2010db}. Some
 representative results are shown in Fig~\ref{hqmdep}: 
when the heavy quark masses are varied in a range which is
 representative of their uncertainty, the heavy PDFs vary by an amount
 (more than 5\%) which is of the same order or larger than the PDF
 uncertainty. This variation then propagates onto all other PDFs and
 especially the gluon, both due to mixing upon perturbative
 evolution, and to sum rules.

It seems clear that for accurate phenomenology the values of the heavy
quark masses will have to be treated analogously to what is now being
done for the strong coupling: PDF sets with varying
heavy quark masses will have to be provided, and the value of the
masses will have to be varied simultaneously in the partonic cross section
and in the PDF sets.

Finally, it should be recalled that there is at present no reliable
estimate of the effect on PDFs of the uncertainty due to the
truncation of the perturbative expansion. This is possibly a
relatively smaller
effect in comparison to those we discussed so far, but this too will
have to be included in PDF sets, for example by providing sets which
correspond to different values of the renormalization and
factorization scales, whose variation is a way of estimating unknown
higher order corrections.
All in all, a proper treatment of theoretical uncertainty is the
current frontier in PDF uncertainties.

\subsection{PDFs for the LHC}
\label{lhcpd}

\begin{figure}\begin{center}
\includegraphics[width=.45\linewidth,clip]{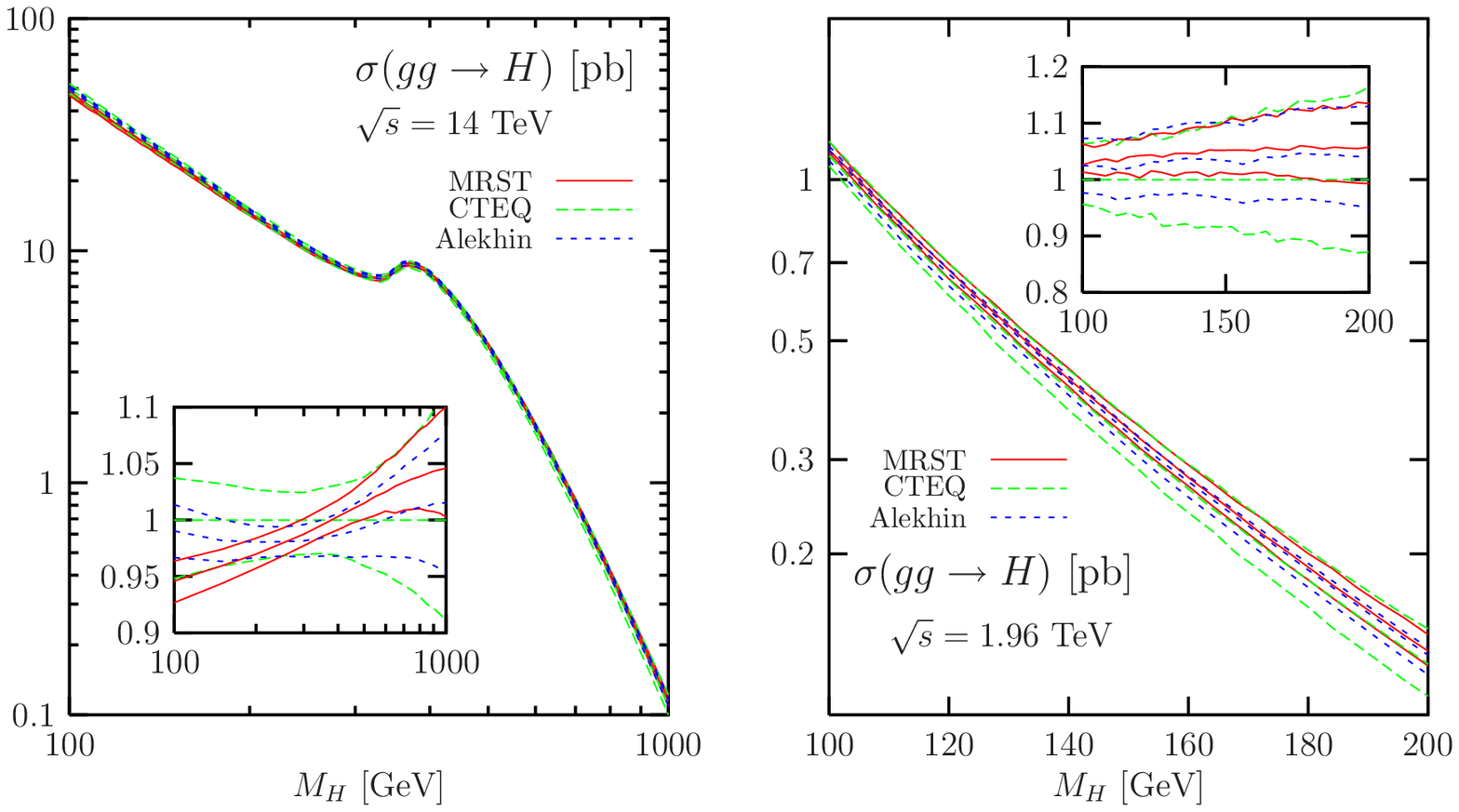}
\includegraphics[width=.45\linewidth]{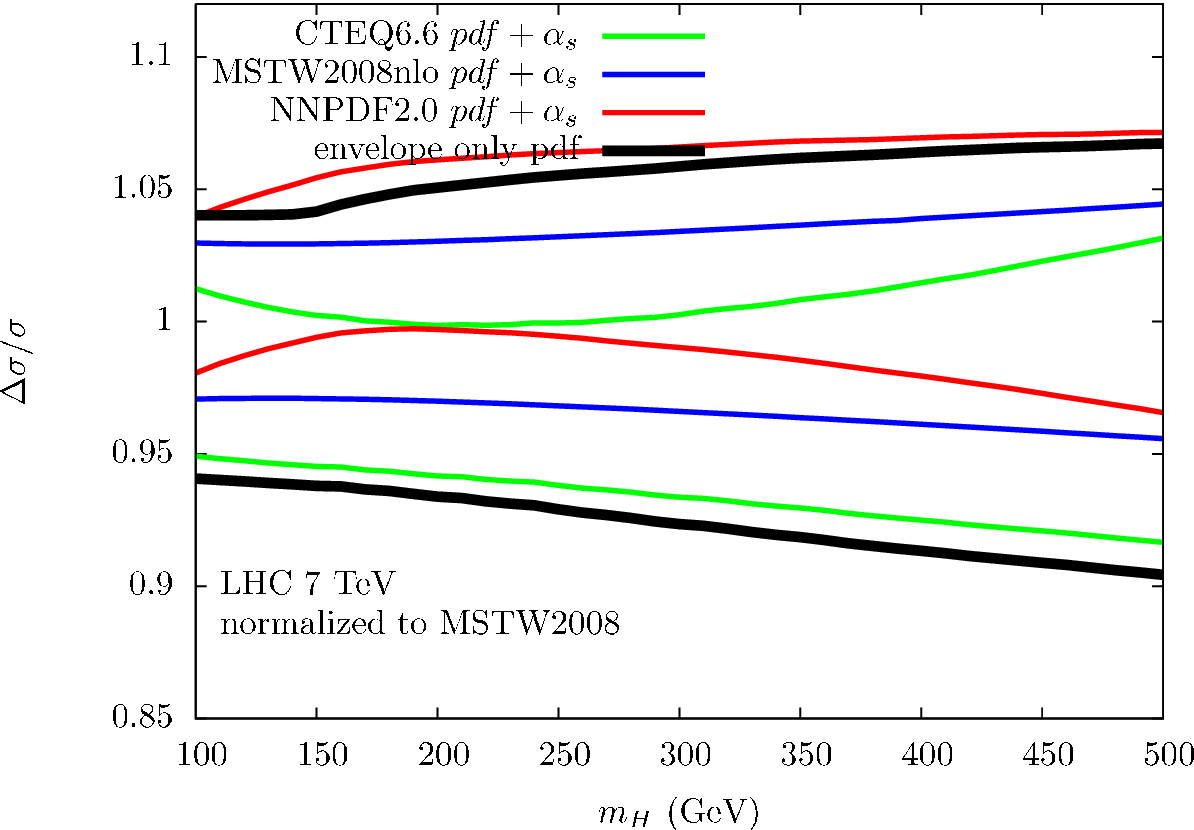}
\caption{Comparison of the NLO Higgs cross section at the LHC as a
  function of the Higgs mass computed using various PDF sets. Left:
  status 2004 (from Ref.~\cite{Djouadi:2003jg}) using
  Alekhin2002~\cite{Alekhin:2002fv}, CTEQ6~\cite{Pumplin:2002vw} and
  MRST2001E~\cite{Martin:2002aw} PDFs ($\sqrt{s}=14$~TeV). Right: status 2010 (from
  Ref.~\cite{Demartin:2010er}) using
  CTEQ6.6~\cite{Nadolsky:2008zw}, 
  MSTW2008~\cite{Martin:2009iq} and NNPDF2.0~\cite{Ball:2008by} 
PDFs ($\sqrt{s}=7$~TeV).} 
\label{higgsxs}
\end{center}
\end{figure}
An ideal PDF determination should include all of the following
features:
\begin{itemize}
\item It should be based on a dataset which is as wide as possible in
  order to ensure that all relevant experimental information is
  retained; in particular, all processes discussed in Sect.~\ref{data}
  should be used.
\item It should be based on a sufficiently general and unbiased parton
  parametrization and/or  it should include a careful
  estimate of the effect of varying the parton parametrization.
\item It should provide PDF uncertainty bands which have been either a
  priori (tolerance) or a posterior (Monte Carlo) checked
  to provide consistently-sized confidence levels for individual
  experiments.
\item It should include heavy quark mass effects through a GM-VFN
  scheme, and provide an estimate of the uncertainties due to
  subleading terms not included in the scheme which has been adopted.
\item It should be based on computations 
performed at the highest available perturbative
  order, namely NNLO for evolution equations and for most or all of
  the processes used for PDF determination.
\item It should provide PDFs for a variety of values of $\alpha_s$,
  reasonably thinly spaced and in a range which is representative of
  the uncertainty on this parameter.
\item {\it Ditto} for the values of heavy quark masses.
\item It should include an estimate of uncertainties related to the
  truncation of the perturbative expansion.
\end{itemize}

At present, there exists no PDF determination which has all these
features simultaneously. Which of these features is most important for
accurate results it will be possible to say with certainty only after
such a PDF determination is constructed. However, the various
features have been listed in the approximate (decreasing) likely order
of importance, at least in the opinion of this author, based on the
arguments presented so far.

Therefore, existing sets satisfy only some of these
requirements. Current PDF sets are provides through a standard
interface, LHAPDF~\cite{Bourilkov:2006cj}, which is regularly updated
for the inclusion of new sets and updates.
Current PDF sets and their salient features include the following
(listed in order of decreasing number of datasets included)
\begin{itemize}
\item {\bf MSTW08}~\cite{Martin:2009iq,Martin:2009bu,Martin:2010db} 
Latest in the MRS-MRST-MSTW series of
  fits (Ref.~\cite{Martin:1987vw} and subsequent papers). 
All data of Sect.~\ref{data}, plus HERA DIS
  jets. Hessian approach with parametrization Eq.~(\ref{pdfparmgen})
  for seven independent PDFs (three lightest flavour and antiflavours
  and the gluon);
  28 free parameters, 8 of which are held fixed in the
  determination of uncertainties; dynamical tolerance uncertainties. GM-VFN scheme. Results available for
  various values of $\alpha_s$, $m_b$ and $m_c$. NNLO perturbative
  order (whenever available).
\item {\bf CT10}~\cite{Lai:2010vv} 
Latest in the CTEQ series of fits (Ref.~\cite{Botts:1992yi} and
  subsequent papers, see also ~\cite{Morfin:1990ck}). All data of Sect.~\ref{data}. 
 Hessian approach with parametrization Eq.~(\ref{pdfparmgen}) for six
independent PDFs (two lightest flavour and antiflavours, total
strangeness and the gluon);
  26 free parameters; dynamical tolerance uncertainties. GM-VFN scheme. Results available for
  various values of $\alpha_s$. NLO perturbative
  order.
\item {\bf NNPDF2.0}~\cite{Ball:2010de} 
Latest in the NNPDF series of fits (Ref.~\cite{DelDebbio:2007ee} and
subsequent papers). All data of Sect.~\ref{data}. 
 Monte Carlo  approach with neural network parametrization for seven
independent PDFs (the three lightest flavours and antiflavours, total
strangeness and the gluon);
   259 ($37\times 7$) free parameters; cross-validation
   uncertainties. 
ZM-VFN scheme. Results available for
  various values of $\alpha_s$. NLO perturbative
  order.
\item {\bf JR}~\cite{JimenezDelgado:2008hf} 
Latest in the GR-GRV-GJR series of fits (Ref.~\cite{Gluck:1977ah} and subsequent papers). All data of Sect.~\ref{data} except $W$ and $Z$ production. 
 Hessian approach with parametrization Eq.~(\ref{pdfparmgen}) for five
independent PDFs (two lightest flavour and antiflavours and the gluon);
  20 free parameters; fixed tolerance uncertainties. Results available for
  single value of $\alpha_s$, but Hessian includes $\alpha_s$. FFN scheme.  
NNLO perturbative
  order.
\item {\bf ABKM}~\cite{Alekhin:2009ni} Latest in the Alekhin series of
  fits (Ref.~\cite{Alekhin:1996za} and subsequent papers).
All DIS data of Sect.~\ref{data} and fixed-target
  virtual photon Drell-Yan  production. 
 Hessian approach with parametrization Eq.~(\ref{pdfparmgen}) for six
independent PDFs (two lightest flavour and antiflavours, total
strangeness 
and the gluon);
  21 free parameters; no tolerance ($\Delta\chi^2=1$). Results available for
  single value of $\alpha_s$, but Hessian includes $\alpha_s$, and
  also heavy quark masses. FFN scheme.  
NNLO perturbative
  order.
\item {\bf HERAPDF1.0}~\cite{:2009wt} HERA only  DIS data 
 Hessian approach with parametrization Eq.~(\ref{pdfparmgen}) for five 
independent PDFs (two lightest flavour and antiflavours,
and the gluon);
  10 free parameters; no tolerance($\Delta\chi^2=1$)
 but inclusion of parametrization
  uncertainties.  GM-VFN scheme. Results available for
  various values of $\alpha_s$. 
NNLO perturbative
  order.
\end{itemize}

Detailed comparisons of these PDF sets is  the subject of
ongoing benchmarking exercises, with the aim of arriving at the most
accurate common determination of PDFs. The successfulness of the
enterprise can be gauged by putting side by side (see Fig.~\ref{higgsxs})
predictions
for  Higgs production via gluon-gluon fusion
at the LHC  with different PDF sets obtained as the first PDF with
uncertainties were published~\cite{Djouadi:2003jg}, with those
obtained more recently as first collisions were taking place at the
LHC~\cite{Demartin:2010er} (see Fig.~\ref{higgsxs}). 
The improvement is quite clear, and
convergence between PDF sets has further improved since. 

\begin{figure}[t]\begin{center}
\includegraphics[width=.4\linewidth]{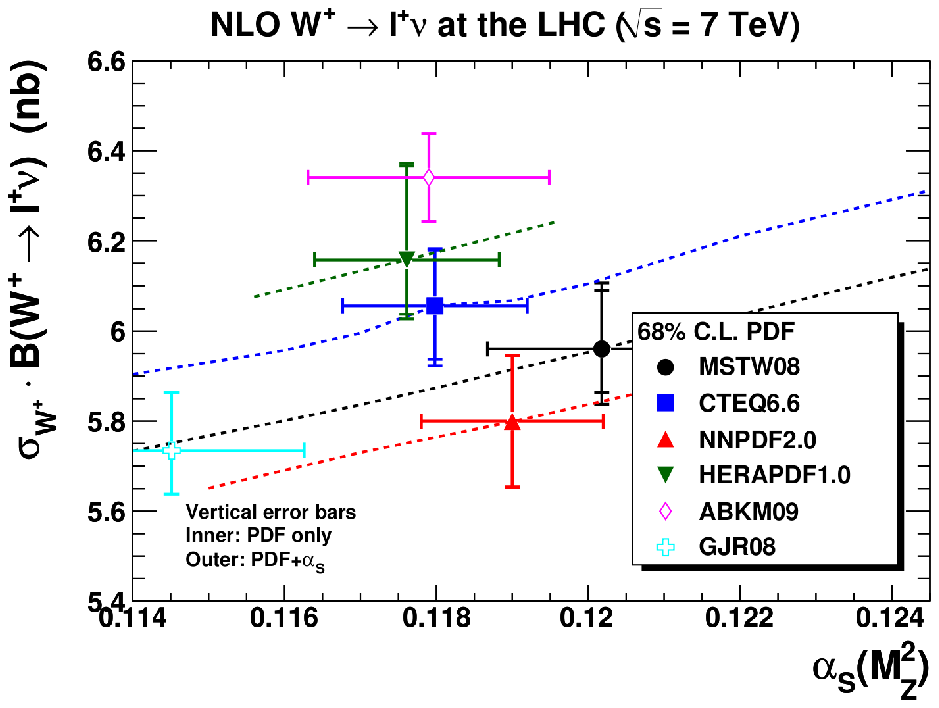}
\includegraphics[width=.4\linewidth]{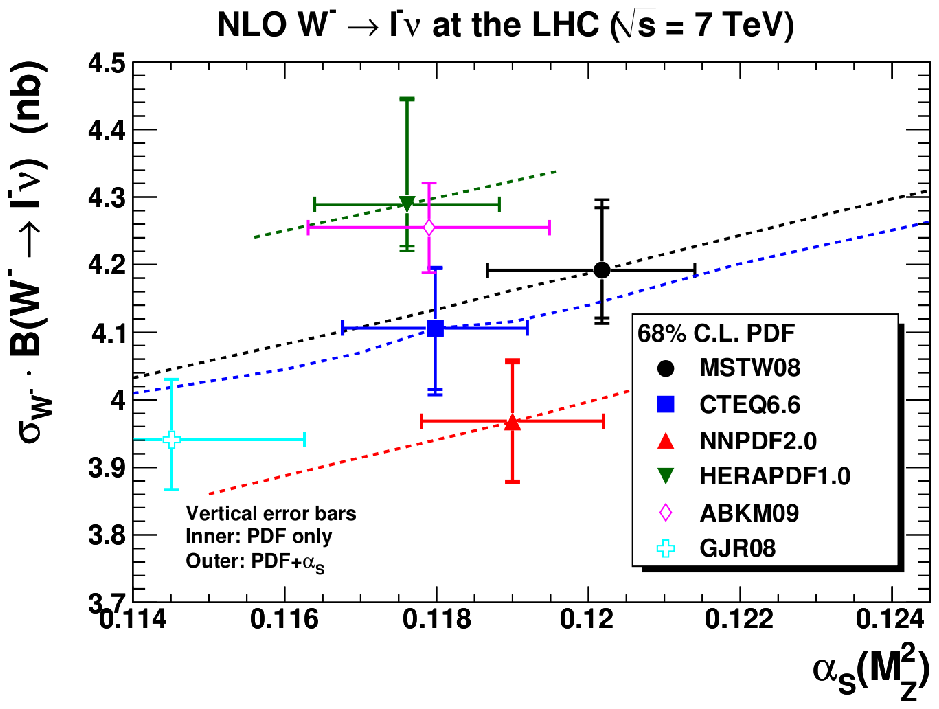}\\
\includegraphics[width=.4\linewidth]{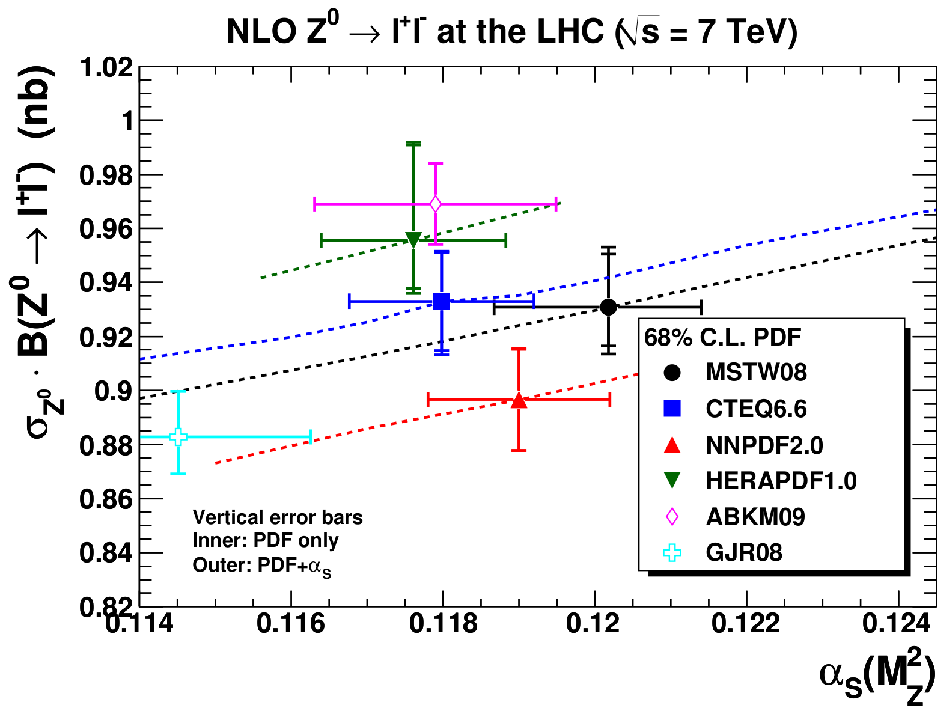}
\includegraphics[width=.4\linewidth]{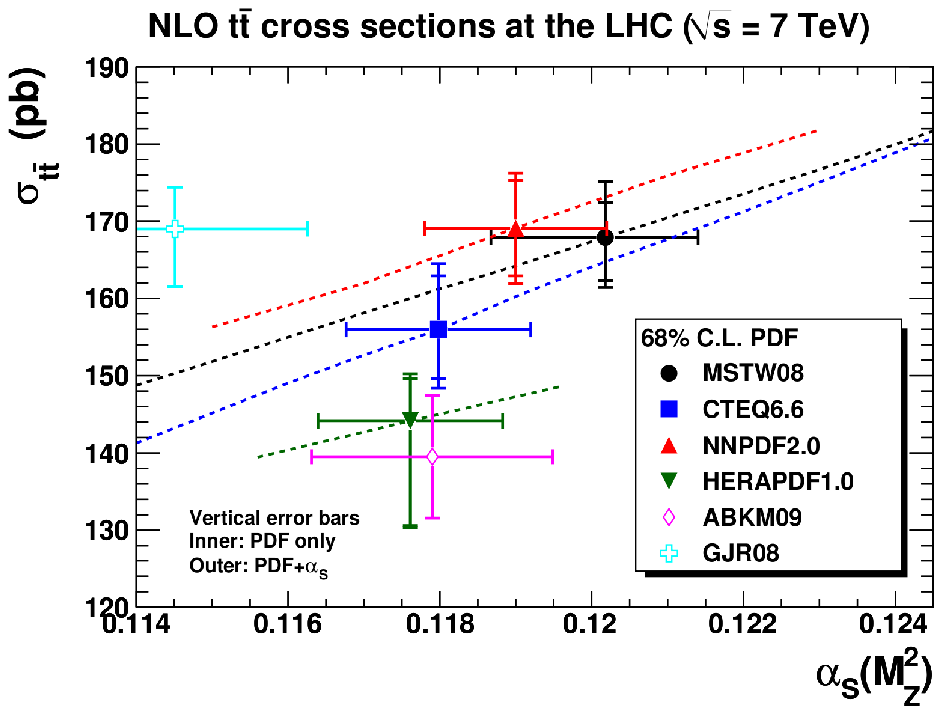}
\caption{LHC ``standard candles'' computed using various PDF sets:
  total cross section  for the production of $W^+$ (top left), $W^-$
  (top right), $Z$ (bottom left) $t\bar t$ (bottom right)
  (from Ref.~\cite{Watt}).} 
\label{lhcstcand}
\end{center}
\end{figure}
The status of the computation of the simplest LHC processes (which
have been suggested as ``standard candles'', e.g. as a means to
measure the machine luminosity~\cite{Anderson}) is summarized in the
plots of Fig.~\ref{lhcstcand}, where predictions for $W^\pm$, $Z$ and
top total cross sections obtained at NLO using the ABKM09~\cite{Alekhin:2009ni}, CTEQ6.6~\cite{Nadolsky:2008zw},
HERAPDF1.0~\cite{:2009wt}, GJR08~\cite{Gluck:2007ck}, MSTW08~\cite{Martin:2009iq}
and NNPDF2.0~\cite{Ball:2010de} sets are plotted as a
function of $\alpha_s$. The dotted lines show how each prediction
can be extrapolated to different values of $\alpha_s$ (the
extrapolation is not available for the ABKM and GJR sets since these
are only published for a single value of $\alpha_s$, though in
principle the information on the $\alpha_s$ dependence
is contained in their covariance matrix). It is clear that first, once
brought to a common value of $\alpha_s$ the various sets are in fair
agreement, and second, the agreement further improves when restricted
to PDF sets that are based on more similar assumptions (such as common
dataset, number of independent PDFs etc.).

An interesting question that remains is how one can proceed when some
disagreement remains, and there is no clear reason to favor one set
over the other. In this case, Bayesian statistics provides an
answer~\cite{D'Agostini:1995fv}: 
in Bayesian terms, the probability distribution for PDFs is a distribution of
true values, i.e. $P(f)$  expresses the degree of belief that the true
value is indeed $f$. Then, given two different, but a priori equally reliable
determinations  $P_1(f)$ and $P_2(f)$ of the probability distribution,
the combined probability is just $P(f)=\half( P_1(f) +P_2(f))$, with
obvious generalizations if the determinations of the probability
distribution are more than two or not all equally likely. In a Monte
Carlo approach this is especially easy to implement: the combined
probability distribution is obtained by simply taking a Monte Carlo
sample in which half of the replicas come from either distribution.
A
68\%c.l. of the combined probability is then simply the region which
contains the central 68\% of all the given distributions, i.e. 68\% of
the combined replica set. 

In practice, a reasonable approximation to the Bayesian estimate may well
consist of simply taking the envelope (i.e. the union) 
of the 68\% intervals
of the probability distributions which are being combined:
this typically leads to a slight overestimate of the error band,
because some of the replicas in the outer 32\% of one of the
distribution may fall within the central 68\% of another distribution,
but this in practice is often a small correction, hence the envelope
prescription can be a simple and effective way of combining
probability distributions.

\section{Conclusion}
\label{concl}

The physics of parton distributions is now close to a beginning,
rather than a conclusion: LHC data for many of the processes discussed
in Sect.~\ref{data} are being collected and will soon be
published. The availability of LHC data is likely to change
significantly our perspective on the subject. The kinematic range  of 
``old'' processes such as Drell-Yan will be extended and their
accuracy will improve. Processes which are at present not competitive
will become important
(such as perhaps prompt photon production). Entirely new processes
will play a role, such as Higgs production. Hopefully, classes of new
physical processes will be discovered: whatever their nature, they will
be observed in proton collisions, and thus  pose new challenges to
our understanding of the nucleon. It may turn out that an interplay
with new machines, such as an electron-proton LHeC
collider~\cite{Newman:2009mb},  or perhaps
even a neutrino factory~\cite{Mangano:2001mj} may be necessary in
order to exploit fully the LHC potential.  A review of this subject
written ten years from now is likely to be quite different from the
present one, and possibly much more interesting.

{\bf Acknowledgements:} I am grateful to the organizers of the
Zakopane school and especially Micha\l\ Prasza\l owicz for giving me
the privilege to present these lectures, and I thank all participants
and lecturers, in particular Paul Hoyer, for their interest and
critical input.
My understanding of this subject was shaped by
many discussions, especially within the PDF4LHC
workshop~\cite{deroeck}: I would like to thank in particular
A.~de~Roeck, A.~Glazov, J.~Huston, P.~Nadolsky, J.~Pumplin,
R.~Thorne. I also thank all the  members of the NNPDF collaboration
and especially J.~Rojo for innumerable discussions on the subject of
these lectures. This work was partly supported
by the European network HEPTOOLS under contract
MRTN-CT-2006-035505.

\end{document}